\documentclass[12pt]{article}
\usepackage{subfigure,graphicx,amsmath,euscript,amssymb,psfrag,epsfig}

\allowdisplaybreaks

\newcommand{\be}{\begin{equation}}
\newcommand{\ee}{\end{equation}}
\newcommand{\bea}{\begin{eqnarray}}
\newcommand{\eea}{\end{eqnarray}}
\newcommand{\nl}{\nonumber\\}
\newcommand{\order}{{\cal O}}

\newcommand{\hcn}{hc}
\newcommand{\hcnb}{\overline{hc}}
\newcommand{\sbr}{{\bar s}}

\newcommand{\muQCD}{\mu_{\mbox{\tiny{QCD}}}}

\newcommand{\sla}[1]{\rlap{\hspace{0.02cm}/}{#1}}
\newcommand{\ov}[1]{\overleftarrow{#1} }
\def\nb{\bar{n}}

\def\calAslash{\rlap{\hspace{0.08cm}/}{{\EuScript A}}}
\def\Dslash{\rlap{\hspace{0.07cm}/}{D}}
\def\nslash{\rlap{\hspace{0.02cm}/}{n}}
\def\nbslash{\rlap{\hspace{0.02cm}/}{\bar n}}

\def\Aslash{\rlap{\hspace{0.07cm}/}{A}}

\def\A{{\EuScript A}}
\def\H{{\EuScript H}}

\def\Q{{\EuScript Q}}
\def\X{{\EuScript X}}

\begin{document}

\begin{titlepage}

\begin{flushright}
CLNS~05/1913\\
FERMILAB-PUB-05-032-T \\
SLAC-PUB-11045\\
{\tt hep-ph/0503263}\\[0.2cm]
March 25, 2005 
\end{flushright}

\vspace{0.7cm}
\begin{center}
\Large\bf 
\boldmath Factorization in $B\to V\gamma$ Decays\unboldmath
\end{center}

\vspace{0.8cm}
\begin{center}
{\sc T.~Becher$^{(a)}$,  R.J.~Hill$^{(b)}$, and M.~Neubert$^{(c)}$ }\\
\vspace{0.7cm}
$^{(a)}${\sl Fermi National Accelerator Laboratory,\\
P.~O.~Box 500, Batavia, IL 60510, U.S.A.} \\
\vspace{0.3cm}
$^{(b)}${\sl Stanford Linear Accelerator Center, Stanford University\\
Stanford, CA 94309, U.S.A.} \\
\vspace{0.3cm}
$^{(c)}${\sl Institute for High-Energy Phenomenology\\
Newman Laboratory for Elementary-Particle Physics, Cornell University\\
Ithaca, NY 14853, U.S.A.}
\end{center}

\vspace{1.0cm}
\begin{abstract}
\vspace{0.2cm}\noindent 
The factorization properties of
the radiative decays $B\to V\gamma$ are analyzed at leading order in
$1/m_b$ using the soft-collinear effective theory. 
It is shown that the decay amplitudes can be expressed in terms of a
$B\to V$ form factor evaluated at $q^2=0$, light-cone distribution
amplitudes of the $B$ and $V$ mesons, and calculable hard-scattering
kernels.  The renormalization-group equations in the effective
theory are solved to resum perturbative logarithms of the different
scales in the decay process. Phenomenological implications for 
the $B\to K^*\gamma$ branching ratio, isospin asymmetry, and CP asymmetries 
are discussed, with particular emphasis on possible effects from physics
beyond the Standard Model. 
\end{abstract}
\vfil

\end{titlepage}

\section{Introduction}

In the Standard Model, the radiative $b\rightarrow s \gamma$
transition is suppressed and it is therefore a sensitive probe for the
effects of New Physics. The total $B\to X_s\gamma$ decay rate can be
calculated in an expansion about the heavy-quark limit using the
operator product expansion (OPE).  At leading order in the heavy-quark
expansion, the total rate can be calculated in perturbation theory and
it is therefore known rather precisely.

However, the OPE is only valid for sufficiently inclusive observables.
It cannot be used if the photon energy in the inclusive process is
restricted to the endpoint region, much less to analyze the exclusive
decay $B\rightarrow K^*\gamma$. Restricted to the region of large
photon energy, the $b\rightarrow s$ transition involves
non-perturbative strong-interaction physics, even in the heavy-quark
limit.  The factorization analysis retains predictive power by organizing 
these non-perturbative
contributions in a universal and process-independent manner. 
An efficient way to study these
decays in the heavy-quark expansion is to use soft-collinear
effective theory (SCET)
\cite{Bauer:2000ew,Bauer:2000yr,Bauer:2001ct,Chay:2002vy,Beneke:2002ph,Hill:2002vw}.
In this approach, the relevant momentum regions in the pertinent Feynman
diagrams are represented by fields in the effective theory, and the
expansion of the diagrams in momentum space translates into a
derivative expansion of the effective Lagrangian. The use of a
Lagrangian makes the structure of the interactions and the resulting
factorization properties of the amplitudes more transparent and allows
identification of the remaining non-perturbative parts of a given
process at the operator level. It provides a simple way to resum large
perturbative logarithms associated with the different scales in the
problem, by solving the renormalization-group (RG)
equations obeyed by the effective-theory operators. For inclusive
decay distributions in the endpoint region, SCET has been used to
obtain a factorization theorem for the decay at next-to-leading order
in the heavy-quark expansion \cite{Lee:2004ja,Bosch:2004cb,Beneke:2004in}, 
an analysis which seems prohibitively difficult on a purely diagrammatic 
level.

In the present paper, we use SCET to analyze the exclusive decay 
$B\to K^*\gamma$, or more 
generally $B\to V\gamma$, where $V$ is a light vector meson.  
We demonstrate that, at leading order in $1/m_b$ and
to all orders in $\alpha_s$, the matrix elements of the operators
$Q_i$ in the effective weak Hamiltonian governing the decay
obey the generalized factorization formula
\begin{equation}
\label{eq:factorization}
\langle V \gamma |\, Q_i\, | B \rangle 
=  T^{I}_i\, F^{B\to V_\perp} 
+ \int_0^\infty{d\omega\over\omega}\,\phi_{B}(\omega)\, \int_0^1\,du\, 
\phi_{V_\perp}(u)\,T^{II}_i(\omega,u)\,. 
\end{equation}
The quantities $T^{I}$ and $T^{II}$ are perturbatively calculable functions, 
appearing as Wilson coefficients of effective-theory operators.   
$F^{B\to V_\perp}$ is a form factor evaluated at maximum recoil ($q^2=0$), and
$\phi_{B}$, $\phi_{V_\perp}$ are light-cone distribution amplitudes
(LCDAs) for the heavy and light mesons, respectively.  

As is manifest from the factorization theorem
(\ref{eq:factorization}), the $B\to V$ form factors at large recoil
energy are an important ingredient in the analysis of rare exclusive
$B\to V\gamma$ decays. These form factors have been analyzed in the
effective theory in \cite{Bauer:2002aj,Beneke:2003pa,Lange:2003pk}. It
was found that the form factors in this energy regime contain a
non-factorizable piece, which however is independent of the Dirac
structure of the current in the heavy-quark limit. A single function
$\zeta_{V_\perp}(E)$ then suffices to describe all $B\rightarrow
V_\perp$ form factors up to factorizable corrections
\cite{Charles:1998dr,Beneke:2000wa}. The proof of the factorization
theorem (\ref{eq:factorization}) is achieved after showing that the
non-factorizable piece of the $B\to V\gamma$ decay amplitude is given
by the same function. We will show that diagrams in which the
photon is emitted from one of the current quarks have the same
structure as those encountered in the study of heavy-to-light form
factors.  Using the SCET formalism and the results from the
form-factor analysis, it is then straightforward to establish
(\ref{eq:factorization}) for these contributions.  Photon emission
from the $B$-meson spectator quark has a more complicated structure.
Diagrams of this type develop singularities for momentum configurations
where some of the quarks and gluons are collinear to the photon
(instead of being collinear to the meson $V$).  Such configurations do
not appear in the form-factor analysis, and to describe them it is
necessary to include additional collinear fields in the effective
theory.  We introduce a counting scheme to systematically list all
operators that may contribute at a given order in the power counting,
and show that at leading power the operator matrix elements obey
(\ref{eq:factorization}).  The matching of QCD onto the effective
theory is performed in two steps: at the hard scale $\mu\sim m_b$ the
operators are matched onto an intermediate theory, SCET$_{\rm I}$.
The matching of SCET$_{\rm I}$ onto the final effective theory,
SCET$_{\rm II}$, is then performed at the hard-collinear scale
$\mu\sim\sqrt{\Lambda m_b}$, where $\Lambda$ is a typical hadronic scale. 
The two-step matching procedure makes it
simpler to identify the operators appearing in SCET$_{\rm II}$ and to
separate the part proportional to the form factor from the remainder.
Such a two-step matching is also required in order to resum large
logarithms in perturbative expansions 
involving both the hard and the hard-collinear scales.  
The term $T^I$ in (\ref{eq:factorization}) involves only the hard scale, so  
that large logarithms are avoided by taking the scale $\mu\sim m_b$.   
The term $T^{II}$ however involves both the hard and hard-collinear scales. 
In this case no scale choice is possible that avoids all large logarithms,   
and we perform the necessary resummation for this term. 

Operators describing spectator emission in radiative $B$ decays appear
at leading power in the effective theory, but in the Standard Model,
the corresponding matrix elements between pseudoscalar $B$ mesons and
transversely-polarized vector mesons vanish.  Such operators would
contribute at leading power to the radiative decay $B^*\to P\gamma$ of
$B^*$ into light pseudoscalar mesons.  While this decay mode is not of
phenomenological importance, it is interesting to note that, as we
will show, the formula (\ref{eq:factorization}) extends without
essential modification to the general case of radiative $B$ or $B^{*}$
decay to a flavor non-singlet light meson $M$: 
\begin{equation}\label{eq:factorgeneral} 
\langle M \gamma |\, Q_i\, | B^{(*)} \rangle = T^{I}_i\, F^{B\to M} +
\int_0^\infty{d\omega\over\omega}\,\phi_{B}(\omega)\, \int_0^1\,du\,
\phi_{M}(u)\,T^{II}_i(\omega,u)\,.  
\end{equation}
Note that the form factors and heavy-quark LCDAs for the $B$ and
$B^*$ mesons are related in the heavy-quark limit. In the presence of New
Physics operators with a chirality structure different from those of
the Standard Model, the $B\rightarrow V \gamma$ decay amplitude
receives leading-power contributions associated with spectator photon
emission \cite{Kagan:2001zk}, with both left- and right-circular photon 
polarization.  We consider a general class of such
operators and show that they also obey the factorization formula
(\ref{eq:factorization}). Even at leading power in $1/m_b$, these  
operators break isospin symmetry, and they can give rise to a 
non-vanishing time-dependent CP asymmetry. Measurements of these asymmetries 
in exclusive radiative $B$ decays can thus provide useful constraints on the 
Wilson coefficients of the associated New Physics effective operators.     
For completeness, we consider also the case
of flavor-singlet final-state mesons.  Here we find new classes of
operators not appearing in the form-factor analysis.  These operators
have vanishing matrix elements between pseudoscalar $B$ mesons and
transversely-polarized final-state vector mesons, but would in
principle contribute to radiative $B^*$ decays to pseudoscalar final states.  
One class of new operators is non-factorizable.  Since these contributions 
cannot be related to form factors, we conclude that a factorization formula 
such as (\ref{eq:factorgeneral}) does not hold for flavor singlet 
$B^*\to P\gamma$ decays.  

For the case of $B\to K^*\gamma$, we assess the impact of strange-quark
mass effects on the predictions of the factorization theorem
(\ref{eq:factorization}).  We first demonstrate the theorem for the
massless case, and then consider the perturbation caused by a small
but finite strange-quark mass, $m_s$.  We argue that the leading
corrections, linear in $m_s/\Lambda$, simply contribute to the
universal LCDA $\phi_{K^*_\perp}$ and non-factorizable form
factor $\zeta_{K^*_\perp}$, breaking the SU(3) flavor symmetry that
one obtains for the purely massless case.  Possible corrections to the
form of the factorization formula (\ref{eq:factorization}) itself
could only appear starting at quadratic order, $(m_s/\Lambda)^2$. 

Factorization for the $B\to V\gamma$ decay process has received
considerable attention in the literature. Extending the QCD
factorization formalism \cite{Beneke:1999br} to the case of exclusive
radiative decays, the formula (\ref{eq:factorization}) was proposed in
\cite{Beneke:2001at,Bosch:2001gv}, where $T^{I}$ and $T^{II}$ were
calculated through $\order(\alpha_s)$.  Diagrams contributing one-loop
matching corrections to $T^{II}$ were studied in
\cite{Descotes-Genon:2004hd}, and (\ref{eq:factorization}) was shown
to hold for such spectator interactions through one-loop order.  Other
studies include \cite{Ali:2001ez}, and recent updates in
\cite{Ali:2004hn,Bosch:2004nd,Beneke:2004dp}.  Such explicit
demonstrations give important insight into the structure of the decay
process.  However, an all orders proof of the validity of
(\ref{eq:factorization}) is still lacking, an issue we address in this
paper using the language of SCET.  In addition to this formal aspect
of establishing factorization, the effective field-theory language has
the advantage of systematically separating higher-order perturbative
corrections that contribute to $T^I$ or $T^{II}$.  SCET achieves this
simplification by identifying the separate contributions on the
operator level, a feature which also allows the resummation of large
logarithms appearing in the perturbative kernels.  A previous analysis
of radiative $B\to V\gamma$ decay using SCET~\cite{Chay:2003kb}
identified the SCET$_{\rm I}$ operators corresponding to the
contributions calculated in \cite{Beneke:2001at,Bosch:2001gv}.
However, to establish factorization one needs
to construct a complete basis of effective-theory operators that can
contribute to the process under consideration in the heavy-quark limit, an
issue that was not addressed in \cite{Chay:2003kb}.  Furthermore, this
analysis suffers from an incomplete treatment of the low-energy
theory, SCET$_{\rm II}$.  As we will emphasize, a demonstration of
factorization must deal with the soft-collinear messenger modes that
can potentially spoil factorization~\cite{Becher:2003qh}. 
In particular, the decoupling of
soft gluons from hard-collinear fields in SCET$_{\rm I}$ is not
sufficient to ensure factorization; for example, the matrix element
of the operator $T_0^F$ in \cite{Bauer:2002aj,Chay:2003kb} is
``factorizable'' in this sense, but it cannot be written as a convergent
convolution of a perturbative kernel with meson
LCDAs~\cite{Lange:2003pk}.

The paper is organized as follows. In Section~\ref{sec:whale}, we
discuss the diagrammatic analysis of the decay. After identifying the
necessary momentum regions we introduce the corresponding fields and
set up the effective Lagrangian.  In Section~\ref{sec:tableology}, we
then find the SCET operators needed to analyze the decay at leading
power.  This point needs special consideration, since the matching of
SCET$_{\rm I}$ onto SCET$_{\rm II}$ involves inverse derivatives
counting as negative powers of the expansion parameter.  The most
general operator at a given power is determined by dimensional
analysis and longitudinal boost invariance.  Aside from the
contribution of messenger modes, which communicate between the soft
and collinear sectors, this issue was addressed by Beneke and Feldmann
in their analysis of heavy-to-light form factors~\cite{Beneke:2003pa}.
We will extend their discussion to cover the more complicated case of
photon emission from the spectator quark, involving two different
types of collinear fields defined with respect to opposite light-cone
directions.  Using the same power-counting arguments we analyze the
infrared messenger modes that can potentially spoil factorization in
the matrix elements defining the second term of
(\ref{eq:factorization}).  In Section~\ref{sec:Matching}, we match the
effective weak Hamiltonian onto the list of operators derived in
Section~\ref{sec:tableology}.  The resulting $B\to V\gamma$ matrix
elements can be written in the form of the factorization theorem
(\ref{eq:factorization}).  We show that the infrared messenger modes
cannot contribute to the matrix elements defining the second term of
(\ref{eq:factorization}), thus demonstrating that the hard-scattering
kernels are free of infrared divergences to all orders in perturbation
theory.  In Section~\ref{sec:phenomenology} we consider the
phenomenological implications of our analysis by computing the $B\to
K^*\gamma$ branching fraction, isospin asymmetry and CP asymmetry.  We
include the first complete treatment of the hard-scattering terms at
leading order in RG improved perturbation theory.  We discuss the
phenomenological impact of the resummation of the leading single and
double logarithms and of the inclusion of one-loop matching
corrections to the hard-scattering kernel.  In
Section~\ref{section:summary} we summarize our results and present our
conclusions.

\section{Perturbative analysis of \boldmath $B\to V\gamma$ \unboldmath}
\label{sec:whale}

In the Standard Model, the effective weak Hamiltonian mediating
flavor-changing neutral current (FCNC) transitions of the type
$b\rightarrow s$ has the form
\begin{equation}\label{eq:Hweak}
\begin{aligned} 
{\cal H}_W  &= {G_F\over \sqrt{2}}\sum_{p=u,c} V^*_{ps} V_{pb}\, 
\bigg[ { C}_1 Q_1^p + { C}_2 Q_2^p + \sum_{i=3}^8 { C}_i Q_i \bigg] \,,
\end{aligned}
\end{equation}
with 
\be
\begin{aligned}
  Q^p_1 &= \bar{s}\gamma^\mu(1-\gamma_5) p \,\, \bar{p}
  \gamma_\mu(1-\gamma_5) b \,, &
  Q^p_2 &= \bar{s}^i\gamma^\mu(1-\gamma_5) p^j \,\, \bar{p}^j
  \gamma_\mu(1-\gamma_5) b^i \,, \\[0.3cm] 
  Q_3 &= \bar{s}\gamma^\mu(1-\gamma_5) b \sum_q \bar{q} 
  \gamma_\mu (1-\gamma_5) q \,, & 
Q_4 &= \bar{s}^i \gamma^\mu(1-\gamma_5) b^j \sum_q \bar{q}^j
  \gamma_\mu (1-\gamma_5) q^i \,, \\ 
Q_5 &= \bar{s}\gamma^\mu(1-\gamma_5) b
 \sum_q \bar{q} \gamma_\mu (1+\gamma_5) q \,, & 
Q_6 &= \bar{s}^i \gamma^\mu(1-\gamma_5) b^j
 \sum_q \bar{q}^j \gamma_\mu (1+\gamma_5) q^i \,, \\ 
Q_7 &= -{e\over 8\pi^2} m_b
  \bar{s} \sigma^{\mu\nu}(1+\gamma_5) b \, F_{\mu\nu} \,, & 
Q_8 &= -{g\over 8\pi^2} m_b \bar{s} \sigma^{\mu\nu}(1+\gamma_5)
  T^a b \, G^a_{\mu\nu} \,.
\end{aligned}
\ee
Here $i$ and $j$ are color indices. 
The effective weak Hamiltonian for $b\rightarrow d$
transitions is obtained by replacing $s\rightarrow d$ in the above
expressions. Our sign
conventions are such that the covariant derivative acting on a
down-type quark is $iD_\mu = i\partial_\mu - \frac{1}{3} e A_\mu + g T^a
A_\mu^a$.  

Our task is to analyze the factorization properties of the matrix elements
involving the above operators. The
factorization theorem (\ref{eq:factorization}) holds trivially for the
operator $Q_7$, which directly maps onto the QCD tensor current. The
goal of the present paper is to show that the matrix elements of the
remaining operators can also be brought
into this form. To analyze the factorization properties of these matrix
elements, we use the reduction formula
\begin{multline}
  \int\! d^4x \int\! d^4y\, e^{i p_V\cdot x-i p_B \cdot y}
  \langle \gamma(p_\gamma,\lambda)\,|\,T\left\{
    J^\dagger_B(x) Q_i(0) J_V^\mu(y)\right\}\, |\,0 \rangle \\
  =\sum_{\lambda'}
  \frac{i f_B^{(J)}}{p_B^2-m_B^2} \frac{i
    f_V^{(J)} \epsilon_{\lambda'}^\mu}{p_V^2-m_V^2}\,\langle
    V(p_V,\lambda')\,\gamma(p_\gamma,\lambda)|\,Q_i(0)\,| B(p_B)
  \rangle+\dots \label{eq:LSZ}\,,
\end{multline}
where the currents $J_B$ and $J_V$ have the quantum numbers of the
$B$ meson and the vector meson, respectively, with associated decay constants
$f_B^{(J)}$ and $f_V^{(J)}$. The ellipsis stands for terms that do not
have a pole at $p_V^2=m_V^2$ with $p_V^0>0$ and at $p_B^2=m_B^2$ with
$p_B^0>0$. We then analyze the correlator on the left-hand side
perturbatively. In this analysis, we assume that the external
momenta are close to their mass shell, $p_B^2-m_b^2\sim m_b\Lambda$ and
$p_{V}^2\sim \Lambda^2$, and that the momentum transfers scale as
$p_B\cdot p_V \sim p_B\cdot p_\gamma\sim p_V\cdot p_\gamma \sim
m_b^2$, where $\Lambda$ remains fixed in the heavy-quark limit. The
correlator is then expanded about the heavy-quark limit. Perturbative
factorization relies on the assumption that if one finds that the
double spectral density of the correlator in (\ref{eq:LSZ}) 
with respect to the variables $p_B^2$ and
$p_V^2$ has certain factorization properties to a given order in
$\Lambda/m_b$ and to all orders in perturbation theory, then
the same is true of the amplitude on the right-hand side of
the reduction formula. For operators such as $Q_{1,2}^c$ containing charm 
quarks, we need to specify how
the charm-quark mass is treated in the heavy-quark limit. We take the
limit holding the ratio $m_c/m_b$ fixed.

\subsection{Diagrammatic analysis and momentum regions\label{sec:region}}

The correlator in (\ref{eq:LSZ}) is a function of Lorentz-invariant scalar 
products of the external momenta. However, to obtain
its expansion in powers of $1/m_b$ 
it is advantageous to introduce reference vectors.  We
introduce a unit four-vector $v_\mu$ in the direction of the
$B$ meson and a light-like vector $n_\mu$ in the direction of the
outgoing vector meson, and define 
\be
\label{eq:nbardef}
\nb^\mu={1\over n\cdot v} \left( 2v^\mu-{n^\mu \over n\cdot v}\right)\,,
\ee
so that $n^2=\nb^2=0$ and $n \cdot \nb =2$. We decompose all
momenta into their light-cone components,
\begin{equation}
p^\mu =  n\cdot p\, \frac{\bar
  n^\mu}{2}+\bar{n}\cdot p\, \frac{n^\mu}{2}+p^\mu_\perp\, =
  p_+^\mu+p_-^\mu+p_\perp^\mu \,.
\end{equation}
Note that our definition implies $v_\perp=0$.  Working in dimensional
regularization, we then employ the strategy of regions
\cite{Beneke:1997zp,Smirnov:2002pj} to expand diagrams about the
heavy-quark limit.  With this technique, the integrands are expanded
in a number of different momentum regions.  The expansion of the full
integral is recovered after integrating the expanded integrands and
summing the contributions from the different regions. 

The momenta in the relevant regions differ by the scaling of their 
components $(p_+ , p_- , p_\perp )$. Not surprisingly,
momentum regions in which a loop momentum has the same scaling as an
external momentum give a non-zero contribution. These regions are:
\begin{center}
\begin{tabular}{lc}
soft: & $(\lambda,\lambda,\lambda)\, m_b$ ,\\
$n$-collinear: &  $(\lambda^2, 1, \lambda)\, m_b$ ,\\
\end{tabular}
\end{center}
where we introduce a dimensionless expansion parameter 
$\lambda\sim\Lambda/m_b$.  
The loop momentum scales like $p_B-m_b v$ in the soft region, and in
the same way as $p_V$ in the $n$-collinear region. An $\bar
n$-collinear region does not appear, since $p_\gamma^2=0$ (it would be
present for $B\rightarrow K^* \ell^+ \ell^-$ if $q^2\sim\Lambda^2$).
In addition to these scalings, regions with $p^2\gg \Lambda^2$ arise:
\begin{center}
\begin{tabular}{lc}
hard: &  $(1, 1, 1)\, m_b$ , \\
$n$-hard-collinear: &  $(\lambda, 1, \sqrt{\lambda})\, m_b$ , \\
$\bar n$-hard-collinear: &  $(1, \lambda, \sqrt{\lambda})\, m_b$ . \\
\end{tabular}
\end{center}
The presence of two large perturbative scales --- the hard scale
$p^2\sim m_b^2$ and the hard-collinear scale $p^2\sim m_b \Lambda$ ---
manifests itself in the factorization formula: the hard-scattering
kernels have the schematic form $T\sim C\otimes J$, where $C$ and $J$
include hard and hard-collinear contributions, respectively, and the 
symbol ``$\otimes$'' denotes a convolution over momentum fractions.
The $\bar n$-hard-collinear region arises in diagrams where the photon
attaches to the spectator.  Offshell propagators also appear; for example,
a momentum scaling $(\lambda,1,\lambda)m_b$ arises when a
soft and $n$-collinear momentum flow into the same vertex.  This will
be discussed in more detail in Section~\ref{sec:scetii}.
Finally, the following low-energy region appears in the expansion 
of the diagrams~\cite{Becher:2003qh}:
\begin{center}
\begin{tabular}{lc}
$n$-soft-collinear: &  $(\lambda^2, \lambda, \lambda^{3/2})\, m_b$ . \\
\end{tabular}
\end{center}
This momentum region appears in interactions with soft and collinear
lines. The scaling of its components is the largest compatible with
both the soft and the collinear scaling: $(p_s+p_{sc})^2\sim p_s^2$
and $(p_c+p_{sc})^2\sim p_c^2$. Note that $(p_s+p_{c})^2\sim p_{hc}^2$
so that collinear lines cannot emit or absorb soft momenta and remain
collinear, or vice versa. 
Since it is the only low-energy interaction connecting the
soft and collinear sectors, proving factorization
to a given order in $\lambda$ amounts to showing that there is no 
contribution from the soft-collinear region to this order.

\begin{figure}[t]
\begin{center}
\includegraphics[width=\textwidth]{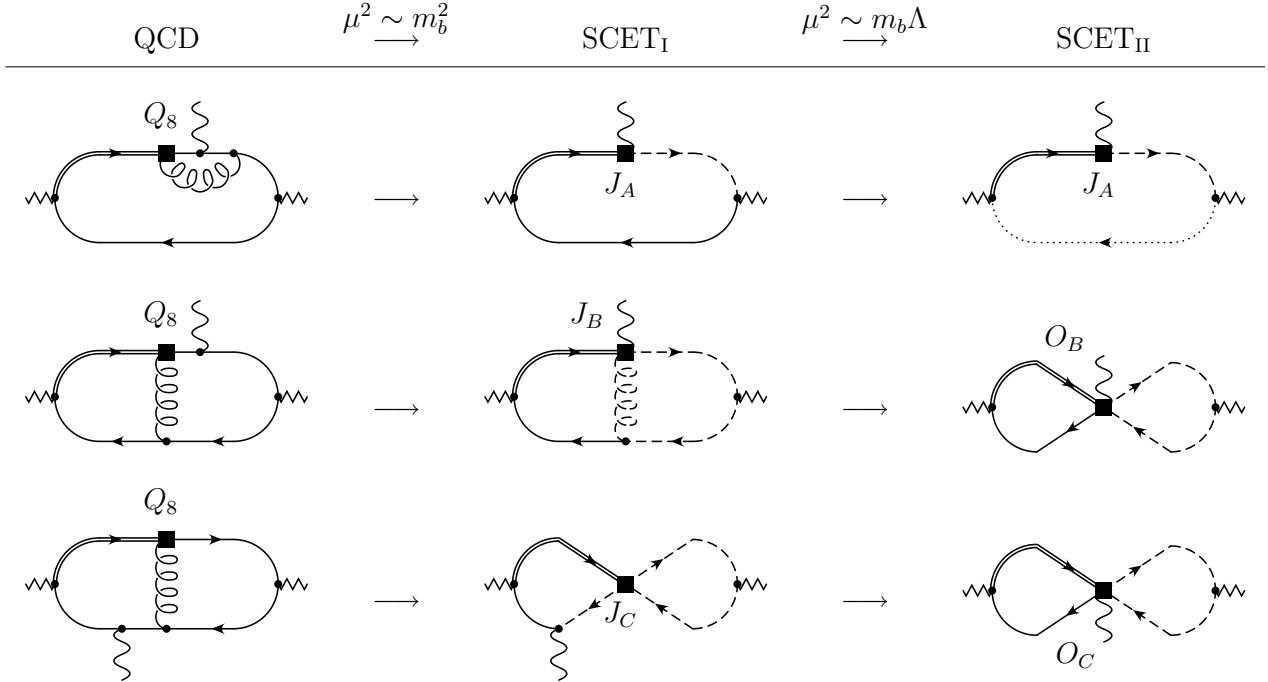}
\end{center}\vspace*{-0.5cm}
\caption{Three QCD Feynman diagrams for the contributions of $Q_8$ 
  and their leading-order representation in the effective theory. The
  double line denotes the heavy-quark field. The dashed lines denote
  hard-collinear fields in SCET$_{\rm I}$ and collinear fields in
  SCET$_{\rm II}$. Solid lines in the effective-theory diagrams
  denote soft fields and the dotted line denotes a soft-collinear
  field.\label{fig:matching}}
\end{figure}

Figure~\ref{fig:matching} shows three typical contributions to the 
decomposition of the correlator (\ref{eq:LSZ}), for the chromomagnetic 
operator $Q_8$.  
The three rows in the figure illustrate the soft-overlap ($A$), 
hard-scattering ($B$), and spectator-emission ($C$) mechanisms. 
The two-step matching procedure QCD $\to$ SCET$_{\rm I}$ $\to$ SCET$_{\rm II}$ 
is described in the following Sections~\ref{sec:intermediate} and 
\ref{sec:scetii}.   

Note that the soft-collinear region has $p^2\sim\Lambda^3/m_b \ll \Lambda^2$. 
It has sometimes been argued that it is ``unphysical'' to allow for momentum
regions with $p^2$ parametrically below $ \Lambda^2$, since
non-perturbative effects would modify physics below this scale, and
that it would therefore be more natural to perform the perturbative
factorization analysis with a hard infrared cut-off in QCD. Since the
key point in factorization proofs is precisely to show that such
infrared regions are either absent or cancel in the sum over diagrams,
simply ignoring such modes is clearly not an option. If one chooses to
introduce an infrared cut-off in QCD, the proof of factorization
becomes equivalent to the demonstration of insensitivity to this
regulator.
However, it is difficult to introduce such a cut-off in a gauge invariant 
way,%
\footnote{The only known way is to quantize in a finite volume and use
  twisted boundary conditions for the gauge fields to eliminate the
  zero mode.}  
and it is also doubtful whether the diagrammatic
analysis with a cut-off can be reformulated in effective-theory
language.  Since the messenger fields do not contribute (by definition) to
factorizable quantities, and since non-factorizable quantities are
categorized as non-perturbative, nothing is gained by removing these
fields in favor of an infrared cut-off.  One of the advantages of 
the effective-theory approach in dimensional regularization is precisely
that the analysis can be performed without explicit momentum cut-offs.

While it is easy to see that all of the above regions are required
to obtain the expansion of the correlator diagrams, we do not have a
proof that they are sufficient.%
\footnote{The same is true for
  traditional diagrammatic factorization proofs. Additional momentum
  regions could invalidate the analysis also in these cases.}
Two-loop applications in similar kinematic situations
\cite{Smirnov:1998vk} suggest that no additional regions are needed.
The above list of momentum scalings is natural in that it 
contains all onshell modes whose components $n\cdot p$
and $\nb\cdot p$ scale with powers of $\lambda$ equal to
the scaling of the components of external momenta. 

Finally, let us note that the analysis of regions presented above assumes 
exactly massless light quarks.  A systematic inclusion of quark mass terms
presents a challenge, since the mode structure in the low-energy
theory is then drastically altered.  For instance, including
$\order(\Lambda)$ masses would eliminate the soft-collinear mode, but
the resulting diagrams for the soft and collinear regions would no longer
be separately well-defined in dimensional regularization, requiring
additional unconventional (e.g., analytic) regulators.  
We will return to this issue in
Section~\ref{sec:factor} and address the more modest question of the
leading corrections for light-quark masses $m_q\ll \Lambda$.  We argue
that contributions linear in the light mass may be absorbed into the
hadronic parameters appearing in the factorization formula, while any terms
that could potentially spoil factorization appear first at quadratic
order.

\subsection{Intermediate effective theory: SCET$_{\rm I}$
\label{sec:intermediate}}

In the construction of the SCET Lagrangian, an effective-theory field
is introduced for each momentum region.  The integrands of the QCD
Feynman diagrams expanded in the various regions are then
reinterpreted as arising from the Feynman rules of the effective
theory.  Furthermore, in order to ensure that the amplitudes 
are appropriately expanded in momentum space, a derivative 
(``multipole'') expansion is performed in the effective action.

Note that if we had chosen the momentum of the vector meson to be
$n$-hard-collinear, then only the hard, $n$-hard-collinear,
$\nb$-hard-collinear, and the soft region would appear in the expansion
of (\ref{eq:LSZ}). It is simpler to first consider the situation where 
we count the external momenta 
in this way and to introduce fields only for these regions. 
This is illustrated by the middle column of Figure~\ref{fig:matching}.
The corresponding effective theory, called SCET$_{\rm I}$, describes QCD 
at or below the hard-collinear scale, and contains the quark and
gluon fields
\be\label{eq:fieldscale}
\begin{aligned}
  \xi_{\hcn}&\sim \lambda^{1/2}, & A_{\hcn}^\mu &\sim
  (\lambda,1,\sqrt{\lambda}), & \xi_{\hcnb}&\sim \lambda^{1/2}, &
  A_{\hcnb}^\mu &\sim(1,\lambda,\sqrt{\lambda}) \,, \\
  q_s&\sim \lambda^{3/2},  &  A_s^\mu &\sim
  (\lambda,\lambda,\lambda) \,, & h&\sim \lambda^{3/2}.
\end{aligned}
\ee
The hard-collinear quarks are described by two-component spinors satisfying 
$n\!\!\!/\,\xi_{\hcn}=\nb\!\!\!/\,\xi_{\hcnb}=0$.  
We have indicated in (\ref{eq:fieldscale}) 
the scaling of the field components, which can be derived from the
scaling of the corresponding propagators. No field is introduced for
the hard region, as this contribution will be absorbed into the Wilson
coefficients of operators in the effective theory. As was shown in
\cite{Beneke:2002ph}, for diagrams involving only a single type of 
hard-collinear field, in dimensional regularization there is no hard
contribution in the pure QCD sector, since the corresponding diagrams
are scaleless and vanish. The effective Lagrangian can then be constructed
exactly, to all orders in perturbation theory. This was done to
next-to-next-to-leading power in \cite{Beneke:2002ni}. The same 
may be done in our case with two types of hard-collinear fields, 
which we denote generically as $\phi_{hc}$ and $\phi_{\hcnb}$.
In fact, the Lagrangian for this case is simply
\begin{equation}\label{eq:lagrangian}
{\cal L}={\cal L}_{\hcn}+{\cal L}_{\hcnb}+{\cal L}_s
\end{equation}
where ${\cal L}_s$ is given by the HQET Lagrangian for heavy quarks, 
and by the restriction of the QCD Lagrangian to soft momentum modes for
light quarks and gluons. ${\cal L}_{\hcn}$ denotes the remainder containing 
the hard-collinear fields and their interactions with the soft fields, and
${\cal L}_{\hcnb}$ is obtained from ${\cal L}_{\hcn}$ by interchanging
$n$ and $\nb$. 

\begin{figure}
\begin{center}
\epsfbox{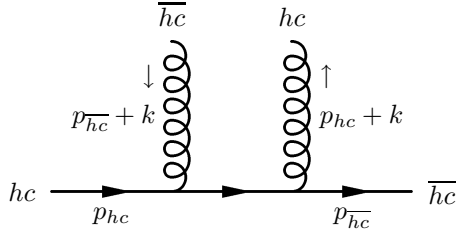}
\end{center}
\caption{
Example of an exceptional momentum configuration giving rise to an 
interaction in SCET$_{\rm I}$ involving $hc$ and $\hcnb$ fields. 
\label{fig:exceptional}}
\end{figure}

Note that we did not write down a Lagrangian containing interactions
with both $n$- and $\nb$-hard-collinear fields.  
By the Coleman-Norton theorem~\cite{Coleman:1965xm}, pinch singularities 
can only occur in momentum configurations that can be
interpreted as classical scattering processes.  By momentum conservation, 
this cannot occur in interactions with both $n$-hard-collinear and 
$\nb$-hard-collinear particles unless both types of particles are present in 
the initial and final states. An example is illustrated in 
Figure~\ref{fig:exceptional}, where the momentum $k$ is restricted to the 
region $k\sim (\lambda,\lambda,\sqrt{\lambda})$. When this skeleton diagram is 
inserted into loop diagrams, an integration over $k$ in this region involves 
denominators of the form $(p_{hc\perp}+k_\perp)^2+ \nb\cdot p_{hc} n\cdot k$ 
and $(p_{\hcnb\perp}+k_\perp)^2+ n\cdot p_{\hcnb}\,\nb\cdot k$.  The contour 
integrals 
in $n\cdot k$ ($\nb\cdot k$) vanish if the external particles all have the
same sign of $\nb\cdot p_{hc}$ ($n\cdot p_{\hcnb}$), and such exceptional 
configurations are therefore not relevant in cases where collinear particles 
are present only in the final state. The absence of such exceptional 
momentum configurations is 
encoded automatically in the usual strategy of regions 
applied to the perturbative expansion of the correlator (\ref{eq:LSZ})
in dimensional regularization. For an $N$-loop diagram, this strategy 
assigns onshell momentum scaling to $N$ internal lines, 
with the scaling of all remaining lines fixed by momentum conservation.  
The full amplitude is recovered by performing this assignment 
in all possible ways, with an unrestricted integration over the $N$ onshell 
loop momenta. With these rules, an isolated offshell line such as in 
Figure~\ref{fig:exceptional} (with momentum $p_{hc}+p_{\hcnb}+k$) cannot 
occur.  

It is convenient to use the SCET$_{\rm I}$ operators to classify the
different mechanisms through which the decay $B\rightarrow V\gamma$
can proceed. In contrast to the Lagrangian interactions, there are
hard matching corrections to the weak-interaction operators in the
effective theory. After performing the matching of QCD onto
SCET$_{\rm I}$, the remaining problem is to examine in each case all
possible SCET$_{\rm II}$ operators that can result.  We now turn to
this problem and discuss the issues involved in integrating out the
hard-collinear components of the SCET$_{\rm I}$ fields.

\subsection{Final effective theory: SCET$_{\rm II}$\label{sec:scetii}}

Counting the external momenta as collinear, instead of
hard-collinear, the full list of regions in Section~\ref{sec:region}
needs to be considered. The dynamical fields in this case are the
collinear and soft fields
\begin{align}
  \xi_c &\sim \lambda \,, & A_{c}^\mu &\sim
  (\lambda^2,1,\lambda) \,,  &
  q_s&\sim \lambda^{3/2},  &  A_s^\mu &\sim
  (\lambda,\lambda,\lambda) \,, & h&\sim \lambda^{3/2} ,
\end{align}
as well as the soft-collinear quark and gluon fields
\begin{align}
 \theta&\sim \lambda^2, & 
  A_{sc}^\mu &\sim(\lambda^2,\lambda, \lambda^{3/2}) \,.
\end{align}
The collinear and soft-collinear fields are again described by
two-component spinors satisfying $\nslash\,\xi_c=\nslash\,\theta=0$.  
The small-component projection of the soft-collinear fermion field, 
satisfying $\nbslash\sigma=0$, is given by
\be \label{eq:sigma}
\sigma=-\frac{\nb\!\!\!/}{2}\frac{1}{i\nb\cdot D_{sc}}{iD\!\!\!\!/}_{sc\perp}
\theta \,. 
\ee 
In SCET$_{\rm II}$, both the hard and hard-collinear contributions
are absorbed into the Wilson coefficients of the operators built from
the above fields. The hard-collinear contributions appear in the
matching step from SCET$_{\rm I}$ onto SCET$_{\rm II}$. As in the
first matching step, the pure QCD part of the effective Lagrangian can
be obtained exactly. It was constructed at next-to-leading power in
\cite{Becher:2003qh}.

\begin{figure}
\begin{center}
\psfrag{s}[r]{$(\lambda,\lambda,\lambda)$}
\psfrag{c}[l]{$(\lambda^2,1,\lambda)$}
\psfrag{o}[l]{$(\lambda,1,\lambda)$}
\includegraphics[width=0.25\textwidth]{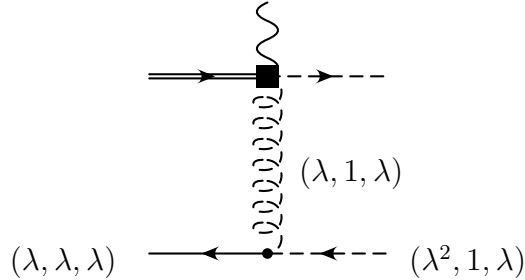} \\
\vspace*{-0.5cm}
\end{center}
\caption{Example of a SCET$_{\rm I}$ diagram relevant for $B\to V\gamma$ 
decay with an offshell-collinear gluon. The same contribution is also
depicted in the second line of Figure \ref{fig:matching}.\label{fig:oc}}
\end{figure}

In preparation for the discussion of general operator 
bases to be considered in Section~\ref{sec:tableology}, we review here 
the procedure employed in integrating out the fields at the hard-collinear 
scale.  We begin by restricting attention to the sector of the SCET$_{\rm I}$ 
Lagrangian (\ref{eq:lagrangian}) involving $n$-hard-collinear and soft fields. 
Just as matching QCD to SCET$_{\rm I}$ involved decomposing the QCD fields 
into their hard, hard-collinear, and soft components, so matching onto 
SCET$_{\rm II}$ involves the decomposition of SCET$_{\rm I}$ fields.  
In particular, a generic hard-collinear field is decomposed as
\be
\label{eq:collinear_dec}
\phi_{hc} \to \phi_{hc} + \phi_{c} + \phi_{oc} \,, 
\ee 
where the hard-collinear, collinear and offshell-collinear momenta scale 
as $(\lambda,1,\sqrt{\lambda})$, $(\lambda^2,1,\lambda)$, and 
$(\lambda,1,\lambda)$, respectively. 
The latter momentum scaling arises from the combination of onshell
soft and collinear momenta, see Figure~\ref{fig:oc}. The
hard-collinear and offshell-collinear fields are integrated out in
passing from SCET$_{\rm I}$ to SCET$_{\rm II}$, just as the hard
components of the QCD fields were integrated out in passing from QCD
to SCET$_{\rm I}$.  Similarly, a generic soft field is decomposed as
\be
\label{eq:soft_dec}
\phi_s \to \phi_s + \phi_{sc} \,,
\ee
where the soft and soft-collinear momenta scale as 
$(\lambda,\lambda,\lambda)$ and $(\lambda^2,\lambda,\lambda^{3/2})$. 
In contrast to SCET$_{\rm I}$, 
where the soft fields are defined to contain all modes below the soft scale, 
the soft fields in SCET$_{\rm II}$ are defined to contain strictly soft, 
and not soft-collinear
modes.  This interpretation is mandated by the appearance of a new 
region in SCET$_{\rm II}$.
If we were to work with explicit cut-offs, the soft fields would be 
required to have $n\cdot p_s$ of order $\lambda$ 
(thus excluding modes with $n\cdot p_s$ of order $\lambda^2$), and    
collinear fields would be required to have $\nb\cdot p_c$ of order unity 
(excluding modes with $\nb\cdot p_c$ of order $\lambda$).  
The situation is analogous to the passage from QCD to SCET$_{\rm I}$, 
where the hard-collinear
region is defined to contain strictly hard-collinear, and not soft, modes. 

We now expand the SCET$_{\rm I}$ Lagrangian (\ref{eq:lagrangian})
using the decompositions (\ref{eq:collinear_dec}) and
(\ref{eq:soft_dec}).  We split the Lagrangian into two parts, one containing 
the ``light'' degrees of freedom present in the low-energy theory, 
and one containing ``heavy'' modes that are to be integrated out. In the
first part, ${\cal L}_{\rm light}$, we collect all terms that contain
only fields that are part of SCET$_{\rm II}$: soft, collinear, and
soft-collinear fields. ${\cal L}_{\rm light}$ has been derived in
\cite{Becher:2003qh} and is required through $\order(\lambda)$:
\bea
\label{eq:Llight}
{\cal L}_{\rm light} &=& {\cal L}_s + {\cal L}_c + {\cal L}_{sc} + {\cal L}_{s + sc}^{\rm int} 
+ {\cal L}_{c + sc}^{\rm int}  + \dots \nl
  &=& {\cal L}_{\rm light}^{(0)} + {\cal L}_{\rm  light}^{(1/2)}
 + {\cal L}_{\rm light}^{(1)} + \dots \,.  
\eea
Terms with both soft and collinear fields appear at subleading power
in the decomposition of the SCET$_{\rm I}$ Lagrangian, 
both directly in ${\cal L}_{\rm light}$, and via induced interactions
after integrating out offshell modes in ${\cal L}_{{\rm heavy}}$ below,
as discussed in \cite{Hill:2002vw}.   
However, such interactions are not 
relevant to our analysis, as they do not appear in the
expansion of the correlator (\ref{eq:LSZ}).  More generally, they are absent
 in cases
where collinear particles are present only in the final state, by the
same reasoning as for the terms with both $n$-hard-collinear and
$\nb$-hard-collinear fields in the decomposition of the QCD Lagrangian
in Section~\ref{sec:intermediate}.

In the remaining part of the Lagrangian, ${\cal L}_{{\rm heavy}}$, we collect 
all terms that involve at least one hard-collinear or offshell-collinear 
field, which will be integrated out in the construction of SCET$_{\rm II}$.
For simplicity, in the discussion of Lagrangian terms involving such ``heavy'' 
modes, we work in light-cone gauge $\nb\cdot A =0$ for the fields descending 
from $A_{hc}$ (i.e., hard-collinear, collinear, offshell-collinear), 
and $n\cdot A=0$ for the fields descending from $A_s$ (i.e., soft and 
soft-collinear). 
To fully separate the different scales,  interactions involving fields 
with different momentum scaling must be multipole-expanded and
the offshell-collinear fields $\xi_{oc}$ and $A_{oc}$ integrated out. 
The remaining onshell fields can be assigned a definite power counting, and
the offshell fields are expressed in terms of a series (ordered in $\lambda$) 
giving the possible branchings into these onshell fields.  
For interactions of collinear with hard-collinear fields we have 
\be
\label{eq:multipole} \phi_{hc}(x)\phi_c(x)=\phi_{hc}(x)\bigg[
\phi_c(x_+) + x_{\perp}^\mu\partial_\mu \phi_c(x_+) + \big( {x_-^\mu
  \partial_\mu} + {1\over 2} x_\perp^\mu x_\perp^\nu
\partial_\mu\partial_\nu \big) \phi_c(x_+) + \dots \bigg] \,.  
\ee
Similarly, for soft and hard-collinear fields $\phi_{hc}(x)\phi_s(x)
\approx \phi_{hc}(x)\phi_s(x_-)$, while for 
soft and collinear fields, $\phi_s(x)\phi_c(x)\approx \phi_s(x_-+x_\perp)\phi_c(x_++x_\perp)$. 

We first expand ${\cal L}_{{\rm heavy}}$ in powers of $\lambda$.  
We begin with the tree level case  
(i.e., neglecting interactions involving onshell hard-collinear fields),
where we will find that the solutions for the offshell fields scale as 
$\xi_{oc}\sim \lambda^{3/2}$, 
$A_{oc\perp}^\mu\sim \lambda^{3/2}$, and $n\cdot A_{oc}\sim \lambda^2$. 
We will then consider the inclusion of 
hard-collinear fields, finding that the solutions for the offshell fields in this 
case start at one power lower in $\lambda$: 
\be
\label{eq:ocmodes}
\begin{aligned}
\xi_{oc} &= \big( \xi_{oc}^{(1/2)} + \xi_{oc}^{(1)}\big) 
 + \xi_{oc}^{(3/2)} + \dots  \,, \\
A_{oc\perp} &= \big( A_{oc\perp}^{(1/2)} + A_{oc\perp}^{(1)}\big)
 + A_{oc\perp}^{(3/2)} + \dots \,, \\
n\cdot A_{oc} &= \big( n\cdot A_{oc}^{(1)} + n\cdot A_{oc}^{(3/2)} \big) 
+  n\cdot A_{oc}^{(2)} +  \dots   \,.
\end{aligned}
\ee
The terms in parentheses only appear when branchings into hard-collinear 
fields are included. 

The tree-level Lagrangian begins at $\order(\lambda)$, 
and for a complete matching at leading power we require terms through
$\order(\lambda^3)$.  
With the inclusion of hard-collinear fields, the 
Lagrangian begins at one power lower in $\lambda$:  
\be
\label{eq:Lheavyexpand}
{\cal L}_{{\rm heavy}} 
= \left( {\cal L}_{{\rm heavy}}^{(0)}
+ {\cal L}_{{\rm heavy}}^{(1/2)} \right) 
+ {\cal L}_{{\rm heavy}}^{(1)}
+ \dots \,. 
\ee
Omitting the terms involving hard-collinear fields for the moment, 
the leading fermion Lagrangian reads 
\be
{\cal L}_{{\rm heavy}}^{(1)} 
= 
\bar{\xi}_{oc}^{(3/2)} {\nbslash\over 2} in\cdot \partial\, \xi_{oc}^{(3/2)} 
+ \big\{ 
  \bar{\xi}_{oc}^{(3/2)} g\Aslash_{c\perp} q_s 
+ \bar{\xi}_c g\Aslash_{oc\perp}^{(3/2)} q_s   
+ \mbox{h.c.} \big\} \,, 
\ee
and solving the equation of motion yields
\be \label{eq:xiOC}
\xi_{oc}^{(3/2)} =  {-1\over in\cdot \partial}{\nslash\over 2} g\Aslash_{c\perp} q_s  \,. 
\ee
Similarly, from the gluon terms in ${\cal L}_{{\rm heavy}}^{(1)}$ 
we find
\bea\label{eq:AOC}
A_{oc\perp}^{(3/2)\mu} &=& {g\over  i\nb\cdot\partial\,in\cdot\partial} \big\{  
 T^a \bar{q}_s \gamma_\perp^\mu T^a \xi_c +  \mbox{h.c.} \big\}  \,, \nl
n\cdot A_{oc}^{(2)} &=& {2g\over i\nb\cdot\partial} [ A_{c\perp\mu}, A_{s\perp}^\mu ] \,. 
\eea
Having found the leading terms, these solutions may be substituted back into
${\cal L}_{{\rm heavy}}$, and the process iterated at the next power in $\lambda$. 
The complete list of SCET$_{\rm II}$ operators at leading power requires also 
$\xi_{oc}^{(2)}$ and $\xi_{oc}^{(5/2)}$; these terms themselves involve
 $A_{oc\perp}^{(2)}$ and 
$n\cdot A_{oc}^{(5/2)}$. 
The tree-level expressions were obtained in 
\cite{Beneke:2003pa}.%
\footnote{
Expressions for our $\xi_{oc}^{(3/2),(2),(5/2)}$ 
are given by $\xi_{hc}^{(3),(4),(5)}$ in \cite{Beneke:2003pa}. 
In the decomposition of SCET$_{\rm I}$ fields at tree level in
this reference, ``$hc$'' refers to what we call ``$oc$''.   
Explicit expressions are given there for the slightly different 
quantities $\psi^{(3),(4),(5)}$, which include contributions from 
the soft field $q_s$ and from the small-component hard-collinear field 
$\eta_{hc}$ in SCET$_{\rm I}$.  These terms are not part of $\xi_{oc}$;   
in particular, the terms containing $q_s$ from $\psi^{(3)}$, 
$(i\nb\cdot\partial)^{-1}(i\Dslash_{c\perp}
+g\Aslash_{s\perp})(\sla{\nb}/2)\xi_c$ from $\psi^{(4)}$ 
and $(i\nb\cdot\partial)^{-1}g\Aslash_{oc\perp}^{(3/2)}(\sla{\nb}/2)\xi_c$ 
from $\psi^{(5)}$ should not be included in $\xi_{oc}$.  
}

Beyond tree level, we must consider the branching of offshell-collinear
fields into two or more onshell hard-collinear modes.   
At each order in $\lambda$, we expand in powers of the coupling constant $g$.  
Only factors of $g$ associated with hard-scale (i.e., hard-collinear and 
offshell-collinear) gluons are included in this expansion.  
Anticipating that $\xi_{oc}^{(1/2)} \sim g$ and
$A_{oc\perp}^{(1/2)} \sim g$,  
contributions to ${\cal L}_{\rm heavy}$ 
involving offshell-collinear fields 
begin at $\order(g^2)$.  For the fermion Lagrangian,  
\bea
\label{eq:Lheavy0}
{\cal L}_{{\rm heavy}}^{(0)} 
&=& {\cal L}_{hc} + \bar{\xi}_{oc}^{(1/2)} {\nbslash\over 2} in\cdot\partial \xi_{oc}^{(1/2)} 
+ \bigg\{ \bar{\xi}_{oc}^{(1/2)} {\nbslash\over 2}\left( g\, n\cdot A_{hc} 
+ g{\Aslash}_{hc\perp}{1\over i\nb\cdot \partial} i\sla{\partial}_\perp 
\right)\xi_{hc} + \mbox{h.c.} \bigg\} \nl 
&& \quad + \bar{\xi}_{hc}{\nbslash\over 2}\left( 
  g\, n\cdot A_{oc}^{(1)}
+ i\sla{\partial}_\perp {1\over i\nb\cdot\partial} g \Aslash_{oc\perp}^{(1/2)} 
+  g \Aslash_{oc\perp}^{(1/2)} {1\over i\nb\cdot\partial} i\sla{\partial}_\perp
\right) \xi_{hc}  + \order( g^3 )  \,. 
\eea
Solving the equation of motion yields
\be\label{eq:xioc}
\xi_{oc}^{(1/2)} = {-g\over in\cdot\partial} 
\left( n\cdot A_{hc} 
+ {\Aslash}_{hc\perp}{1\over i\nb\cdot \partial} i\sla{\partial}_\perp 
\right)\xi_{hc} + \order(g^2) \,. 
\ee
From the gluon terms in ${\cal L}_{{\rm heavy}}^{(0)}$ we find in the same 
manner
\begin{align}
\label{eq:Aoc}
A_{oc\perp}^{(1/2)\mu} &=
{g \over i\nb\cdot\partial\,in\cdot\partial} \bigg\{  
T^a \bar{\xi}_{hc}{\sla{\nb}\over 2}
\bigg( \gamma_\perp^\mu T^a {1\over i\nb\cdot\partial} i\sla{\partial}_{\perp} 
 +  i\ov{\sla{\partial}}_{\perp} {1\over i\nb\cdot\ov{\partial}}
 \gamma_\perp^\mu T^a \bigg) \xi_{hc}  \nl
& \!\!
+{1\over 2}  [A_{hc\perp}^\mu, i\nb\cdot\partial n\cdot A_{hc} ]
 - [n\cdot A_{hc}, i\nb\cdot\partial A_{hc\perp}^\mu]  
- [A_{hc\perp\nu}, i\partial_{\perp}^\nu A_{hc\perp}^\mu
 - i\partial_\perp^\mu A_{hc\perp}^\nu ] 
\bigg\} + \order(g^2) \,,
\nl
n\cdot A_{oc}^{(1)} &= 
{-4g\over (i\nb\cdot\partial)^2} 
\bigg\{ 
\frac12 \left[ A_{hc\perp}^\mu, i\nb\cdot\partial A_{hc\perp\mu} \right] 
+ T^a \bar{\xi}_{hc} {\nbslash\over 2} T^a \xi_{hc} 
\bigg\}  + \order(g^2) \,. 
\end{align}
Note that  
the hard-collinear fields appearing on the right-hand side
in (\ref{eq:xioc}) and (\ref{eq:Aoc})
must have total transverse momentum of order $\lambda$, 
even though the 
individual fields 
have transverse momentum of order $\lambda^{1/2}$. 
This constraint is automatically enforced by the scaling of external soft and 
collinear momenta in the evaluation of diagrams corresponding to these
interactions.
Having derived $\xi_{oc}^{(1/2)}$, $A_{oc\perp}^{(1/2)}$, and $n\cdot A_{oc}^{(1)}$ 
to the desired order in $g$, these solutions can be substituted back into  
${\cal L}_{{\rm heavy}}$, and the procedure iterated at the next order in $\lambda$. 
This process can be carried out to any order in the power expansion. 
For offshell-collinear fields branching into hard-collinear fields, 
the complete list of SCET$_{\rm II}$ operators at leading power requires also
$\xi_{oc}^{(1)}$ and $\xi_{oc}^{(3/2)}$,  which themselves involve
 $A_{oc\perp}^{(1)}$ and $n\cdot A_{oc}^{(3/2)}$.  
Since further subleading Lagrangian interactions are 
required to convert the remaining onshell hard-collinear
fields into soft and collinear partons, 
such operators are required only up to 
one power in $\lambda$ lower than in the tree-level case. 

Substituting the expressions (\ref{eq:ocmodes}) 
for the offshell fields into 
${\cal L}_{{\rm heavy}}$ in (\ref{eq:Lheavyexpand}), and inserting 
appropriate gauge strings to relate the expressions in light-cone gauge to 
those valid in an arbitrary gauge, yields the final result for the 
decomposition of the SCET$_{\rm I}$ Lagrangian in the $n$-hard-collinear sector. 
The sector of SCET$_{\rm I}$ involving $\nb$-hard-collinear modes may
be treated similarly. It is convenient to treat the photon as being
an $\nb$-hard-collinear field in the intermediate effective theory and
as an $\nb$-collinear field, $A^{\rm (em)}_{\bar{c}\perp} \sim \lambda$, in
SCET$_{\rm II}$. The final results are independent of any power
counting assigned to this field, since we work to first order in the
electromagnetic coupling.  As discussed in Section~\ref{sec:region}, no
other $\nb$-collinear fields appear in the $B\to V\gamma$ analysis.
The decomposition of the SCET$_{\rm I}$ Lagrangian in this sector is
obtained from (\ref{eq:Lheavyexpand}) 
simply by replacing $n\leftrightarrow \nb$ and dropping all other
$\nb$-collinear fields.  The same manipulations as in the previous
case yield the final result for the explicitly gauge-invariant and
multipole-expanded Lagrangian in the sector involving
$\nb$-hard-collinear and soft fields.
The matching of SCET$_{\rm I}$ onto SCET$_{\rm II}$ is completed 
by substituting the solutions for the offshell fields into external 
current operators, and integrating out the remaining onshell hard-collinear
fields.  
The hard-collinear modes do not contribute additional 
renormalizations to the 
relevant part of the 
low-energy QCD Lagrangian, but result in non-trivial matching conditions for 
external currents. This matching is discussed in detail in 
Section~\ref{sec:tableology}.

As an illustration of the passage from SCET$_{\rm I}$ to SCET$_{\rm II}$, we 
may consider the representation of operators contributing
to form-factor matrix elements.  The leading-power SCET$_{\rm I}$
current operators are of the schematic form $\bar\xi_{hc} h$.  Using
the decomposition (\ref{eq:collinear_dec}) and enforcing momentum
conservation to drop the term involving a single hard-collinear ($hc$)
field, the mapping onto SCET$_{\rm II}$ operators is given by:
\be\label{eq:JAdecomp}
\begin{aligned}
\bar{\xi}_{hc}  h 
&\to \bar{\xi}_c h + \bar{\xi}_{oc} h \,. 
\end{aligned}
\ee
Expanding the solution of the equation of motion for the field 
$\bar{\xi}_{oc}$ as in (\ref{eq:ocmodes}), we then have diagrammatically at 
tree-level: 
\be
\label{eq:ocdiagrams}
\begin{aligned} 
\bar{\xi}_{oc}^{(3/2)} h &= 
\parbox{23mm}{\epsfbox{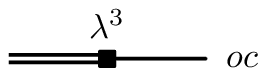}}
&&\equiv 
\parbox{23mm}{\epsfbox{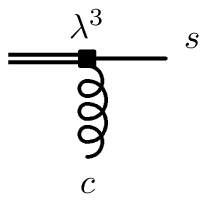}} \\
\bar{\xi}_{oc}^{(2)} h &=  
\parbox{23mm}{\epsfbox{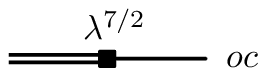}}
&&\equiv 
\parbox{23mm}{\epsfbox{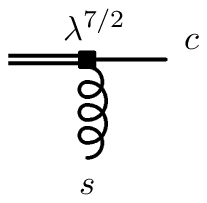}}
+
\parbox{23mm}{\epsfbox{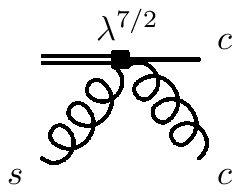}} 
+ 
\parbox{23mm}{\epsfbox{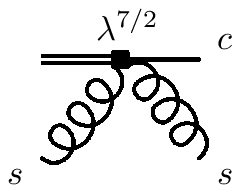}} 
+
\parbox{23mm}{\epsfbox{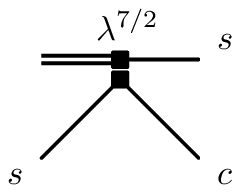}} \\
\bar{\xi}_{oc}^{(5/2)} h &= 
\parbox{23mm}{\epsfbox{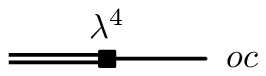}}
&&\equiv 
\parbox{23mm}{\epsfbox{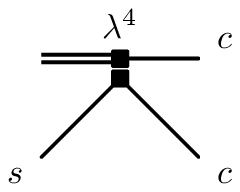}}
+\quad
\parbox{23mm}{\epsfbox{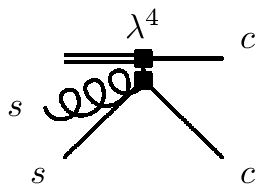}}
+
\parbox{23mm}{\epsfbox{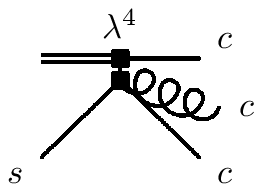}}
\quad 
+ \quad \dots
\end{aligned}
\ee
The contribution of the leading operator $\bar{\xi}_c h\sim\lambda^{5/2}$ 
receives an additional suppression when inserted into correlator diagrams
(analogous to Figure~\ref{fig:matching}) for the form factor, because these 
diagrams will always involve soft-collinear quark lines. For example
\be
\label{eq:softover}
\begin{aligned} 
\bar{\xi}_c h &=  
\parbox{23mm}{\epsfbox{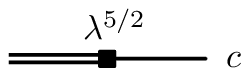}}
&& \to 
\parbox{43mm}{\epsfbox{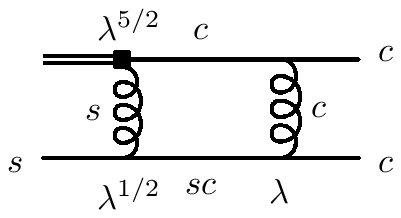}}
+ 
\parbox{43mm}{\epsfbox{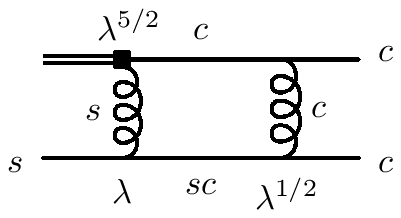}}
+ \quad \dots 
\end{aligned}
\ee
With the additional $\lambda^{3/2}$ suppression from subleading
Lagrangians terms in ${\cal L}_{\rm light}$ describing the coupling to
soft-collinear quarks, the contribution ends up being of order
$\lambda^4$, the same order as the contribution of
$\bar{\xi}_{oc}^{(5/2)} h$. Another example of this additional suppression 
relating to $B\rightarrow V\gamma$ is illustrated by the 
first row in Figure \ref{fig:matching}, where the operator $\bar{\xi}_c h$ 
appears in combination with subleading interpolating current operators 
for the initial- and final-state mesons that contain soft-collinear quarks. 
We will discuss this in more detail in Section~\ref{sec:tableology}.

The operators from $\bar{\xi}_{oc}^{(5/2)} h \sim \lambda^4$ and their
leading-order matching coefficients were given in \cite{Lange:2003pk}.
Additional terms arise at leading power from $\bar{\xi}_{oc}^{(5/2)}
h$ for flavor-singlet final-state mesons and have not been shown in
(\ref{eq:ocdiagrams}).  As discussed in more detail in
Section~\ref{sec:tableology}, these terms contain collinear gluon
fields in place of the collinear fermion bilinear.  For the remaining
terms, $\bar{\xi}_{oc}^{(3/2)} h \sim \lambda^3$ gives rise to
soft-overlap contributions in the flavor-singlet case, connected with
terms arising from $\bar{\xi}_{oc}^{(5/2)} h$. Similar to $\bar{\xi}_c
h$, the operators in $\bar{\xi}_{oc}^{(2)} h \sim \lambda^{7/2}$
receive an $\order(\lambda^{3/2})$ endpoint suppression, and as a
result do not contribute at leading power.

The subleading SCET$_{\rm I}$ operator 
$\bar{\xi}_{hc}\ov{\partial}_{\!\perp} h$ may be decomposed in a similar way: 
\be 
\label{eq:extraderiv}
\begin{aligned}
\bar{\xi}_{hc} \ov{\partial}_{\!\perp} h 
  &\to \bar{\xi}_c \ov{\partial}_{\!\perp} h 
  + \bar{\xi}_{oc} \ov{\partial}_{\!\perp} h  \,.
\end{aligned}
\ee
However, in each case the $\partial_\perp$ derivative gives an additional 
$\order(\lambda)$ suppression relative to the operators in 
(\ref{eq:JAdecomp}). The remaining form-factor contributions of leading power 
arise from subleading SCET$_{\rm I}$ operators of the form 
$\bar{\xi}_{hc} A_{hc\perp} h$. The decomposition in this case is 
\be
\label{eq:xiA_hc}
\begin{aligned}
\bar{\xi}_{hc} A_{hc\perp} h
&\to \bar{\xi}_{hc} A_{hc\perp} h + \bar{\xi}_c A_{oc\perp} h
  + \bar{\xi}_{oc} A_{c\perp} h 
  + \bar{\xi}_c A_{c\perp} h + \bar{\xi}_{oc} A_{oc\perp} h \,. 
\end{aligned}
\ee
Again, from the expansions (\ref{eq:ocmodes}) we find at tree-level: 
\begin{align}
\bar{\xi}_{c} A_{oc\perp}^{(3/2)} h &= 
\parbox{23mm}{\epsfbox{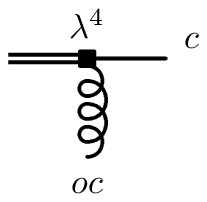}}
\quad\equiv \quad
\parbox{23mm}{\epsfbox{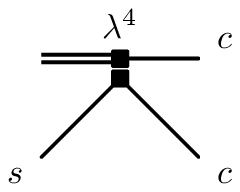}}
\nonumber \\
\bar{\xi}_{oc}^{(3/2)} A_{c\perp} h &=
\parbox{23mm}{\epsfbox{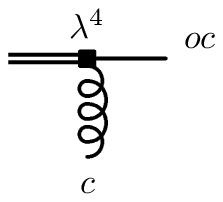}}
\quad\equiv \quad
\parbox{23mm}{\epsfbox{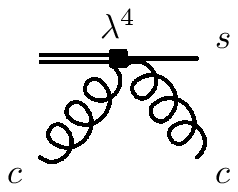}} \nonumber\\
\bar{\xi}_c A_{c\perp} h &=
\parbox{23mm}{\epsfbox{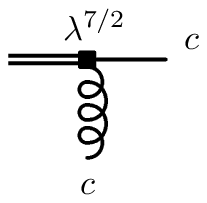}} \nonumber\\
\bar{\xi}_{oc}^{(3/2)} A_{oc\perp}^{(3/2)} h &= \order(\lambda^{9/2})
\end{align}
The operator $\bar{\xi}_c A_{oc\perp}^{(3/2)} h \sim \lambda^4$ contributes 
hard-scattering contributions of leading power, while 
$\bar{\xi}^{(3/2)}_{oc} A_{c\perp} h \sim \lambda^4$ contributes only for 
flavor-singlet final-state mesons.  Because of an additional
$\lambda^{3/2}$ endpoint suppression, 
$\bar{\xi}_c A_{c\perp} h \sim \lambda^{7/2}$ cannot contribute at leading 
power. Likewise, the contributions of 
$\bar{\xi}_{oc}^{(3/2)}A_{oc\perp}^{(3/2)} h\sim \lambda^{9/2}$ are also power-suppressed. 

This procedure can be extended beyond tree-level by integrating out the 
(onshell) hard-collinear modes.  For instance, for the first term on the
right-hand side of (\ref{eq:xiA_hc}), 
\be\label{eq:hcdiagrams}
\begin{aligned}
\bar{\xi}_{hc} A_{hc\perp} h =  
\parbox{23mm}{\epsfbox{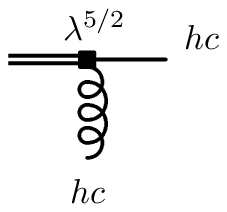}}
\to
\parbox{43mm}{\epsfbox{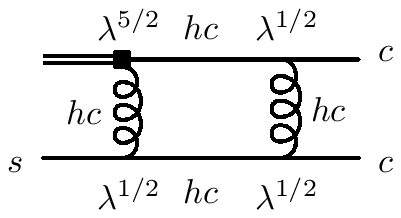}}
+ \quad \dots \quad 
= 
\parbox{23mm}{\epsfbox{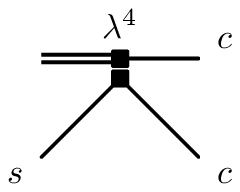}}
\end{aligned}
\ee
There are also contributions where
offshell-collinear
 fields branch into hard-collinear modes.  For instance, from 
the second term on the right-hand side of (\ref{eq:JAdecomp}), 
\be
\label{eq:xihalf}
\begin{aligned}
\bar{\xi}_{oc}^{(1/2)} h = \parbox{23mm}{\epsfbox{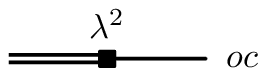}}  
\quad &\to
\raisebox{-0.15cm}{\parbox{43mm}{\epsfbox{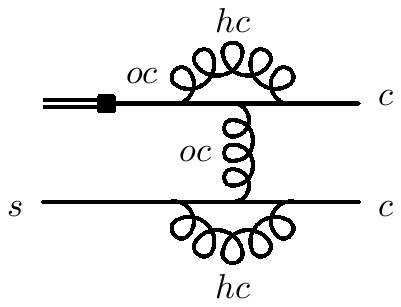}}} \!\!\! + \dots \\ 
&= \!\!
\raisebox{0.1cm}{\parbox{43mm}{\epsfbox{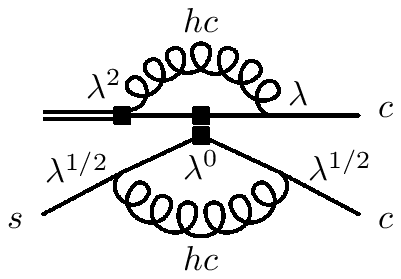}}} \!\!\! + \dots \,\,
= 
\raisebox{0.00cm}{\parbox{23mm}{\epsfbox{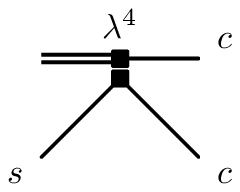}}}
\end{aligned}
\ee 
In (\ref{eq:xihalf}) we have displayed a contribution involving
the solution for the offshell-collinear gluon field
$A_{oc\perp}^{(1/2)}$ substituted back into ${\cal L}_{\rm
  heavy}^{(0)}$ in (\ref{eq:Lheavy0}). While straightforward in
principle, these examples illustrate the nontrivial nature of the
SCET$_{\rm I}$ to SCET$_{\rm II}$ matching. Instead of explicitly
integrating out the hard-collinear modes, in
Section~\ref{sec:tableology} we will arrive at the complete SCET$_{\rm
  II}$ operator basis using only general properties of the
decomposition of SCET$_{\rm I}$ operators. The general form is
required both for explicit computations, and to demonstrate
factorization properties, such as the decoupling of leading-power
soft-collinear interactions.  In contrast to SCET$_{\rm I}$, the power
counting for operators in SCET$_{\rm II}$ cannot be deduced simply by
inspection of the field content. This is illustrated by the third line
of (\ref{eq:ocdiagrams}): the two operators with an additional gluon
field turn out to be of the same order as the four-quark operator, due
to the non-localities introduced by integrating out the hard-collinear
modes, of virtuality $p^2 \sim m_b\Lambda$.  These non-localities
manifest themselves as inverse partial derivatives, counting like
$\lambda^{-1}$ \cite{Hill:2002vw}. The appearance of these derivatives is 
manifest in (\ref{eq:xiOC}) and (\ref{eq:AOC}), the solution of the equations of motion 
for the offshell-collinear fields.  In order to proceed, we require a set of
rules that can restrict the appearance of such factors and, more
generally, allows us to write down the most general SCET$_{\rm II}$ 
operators.

\section{SCET representation of the weak Hamiltonian\label{sec:tableology}}

As discussed in the previous section, the QCD part of the low-energy
effective theory does not receive matching corrections and can be
constructed exactly. This is not true for the operators in the weak
Hamiltonian. For this case we proceed in the usual way: we write down
all operators with the correct quantum numbers built from the
available fields, and perform perturbative matching to the desired
order. Our goal in this section is two-fold: to find the SCET$_{\rm
  II}$ operators that contribute to $B\to V\gamma$ decay at leading
power, and to construct the SCET$_{\rm I}$ operators which match onto
these SCET$_{\rm II}$ operators. The utility in identifying the
SCET$_{\rm I}$ operators lies in the fact that the soft-overlap
contributions can be isolated already at this stage, before further
decomposition into SCET$_{\rm II}$ fields.  The two-step matching
procedure is also required for the resummation of large perturbative
logarithms, which we address in Section~\ref{sec:phenomenology}.
Using building blocks defined below, it is straightforward to write
down all SCET$_{\rm I}$ operators that can contribute up to a given
order in $\lambda$. The situation is more complicated for the
SCET$_{\rm II}$ operators: when integrating out hard-collinear modes,
inverse derivatives $n\cdot \partial$ on the soft fields appear
\cite{Hill:2002vw}, counting as $\lambda^{-1}$.  Despite the presence
of such derivatives, we will see that only a finite number of
operators can appear to a given order in $\lambda$.  A second,
practical difficulty is that the leading SCET$_{\rm II}$ operators
are of a much higher order in $\lambda$ than the leading SCET$_{\rm
  I}$ operators.  For example, treating the photon field as a
hard-collinear field in SCET$_{\rm I}$ and as a collinear field in
SCET$_{\rm II}$, the leading operator in the intermediate theory
contributing to $B\to V\gamma$ counts as $\lambda^{5/2}$, while the
leading operators in the final effective theory count as $\lambda^5$.
Power counting alone does not strongly constrain the possible
SCET$_{\rm I}$ operators, and would leave us with a very large number of
operators in the intermediate theory, most of which would turn out to
be irrelevant upon matching onto SCET$_{\rm II}$. We will find that
counting the mass dimension of the SCET$_{\rm I}$ operators leads to
much stronger restrictions.

\subsection{Building blocks\label{sec:building}}

A characteristic feature of SCET is that derivatives of the
(hard-)collinear fields corresponding to large momentum components
are unsuppressed, and operators with an arbitrary number of such
derivatives can appear at the same order in the power counting.  To
account for this, the operators are allowed to be non-local along a
light ray: for example, the SCET$_{\rm I}$ representation of a QCD
operator at position $x$ can contain the hard-collinear fields
$\phi_{\hcn}(x+s \nb)$, $\phi_{\hcnb}(x+r n)$.  The Wilson
coefficients of the operators are then functions of the light-ray
variables ($r$ and $s$ in our example).

To obtain gauge-invariant operators, the fields at different points on
the light-ray must be connected by light-like Wilson lines (deviations of 
such Wilson lines from the light cone can be expanded and appear as
power-suppressed operators). Instead of 
inserting these Wilson lines for each operator, it is simpler to work with 
building blocks~\cite{Hill:2002vw,Bauer:2001yt} obtained by multiplying the
fields by Wilson lines which run along the light-ray to
infinity.  These building blocks will be invariant under hard-collinear 
gauge transformations in SCET$_{\rm I}$, and under soft and collinear gauge
transformations in SCET$_{\rm II}$.  We choose to work with building blocks 
that have simple transformations, but are not invariant, 
under soft and soft-collinear gauge transformations
in SCET$_{\rm I}$ and SCET$_{\rm II}$, respectively.  
Purely gauge-invariant quantities may be obtained by introducing additional
soft or soft-collinear Wilson lines, but this will not be necessary for our 
arguments, and would require the appearance of residual Wilson-line factors 
in SCET current operators.   The building blocks defined here are also easier 
to work with when performing explicit loop calculations. 
Thus, for SCET$_{\rm I}$, we introduce the fields
\begin{align}\label{eq:scetibb}
  \X_{\hcn}(x)&=W^\dagger_{\hcn}(x) \xi_{\hcn}(x) \,,\\
  \A_{\hcn}^\mu(x)&=W^\dagger_{\hcn}(x)\big[
  i D^\mu_{\hcn}(x)\,W_{\hcn}(x)\big]
+\frac{\nb^\mu}{2}\big[W^\dagger_{\hcn}(x)
  g n\cdot A_s(x_-) W_{\hcn}(x)-g n\cdot A_s(x_-)\big]\, , \nonumber
\end{align}
with $i D^\mu_{\hcn}=i\partial_\mu+g A_{\hcn}$ and Wilson line
\begin{equation}\label{eq:Whc}
W_{\hcn}(x)={\rm\bf P}\,\exp\left(ig \int_{-\infty}^0\! ds\, \nb\cdot
  A_{\hcn}(x+s \bar{n} )\right) .
\end{equation}
Note that $\nb\cdot\A_{\hcn}^\mu(x)=0$. The building blocks for the
hard-collinear fields in the opposite direction, $\xi_{\hcnb}(x)$ and
$A_{\hcnb}(x)$, are obtained by interchanging $n$ and $\nb$ (and $x_-\to
x_+$) in the above expressions.

The building blocks of SCET$_{\rm II}$ are defined in an
analogous way. In this case the role of the soft fields is played by
the soft-collinear fields, and both the soft and the collinear fields
are supplied with Wilson lines: 
\be
\begin{aligned}
  \X_{c}(x)&=W^\dagger_{c}(x) \xi_{c}(x) \,, \\
  \A_{c}^\mu(x)&=W^\dagger_{c}(x)\big[
  i D^\mu_{c}(x)\,W_{c}(x)\big]
  +\frac{\nb^\mu}{2}\big[W^\dagger_{c}(x)
  g n\cdot A_{sc}(x_-) W_{c}(x)-g n\cdot A_{sc}(x_-)\big]\,, \\
  \Q_{s}(x)&=S^\dagger_{s}(x) q_{s}(x) \,, \hspace{2cm}
  \H_{s}(x)=S^\dagger_{s}(x) h(x) \,, \\
  \A_{s}^\mu(x)&=S^\dagger_{s}(x)\big[
  i D^\mu_{s}(x)\,S_{s}(x)\big]
  +\frac{n^\mu}{2}\big[S^\dagger_{s}(x)
  g \nb\cdot A_{sc}(x_+) S_{s}(x)-g \nb\cdot A_{sc}(x_+)\big]\, .
\end{aligned}
\ee
The collinear Wilson line $W_{c}(x)$ is defined in the same way as
$W_{hc}(x)$ in (\ref{eq:Whc}), except that it is constructed with the 
collinear instead of the hard-collinear gluon field. The soft Wilson line is
\begin{equation}\label{eq:softW}
S_{s}(x)={\rm\bf P}\,\exp\left(ig \int_{-\infty}^0\! dt\, n\cdot
  A_{s}(x+t n)\right) .
\end{equation}
For a detailed discussion of the gauge transformation properties of
the SCET$_{\rm II}$ fields and the construction of gauge-invariant
building blocks, we refer the reader to \cite{Becher:2003qh}. Similar
to the building blocks for the $\nb$-hard-collinear fields, 
$\X_{\hcnb}(x)$ and $\A_{\hcnb}(x)$, we will also need
SCET$_{\rm II}$ building blocks for which $n$- and $\nb$-directions
are interchanged.  The only collinear field in the $\nb$-direction is the
photon field. However, we will need the associated soft building
blocks with Wilson lines in the $\nb$-direction and will denote them
by $\Q_{\sbr}(x)$, $\H_{\sbr}(x)$ and $\A_{\sbr}(x)$.%
\footnote{$\A_{\sbr}(x)$ contains a soft-collinear gluon field in the
  opposite direction. However, since there are no collinear quark
  or gluon fields in the $\nb$-direction, and since the messenger
  fields only contribute in exchanges between soft and collinear
  particles, this region does not contribute in $B\to V\gamma$.}

Arbitrary SCET operators are obtained by combining the above building blocks. 
In products involving different momentum modes, a derivative expansion of the 
fields has to be performed \cite{Beneke:2002ph,Becher:2003qh}.  The expansion 
for fields in SCET$_{\rm I}$ is 
as in (\ref{eq:multipole}), and for SCET$_{\rm II}$ we have  
\begin{multline}
  \phi_s(x) \phi_c(x)\rightarrow \phi_s(x_-+x_\perp) \phi_c(x_++x_\perp)\\
  + x_+\cdot\partial_s\phi_s(x_- +x_\perp)
  \phi_c(x_+ +x_\perp)+\phi_s(x_-+x_\perp) x_-\cdot\partial_c
  \phi_c(x_++x_\perp)+\dots\,,
\end{multline}
and similarly $\phi_s(x)\phi_{sc}(x)\approx \phi_s(x)\phi_{sc}(x_+)$, 
$\phi_c(x)\phi_{sc}(x)\approx \phi_c(x)\phi_{sc}(x_-)$.  
For our leading-power analysis the derivative terms can be dropped,
and we will suppress the $x$-dependence of the various fields 
in the following. 

\subsection{Operators in SCET\label{sec:operators}}

We now present a general procedure for matching generic SCET$_{\rm I}$
operators onto SCET$_{\rm II}$.  The SCET$_{\rm I}$ operators are
products of soft fields and hard-collinear fields in the $n$- and
$\nb$-directions; schematically we may write
\begin{equation}\label{eq:scetidec}
{\cal O}= [ \text{``soft''} ]\times [ \text{``$n$-hard-collinear''}]
 \times [ \text{``$\nb$-hard-collinear''}]\,.
\end{equation}
Because the SCET$_{\rm I}$ Lagrangian (\ref{eq:lagrangian})
decomposes into the two hard-collinear sectors, each of the three
brackets can be treated separately and they match as follows:
\begin{equation}\label{eq:sectors}
\begin{matrix}
  [ \text{``$n$-hard-collinear''}] &\longrightarrow& [
  \text{``$n$-collinear''}]\times [\text{``soft''}]
 \times [\text{``soft-collinear''}], \\
  [ \text{``$\nb$-hard-collinear''}]
 &\longrightarrow& A_{\bar{c}\perp}^{\rm (em)}\times [\text{``soft''}]
 \times [\text{``soft-collinear''}],\\
  [ \text{``soft''}] &\longrightarrow& [\text{``soft''}]
 \times [\text{``soft-collinear''}] .
\end{matrix}
\end{equation}
Physically, the reason that the sectors match separately can be
understood by picturing the decay process: at a certain time, the
heavy-quark decays into two energetic partons flying in opposite
directions. Each of these two particles can subsequently emit soft and
collinear particles, but the energetic particles from opposite directions
cannot annihilate each other. This physical picture is formalized by
the Coleman-Norton theorem. The fact that the soft sector matches
separately follows because the soft fields are not integrated out in
the transition to SCET$_{\rm II}$. Each soft field in SCET$_{\rm I}$
is simply replaced by the sum of a soft and a soft-collinear field in 
SCET$_{\rm II}$, see (\ref{eq:soft_dec}).

One complication is that the individual sectors are generally not
invariant under soft gauge transformations, while our SCET$_{\rm II}$
building blocks {\it are} invariant. In most cases we can avoid
matching non-invariant operators by grouping the ``soft'' bracket
together with either the ``$n$-hard-collinear'' or
``$\nb$-hard-collinear'' bracket.  In the general case, we can
introduce soft Wilson lines to make each sector gauge invariant and
remove them after the matching is completed. We shall come back to
this point in Section~\ref{sec:operatorsGeneral}.

\subsubsection{Current operators\label{sec:currentops}}

We first discuss the simplest case, namely SCET$_{\rm I}$ operators of the 
form
\begin{equation}
{\cal O}= [ \text{``soft''} ]\times [ \text{``$n$-hard-collinear''}]
 \times A_{\hcnb\perp}^{\rm (em)}\,,
\end{equation}
where we use the schematic notation of (\ref{eq:scetidec}) for the special 
case where only the photon field appears in the ``$\nb$-hard-collinear'' 
bracket. These operators arise when the photon is emitted from one of the
current quarks, but our discussion does not depend on this fact. The analysis 
for this case is identical to that for the current operators defining 
heavy-to-light form factors. The construction of the general SCET$_{\rm II}$ 
operator basis relevant at leading power has been performed in 
\cite{Beneke:2003pa,Lange:2003pk}. We now rederive these
results as a preparation for the general case, and to introduce our method.
We start by writing out a list of the lowest-dimension current
operators in SCET$_{\rm I}$.  By momentum conservation, the operators must 
contain at least one hard-collinear field.  
We will see below that it is most convenient to classify SCET$_{\rm I}$ 
operators according to mass dimension, rather than power counting in 
$\lambda$. Up to dimension five, we find: 
\begin{align}
\label{eq:scetiff}
  d=3 :\;\;& \bar\X_{hc}\, \Gamma^{\prime}\,h \quad (J^A) \,, \nonumber\\
  d=4 :\;\;& \bar\X_{hc}\, \A_{hc\perp}\,\Gamma^{\prime}\,h \quad(J^B)
  \,,\;\;\;
  \bar{q}_s\, \A_{hc\perp}\,\Gamma^{\prime\prime}\,h \,, \nonumber\\
  \;\;& \bar\X_{hc}\, {\partial}_{hc\perp}
  \,\Gamma^{\prime}\,h\,,\;\;\; \bar\X_{hc}\, D_{s\perp} \,
  \Gamma^{\prime}\,h \,, \nonumber\\
  d=5 :\;\;& \bar\X_{hc}\, \A_{hc\perp}\,
  \A_{hc\perp}\,\Gamma^{\prime}\,h \,, \;\;\;
  \bar{q}_s\,\A_{hc\perp}\, \A_{hc\perp}\,\,\Gamma^{\prime\prime}\,h \,, \nonumber\\
  & \bar\X_{hc}\,
  \nb\cdot\partial_{hc}\,n\cdot\A_{hc}\,\Gamma^{\prime}\,h \,, \;\;\;
  \bar\X_{hc}\,{1\over \nb\cdot\partial_{hc}} {\sla{\nb}\over
    2}\Gamma^{\prime}\,
  \X_{hc}\,\,\bar\X_{hc}\,\Gamma^{\prime}\,h \,, \\
  & \bar\X_{hc}\,{1\over \nb\cdot\partial_{hc}} {\sla{\nb}\over
    2}\Gamma^{\prime}\, \X_{hc}\,\,\bar{q}_s\,\Gamma^{\prime\prime}\,h
  \,, \;\;\; \bar{q}_{s}\,
  \nb\cdot\partial_{hc}\,n\cdot\A_{hc}\,\Gamma^{\prime\prime}\,h \,, \nonumber\\
  &\bar{q}_s\,
  \nb\cdot\partial_{hc}\,\A_{hc\perp}\,\frac{\sla{n}}{2}\Gamma^{\prime}\,h\,,
  \; \dots \,.\nonumber
\end{align}
The symbols in parentheses, $J^A$ and $J^B$, anticipate the notation to be 
introduced in Section~\ref{sec:Matching} for the relevant SCET$_{\rm I}$ 
operators. We do not display transverse Lorentz indices or color indices; 
the former may be contracted with the metric and epsilon tensor in the
transverse plane, 
\be\label{eq:gperp}
g_\perp^{\mu\nu} = g^{\mu\nu}-\frac{1}{2}(\nb^\mu n^\nu+n^\mu\nb^\nu) \,,
\qquad
  \epsilon_\perp^{\mu\nu}=\frac{1}{2}\epsilon^{\mu\nu\alpha\beta}\,
 {\bar n}_\alpha n_\beta\,.
\ee
We use the convention $\epsilon^{0123}=-1$. 
In the schematic notation of (\ref{eq:scetiff}), it is understood that 
hard-collinear derivatives can act on any of the hard-collinear 
fields in the operators, and similarly for soft derivatives. 
We do indicate the Dirac structures that can occur in the above expressions, 
using the following Dirac matrices that are invariant under the ``boost''
$n_\mu \rightarrow \alpha n_\mu$ and $\nb_\mu \rightarrow \alpha^{-1}
\nb_\mu$: 
\begin{equation}\label{eq:Gammaprime}
\begin{aligned}
\Gamma^\prime&=\{1,\gamma_5,\gamma_\perp^\mu\} \,, \\ 
\Gamma^{\prime\prime}&=\Gamma^{\prime}\cup\{\nb\!\!\!/\,n\!\!\!/,
\gamma_\perp^\mu\gamma_5,\gamma_\perp^{\mu}\gamma_\perp^{\nu}
-\gamma_\perp^{\nu}\gamma_\perp^{\mu}\} \,. 
\end{aligned}
\end{equation}
The sixteen matrices $\sla{n}\Gamma'$, $\sla{\nb}\Gamma^{\prime}$, and 
$\Gamma^{\prime\prime}$ form a Dirac basis. We only consider boost-invariant 
operators; such a choice is always possible and is also natural because the 
operators we reproduce with the effective theory are independent of the 
reference vectors $n_\mu$ and $\nb_\mu$.%
\footnote{There are additional constraints arising from the independence
  from the reference vectors. Requiring complete reparameterization
  invariance, also under $n^\mu \to n^\mu + \epsilon_\perp^\mu$,
  yields relations linking operators of different orders in the power
  counting. Since we are concerned only with the leading order, such
  transformations will not be relevant to the present discussion.
}
Operators that are not boost-invariant can be eliminated in favor of
invariant operators obtained by multiplying them with an
appropriate number of derivatives $\nb\cdot\partial_{hc}$.  The last
five operators of dimension five are examples where such derivatives have
been included. The presence of these derivatives can be compensated
by the Wilson coefficients of the non-local operators, for example
\begin{equation}\label{eq:partint}
\int ds\, C(s)\, \nb\cdot\partial_{hc} \phi_{hc}(x+s \nb) \phi_s(x) = 
- \int ds \frac{\partial C(s)}{\partial s}\, \phi_{hc}(x+s \nb) \phi_s(x) \,.
\end{equation}

We did not allow for operators which explicitly involve the vector $v_\mu$ in 
(\ref{eq:scetiff}) because it can be eliminated in favor of $\nb_\mu$, 
cf.~(\ref{eq:nbardef}).  Furthermore, we have used the projection properties 
of the spinors $\sla{v} h = h$, $\sla{n}\X_{hc}= 0$ to eliminate occurrences 
of $\sla{\nb}\,h$. For instance, the third operator for $d=5$ is obtained by 
rearranging an operator with $d=4$:
\begin{equation}
\label{eq:dimeg}
\begin{aligned}
 \bar\X_{hc}\,\Gamma^\prime\,
  {\sla{\nb}\over 2}\,n\cdot\A_{hc}\,h  
&= 
 \bar\X_{hc}\,\Gamma^\prime\,
  \left( \sla{v} - {\sla{n}\over 2 n\cdot v}\right)
  {1\over n\cdot v} \,n\cdot\A_{hc}\,h = \frac{1}{n\cdot v}\bar\X_{hc}\,
  \Gamma^\prime\,n\cdot\A_{hc}\,h \\ 
&\to  \bar\X_{hc}\,
  i\nb\cdot\partial_{hc}\,n\cdot\A_{hc}\,\Gamma^\prime\,h \,.
\end{aligned}
\end{equation}
In the second line, we have absorbed a factor 
$(i\nb\cdot\partial\,n\cdot v)^{-1}$ into the Wilson coefficient of the 
operators, as in (\ref{eq:partint}). To minimize the list of possible 
SCET$_{\rm I}$ operators appearing with a given dimension, it is convenient 
to always make use of such rearrangements.%
\footnote{Another possibility would be to group the $\sla{\nb}$
  appearing on the left-hand side of (\ref{eq:dimeg}) together with
  the heavy-quark field.  Since the heavy-quark does not
  participate in the matching of SCET$_{\rm I}$ onto SCET$_{\rm II}$, 
  the general SCET$_{\rm II}$ operator is
  given by examining the matching of a boost non-invariant operator
  containing $n\cdot\A_{hc}$.  This approach is taken in
  \cite{Beneke:2003pa}, and using a larger set of building blocks,
  such operators can be shown not to contribute at leading power.}

For contributions arising at leading power, the SCET$_{\rm I}$ operators 
should not contain soft fields in addition to those found in the
final SCET$_{\rm II}$ operators. 
Such soft fields would not participate in the matching, 
and result in a power suppression relative to the corresponding operators 
without the additional soft fields.  
Similar arguments apply to power-suppressed soft or collinear derivatives;   
an explicit example of this effect was mentioned in (\ref{eq:extraderiv}) of 
Section~\ref{sec:scetii}. Thus only the first two operators for $d=4$ can be 
relevant for our leading-power analysis.  
We have also not listed operators of any dimension containing additional 
factors of $\nb\cdot D_s/\nb\cdot\partial_{hc}$. The ellipsis for
$d=5$ denotes similarly irrelevant terms.
 
\begin{table}
\begin{center}
\begin{tabular}{|c|c|c|}\hline
 & $d$ & $[\lambda]$ \\[3pt] \hline
$\frac{1}{\nb\cdot \partial_c}\, \bar \X_c \frac{\nb\!\!\!/}{2}
 \Gamma^{\prime} \X_c$ & 2 & 2 \\[3pt]
$\frac{1}{n\cdot \partial_s}\, \bar \Q_s \frac{n\!\!\!/}{2} \Gamma^{\prime}
 \Q_s$ &
2 & 2\\[3pt]
$ \bar \Q_s  \Gamma^{\prime\prime} \Q_s$ & 3 & 3 \\ 
$\!\!{n\!\cdot\! \partial_s}\, \bar \Q_s \frac{\nb\!\!\!/}{2} \Gamma^{\prime}
\Q_s\!\!$ & 4  & 4\\[3pt] 
& $\phantom{\frac{1}{\nb\cdot \partial_c\,n\cdot \partial_s}}$ & \\[5pt]
\hline
\end{tabular} 
\begin{tabular}{|c|c|c|}\hline
 & $d$ & $[\lambda]$ \\[3pt] \hline
$g_\perp^{\mu\nu}$, $\epsilon_\perp^{\mu\nu}$ & 0 & 0 \\[3pt]
$\partial_{c\perp}^\mu$, $\A_{c\perp}^\mu$, $\partial_{s\perp}^\mu$,  
$\A_{s\perp}^\mu$ & 1 & 1 \\[3pt]
$\!\!{n\!\cdot\! \partial_s}\nb\!\cdot\! \partial_s$, 
${n\!\cdot\! \partial_s}\nb\!\cdot\! \A_s\!\!$ & 2 & 2\\[3pt] 
$\!\!{\nb\!\cdot\! \partial_c}n\!\cdot\! \partial_c$, 
${\nb\!\cdot\! \partial_c}\,n\!\cdot\! \A_c\!\!$ & 2 & 2\\[3pt] 
$\frac{1}{\nb\cdot \partial_c\,n\cdot \partial_s}$ & $-2$ & $-1$ \\[3pt]
\hline
\end{tabular}
\end{center}\vspace*{-0.3cm}
\caption{Boost-invariant building blocks for SCET$_{\rm II}$ operators, with
  their dimension $d$ and order $[\lambda]$ in the power expansion. Soft
  derivatives $\partial_s$ can act on any soft field in the operator,
  collinear derivatives $\partial_c$ on any collinear field. 
  $\Gamma^\prime$ and $\Gamma^{\prime\prime}$ are defined in 
  (\ref{eq:Gammaprime}), while $g_\perp^{\mu\nu}$ and 
  $\epsilon_\perp^{\mu\nu}$ 
  are defined in (\ref{eq:gperp}). Additional building blocks are obtained by
  hermitian conjugation or by replacing $\Q_s$ with $\H_s$.
\label{tab:build}}
\end{table}

In order to construct all SCET$_{\rm II}$ operators up to a given power, we 
work with the set of building blocks in Tables~\ref{tab:build} and 
\ref{tab:fermion}, which are invariant under soft and collinear gauge 
transformations.  Again, we choose to work with boost-invariant quantities. 
We will begin by using Table~\ref{tab:build} to describe the matching onto 
SCET$_{\rm II}$ operators corresponding to ``typical'' momentum
configurations in which the partons in the
initial- and final-state mesons all carry $\order(1)$ fractions of 
the total soft and collinear momenta, respectively. 
Such configurations are represented by operators with fermion content 
$\bar{\X}_c(\dots)\X_c \bar{\Q}_s(\dots)\H_s$. We will then consider 
``endpoint'' configurations using the generalization in 
Table~\ref{tab:fermion}. These configurations occur when the momentum 
fraction carried by one of the partons tends to zero, so that the parton may 
be absorbed from the initial into the final state without hard momentum 
transfer.  In particular, we will find configurations represented by operators 
with fermion content $\bar\X_{c}(\dots) \H_s$. In both cases, using the 
counting rules for the building blocks containing soft-collinear fields 
described by the second column in Table~\ref{tab:fermion}, we will show that 
the operators representing 
the weak current at leading power do not contain soft-collinear modes. 
At leading power,  soft-collinear modes appear only in 
time-ordered products of the weak current with subleading SCET$_{\rm II}$ 
Lagrangian interactions, and with subleading terms in the interpolating 
currents for the meson states.   

The presence of the building block 
$(n\cdot\partial_s\,\nb\cdot\partial_c)^{-1}$, which counts as an inverse 
power of $\lambda$, is troubling at first sight. Naively, one could think that
there would be infinitely many operators of a given
dimension and order in $\lambda$. However, this is not the case: if an inverse
derivative $(n\cdot\partial_s\,\nb\cdot\partial_c)^{-1}$ is added to a
given operator, then it is necessary to also add two other
building blocks with $d=1$ or one building block with $d=2$ at the
same time to obtain an operator of the same dimension. As is evident
from the tables, this inevitably makes the resulting operator at least
one power in $\lambda$ higher than the operator without the inverse
derivative.  In fact, from Table~\ref{tab:build} we see that for operators 
involving only soft and collinear fields, with zero fermion number in both 
the soft and collinear sectors, the difference between the dimension of the 
SCET$_{\rm I}$ operators and the order in SCET$_{\rm II}$ power counting is 
given precisely by the number of occurrences of the building block 
$(n\cdot\partial_s\,\nb\cdot\partial_c)^{-1}$. 

We focus first on the case of flavor non-singlet final states and will then 
discuss the modifications necessary for the flavor-singlet case. We begin by 
considering SCET$_{\rm II}$ operators with fermion field content 
$\bar\X_c(\dots)\X_c\,\bar\Q_s(\dots)\H_s$, corresponding to ``typical'' 
partonic configurations inside the initial- and final-state soft and collinear 
mesons. Using Table~\ref{tab:build}, and the fact that SCET$_{\rm I}$ 
operators have dimension $d\ge 3$, it follows that leading-power contributions 
from these configurations are $\order(\lambda^4)$. Later we will discuss other 
possible ``endpoint'' configurations, finding that they also appear at the 
same order in power counting. From Table~\ref{tab:build} we see that 
with the exception of $(n\cdot\partial_s\,\nb\cdot\partial_c)^{-1}$, 
the building blocks satisfy $[\lambda]\ge d$, so that leading-power operators 
of a given dimension must be generated with the 
minimal number of occurrences of this building block.    
Starting with the SCET$_{\rm I}$ operator in (\ref{eq:scetiff}) 
of dimension three, we find that the appropriate fermion field content cannot 
be obtained while 
remaining at $d=3$ without at least one occurrence of the inverse derivative.
The two possibilities at $\order(\lambda^4)$ are then
\begin{equation}\label{eq:fourA}
J^A \to  \frac{1}{\nb\cdot \partial_c\,n\cdot \partial_s} \, 
\bigg[ {1\over \nb\cdot \partial_c}
\bar\X_c \frac{\nb\!\!\!/}{2} \Gamma^{\prime} \X_c\bigg] \;
\bigg[ \frac{1}{n\cdot\partial_s}\, \bar \Q_s \frac{n\!\!\!/}{2}
 \Gamma^{\prime} \H_s\bigg]\, \big\{
  \partial_{c\perp}^\mu,\; \A_{c\perp}^\mu,\; \partial_{s\perp}^\mu,\;
  \A_{s\perp}^\mu\big\} \, 
\end{equation}
and
\begin{equation}\label{eq:fourA2}
J^A \to 
  \frac{1}{\nb\cdot \partial_c\,n\cdot \partial_s} \,
\bigg[{1\over \nb\cdot \partial_c} \bar\X_c \frac{\nb\!\!\!/}{2}
 \Gamma^{\prime} \X_c\bigg]\; 
\bar \Q_s \Gamma^{\prime\prime} \H_s  \,.
\end{equation}
As in the tables, the notation is schematic: it is understood that the
soft derivatives can act on any of the soft fields, and the collinear
derivatives on any of the collinear fields. 
Using the equation of motion for the soft light-quark field,
the above possibilities result in four independent operators, whose explicit 
forms are given in \cite{Lange:2003pk}. Their matrix elements can be expressed
in terms of (endpoint divergent) convolution integrals involving twist-2 and
twist-3, two- and three-particle LCDAs of the $B$ meson and the light meson.
The matching relations (\ref{eq:fourA}) and (\ref{eq:fourA2}) 
are represented by the term $\bar{\xi}_{oc}^{(5/2)} h$, shown at tree level 
in (\ref{eq:ocdiagrams}).

Next, let us consider the SCET$_{\rm I}$ current operators of
dimension four.  First, we observe that the operator $\bar{q}_s {\cal
  A}_{hc\perp} \Gamma^{\prime\prime} h$ does not match onto a leading
order SCET$_{\rm II}$ operator. Its soft bracket $\bar{q}_s
\Gamma^{\prime\prime} h$ is of order $\lambda^3$ and remains unchanged
in the matching. The gluon field ${\cal A}_{hc\perp}$ must then match
onto a $d=1$ operator with collinear field content
$\bar{\X}_c(\dots)\X_c$.  Inspection of the table shows that such an
operator is of order $\lambda^2$, making the
overall operator subleading. The only possibility to obtain a leading
SCET$_{\rm II}$ operator at $d=4$ is
\begin{equation}\label{eq:fourB}
J^B \to 
\bigg[{1\over \nb\cdot \partial_c} \bar \X_c \frac{\nb\!\!\!/}{2}
 \Gamma^{\prime} \X_c\bigg]\;
\bigg[ \frac{1}{n\cdot
    \partial_s}\, \bar \Q_s \frac{n\!\!\!/}{2} \Gamma^{\prime} \H_s\bigg] \,. 
\end{equation}
At tree level, the matching (\ref{eq:fourB}) is represented by the term 
$\bar{\xi}_c A_{oc\perp}^{(3/2)} h$ in (\ref{eq:hcdiagrams}).
At dimension five there are no possibilities for leading-power SCET$_{\rm II}$ 
operators, due to the constraint $[\lambda]\ge d$. We thus need the 
SCET$_{\rm I}$ operators only through $d=4$ for leading-power matching.  
Finally, from the second column in Table~\ref{tab:fermion}, we note that 
replacing any of the soft or collinear fields in (\ref{eq:fourA}), 
(\ref{eq:fourA2}), or (\ref{eq:fourB})
by soft-collinear fields results in power suppression.     

\begin{table}
\begin{center}
\begin{tabular}{|c|c|c|}\hline
 & $d$ & $[\lambda]$ \\[3pt] \hline
$\X_c$ & $\frac32$ & $1$ \\[3pt]
${1\over \nb\cdot\partial_c}{\not{\nb}\over 2}$ & $-1$ & $0$ \\[3pt]
$\Q_s$ & $\frac32$ & $\frac32$ \\[3pt]
${1\over n\cdot\partial_s}{\not{n}\over 2}$ & $-1$ & $-1$ \\[3pt]
$\Gamma^{\prime\prime}$ & 0 & 0 \\[3pt]
\hline
\end{tabular}\hspace*{-0.0cm}
\begin{tabular}{|c|c|c|}\hline
 & $d$ & $[\lambda]$ \\[3pt] \hline
$\theta$ & $\frac32$ & 2 \\[3pt]
$\sigma$ & $\frac32$ & $\frac52$ \\[3pt]
$D_{sc\perp}^{\mu}\!\!$ & 1 & $\frac32$ \\[3pt]
${\nb\!\cdot\!\partial_c} n\!\cdot\! D_{sc}$ & 2 & 2 \\[3pt] 
${n\!\cdot\!\partial_s} \nb\!\cdot\! D_{sc}$ & 2 & 2 \\[3pt] 
\hline
\end{tabular}\hspace*{-0.0cm}
\end{center}
\caption{
Boost-invariant building blocks for SCET$_{\rm II}$ operators containing 
non-zero fermion number and/or soft-collinear fields.  The Dirac structures 
$\Gamma^{\prime\prime}$ are 
defined in (\ref{eq:Gammaprime}).  $\sigma$ is the small-component projection 
of the soft-collinear fermion, as in (\ref{eq:sigma}).
}
\label{tab:fermion}
\end{table}

Our analysis has so far relied on the assumption that the field
content of the SCET$_{\rm II}$ operator is $\bar\X_c(\dots) \X_c\,
\bar\Q_s(\dots) \H_s$, corresponding to ``typical'' parton
configurations.  We now consider possible ``endpoint'' \ 
contributions, corresponding to SCET$_{\rm II}$ operators with fermion
field content $\bar\X_c(\dots)\H_s$.  For this purpose, we consider
the building blocks in the first column of Table~\ref{tab:fermion}, which 
generalize the first column of Table~\ref{tab:build} to allow the possibility 
of non-zero fermion number in the soft and collinear sectors.  Starting
with the SCET$_{\rm I}$ operator in (\ref{eq:scetiff}) of dimension
three, the leading SCET$_{\rm II}$ operator is 
\begin{equation}
\label{eq:twoA} 
J^A \to \bar \X_c\Gamma^{\prime}\H_s \sim \lambda^{5/2} \,.  
\end{equation} 
Again, from the second column in Table~\ref{tab:fermion} we note that 
replacing any of the soft or collinear fields in (\ref{eq:twoA})
by soft-collinear fields results in power suppression.     
The operator in (\ref{eq:twoA}) can yield a leading-power contribution to the 
form-factor analogue of the correlator
(\ref{eq:LSZ}) when combined with leading-power meson currents and
subleading Lagrangian interactions involving the soft-collinear modes
\cite{Lange:2003pk}. Essentially, these interactions are summarized by the 
term $S^{\rm induced(3/2)}_{s+c}$ in the effective action of
\cite{Becher:2003qh}.%
\footnote{
More precisely, in the presence of the external weak current, the vacuum 
correlator of soft-collinear fields defining $S_{s+c}^{\rm induced}$ includes 
an extra soft-collinear Wilson loop 
$S_{sc}^\dagger W_{sc}$~\cite{Lange:2003pk,Becher:2003kh}.
}
Leading contributions can also arise from
subleading meson currents and leading Lagrangian interactions.  A
contribution of this type to the $B\to K^*\gamma$ amplitude is
illustrated in the first line of Figure \ref{fig:matching}.  The
interpolating current for the $B$ meson takes the form 
\be
\begin{aligned}\label{eq:Binter}
\bar{b} \gamma_5 q 
\to\quad
& \bar{\H}_s \Gamma^{\prime\prime} \Q_s
  &+\quad & \bar{\H}_s \Gamma^{\prime\prime} \theta &+\quad & 
\bar{\H}_s \Gamma^{\prime\prime} \sigma &+\quad & \dots \\
\sim\quad & \quad \lambda^3 &+\quad & \lambda^{7/2}
 &+\quad& \lambda^4 &+\quad & \dots\,, 
\end{aligned}
\ee
where the small-component projection $\sigma$ of the soft-collinear fermion 
field is related to $\theta$ as in (\ref{eq:sigma}). 
Similarly for the light meson, taking for example the pseudoscalar case, 
\begin{equation}\label{eq:Minter}
\begin{aligned}
\bar{q}\gamma_5 {\sla{\nb}\over 2} q \to \quad
& \bar{\X}_c {\sla{\nb}\over 2} \Gamma^\prime \X_c &+\quad & 
\bar{\theta} {\sla{\nb}\over 2} \Gamma^\prime \X_c &+\quad & \dots \\ 
\sim\quad & \quad \lambda^2 &+\quad & \lambda^3 &+\quad & \dots\,. 
\end{aligned}
\end{equation}
The subleading currents for both mesons suppress the contribution
of $ \bar \X_c\Gamma^{\prime}\H_s$ by $\lambda^{3/2}$, so that it ends
up being of the same order as the contribution of the four-quark
operators.  Finally, mixed cases can also occur, where an
$\order(\lambda^{1/2})$, or $\order(\lambda)$, suppressed meson
current from (\ref{eq:Binter}) or (\ref{eq:Minter}) is combined with
an $\order(\lambda)$, or $\order(\lambda^{1/2})$, suppressed
Lagrangian interaction, respectively.  The relevant Lagrangian
interactions in this case are given by ${\cal L}_{c+sc}^{\rm int}$ and
${\cal L}_{s+sc}^{\rm int}$ in \cite{Becher:2003qh}.  By the same
reasoning, we find that all such endpoint configurations arising from
the $d=4$ SCET$_{\rm I}$ operator $J^B$ in (\ref{eq:scetiff}) are
power suppressed.  

Before ending our discussion of heavy-to-light form factors, we consider the 
case of flavor-singlet final states.  Operators corresponding to ``typical'' 
partonic configurations again have zero collinear fermion number, but may 
contain collinear gluon degrees of freedom in place of the fermion bilinear 
$\bar{\X}_c(\dots) \X_c$. Requiring also that the collinear fields carry the 
appropriate twist and color quantum numbers  to have overlap with the 
final-state collinear meson, there must be at least two such collinear gluon 
fields. From Table~\ref{tab:build} we see that 
the new operators are obtained by the replacements
\be
\label{eq:singletop}
\bigg[ {1\over \nb\cdot \partial_c} \bar\X_c \frac{\nb\!\!\!/}{2}
 \Gamma^{\prime} \X_c \bigg]
\big\{
  \partial_{c\perp}^\mu,\; \A_{c\perp}^\mu,\; \partial_{s\perp}^\mu,\;
  \A_{s\perp}^\mu\big\}
 \to 
\left\{ 
\begin{matrix} 
\A_{c\perp} \A_{c\perp}
\big\{
  \partial_{c\perp}^\mu,\; \A_{c\perp}^\mu,\; \partial_{s\perp}^\mu,\;
  \A_{s\perp}^\mu\big\}\,, \\
\nb\cdot\partial_c n\cdot\A_c \, \A_{c\perp} 
\end{matrix}
\right\}
\ee
in (\ref{eq:fourA}), and by the replacement
\be
\label{eq:singletop2}
\bigg[ {1\over \nb\cdot \partial_c} \bar  \X_c \frac{\nb\!\!\!/}{2}
 \Gamma^{\prime} \X_c \bigg]
 \to \A_{c\perp} \A_{c\perp}
\ee
in (\ref{eq:fourA2}) and (\ref{eq:fourB}).  From the leading SCET$_{\rm I}$ 
current, there will also be new SCET$_{\rm II}$ operators that combine with 
subleading soft-collinear Lagrangian interactions and meson currents to yield
leading-power contributions. From Tables~\ref{tab:build} 
and \ref{tab:fermion}, we find that at leading power the new operators are 
obtained by the replacement 
\be\label{eq:Xcreplace}
\bar\X_c \to \bar\Q_s {1\over n\cdot\partial_s}{\sla{n}\over 2} \A_{c\perp} 
\ee
in (\ref{eq:twoA}). 
Although the right-hand side of (\ref{eq:Xcreplace}) scales as 
$\lambda^{3/2}$ (compared to the left-hand side, which scales as $\lambda$), 
leading contributions to form-factor matrix elements may still be obtained 
from subleading Lagrangian interactions involving the soft-collinear modes, 
which in this case are essentially summarized by the term 
$S_{s+c}^{{\rm induced}(1)}$ of the effective action in \cite{Becher:2003qh}.  

\subsubsection{General operators \label{sec:operatorsGeneral}}

After this warm-up, we are ready to discuss the general case where the
photon is not necessarily part of the SCET$_{\rm I}$ operator.  
The new operators appearing in this case correspond to photon emission from 
the spectator quark. 
The argumentation will be similar to the previous section; however, we will
have to match also the $\nb$-hard-collinear part in (\ref{eq:sectors}):
\begin{equation}
\begin{matrix}
  [ \text{``$\nb$-hard-collinear''}] \to A_{\bar{c}\perp}^{\rm
  (em)}\times [\text{``soft''}]\times [\text{``soft-collinear''}]\, .
\end{matrix}
\end{equation}
The building blocks needed in this case are obtained from
Tables~\ref{tab:build} and \ref{tab:fermion} by exchanging $n$ and
$\nb$, dropping the collinear quark fields, and replacing the
collinear gluon with the photon field. Note that the definition of the
soft fields then involves Wilson lines in the $\nb$-direction. To
distinguish them from the soft-fields appearing in conjunction with
the $n$-collinear sector, we denote them by $\H_\sbr$, $\Q_\sbr$, and
$\A_\sbr$. We also recall that the SCET$_{\rm I}$ building blocks
introduced in (\ref{eq:scetibb}) are not invariant under soft gauge
transformations; strictly gauge-invariant combinations are given by
\be\label{eq:sgaugeinv} 
\X^{(0)}_{hc}(x)= S_s^\dagger(x_-)\X_{hc}(x) \,,
\quad \A_{hc}^{(0)\mu}(x) = S_s^\dagger(x_-) \A_{hc}^{\mu}(x) S_s(x_-)
\,, \ee with the soft Wilson line defined in (\ref{eq:softW}).  In general, 
the fields contained in the ``$n$-hard-collinear'' and
``$\nb$-hard-collinear'' brackets in (\ref{eq:scetidec}) are
not separately gauge-invariant.  In order to match onto the
building blocks in Tables~\ref{tab:build} and \ref{tab:fermion} in the
general case, we first translate to the gauge-invariant combinations
appearing in (\ref{eq:sgaugeinv}).

Let us again start by writing down a list of the relevant SCET$_{\rm I}$ 
operators. By momentum conservation, they must have at least one
$n$-hard-collinear and one $\nb$-hard-collinear field in addition to the 
heavy-quark field, and therefore start with dimension 
$d=d_n+d_{\bar{n}}\,(+d_{s})= 4$:
\begin{equation}
\label{eq:scetigeneral}
\begin{aligned}
d=4 &= \left\{ 
\begin{matrix}
 \phantom{0} 3+1 :  & & \phantom{123456789}
 \bar\X_{\hcn}\,\A_{\hcnb\perp}\,\Gamma^{\prime}\,h \quad (J^A) \,, \\
  \phantom{0}1+3 : & & \phantom{123456789}
 \bar\X_{\hcnb}\A_{\hcn\perp}\,\Gamma^{\prime}\,h \quad
  (J^D) \,,
\end{matrix}\right. \\
d=5 &= \left\{ 
\begin{matrix}
4+1:&
  \phantom{123456}\bar\X_{\hcn}\,\A_{\hcn\perp}\,
 \A_{\hcnb\perp}\,\Gamma^{\prime}\,h \quad (J^B) \,, &
\\[6pt] 
3+2:&
 \bar\X_{\hcn}\,\A_{\hcnb\perp}\,
 \A_{\hcnb\perp}\,\Gamma^{\prime}\,h\,, &
\bigg[ {1\over n\cdot\partial_{\hcnb}} \bar\X_{\hcnb}\,
 \frac{\not{n}}{2}\,\Gamma^{\prime}\,
   \X_{\hcnb}\bigg] \,\,\bar\X_{\hcn}\,\Gamma^{\prime}\,h \,,
\\[6pt] 
2+3:&
\bigg[ {1\over \nb\cdot\partial_{\hcn}} \bar\X_{\hcn}\,
   \frac{\not{\nb}}{2}\,\Gamma^{\prime}\,\X_{\hcn}\bigg]\,\,
   \bar\X_{\hcnb}\,\Gamma^{\prime}\,h \quad(J^C) \,, &
\bar\X_{\hcnb}\,
  \A_{\hcn\perp}\, \A_{\hcn\perp}\,\Gamma^{\prime} \,h\quad (J^E)\,,
\\[6pt] 
1+4:&
\bar\X_{\hcnb}\,
  \A_{\hcnb\perp}\, \A_{\hcn\perp}\,\Gamma^{\prime}\,h\,, &
\\[6pt] 
1+1+3:&
\bar{q}_s \, \A_{\hcn \perp} \, \A_{\hcnb\perp} \,
 \Gamma^{\prime\prime}\,h \,. &
\end{matrix}\right.
\end{aligned}
\end{equation}
The symbols $J^A, \dots , J^E$ in parentheses anticipate the notation
to be introduced for these operators in Section~\ref{sec:Matching}.
In constructing this list, we made the same simplifications as in
(\ref{eq:scetiff}) for the form-factor case. In the above operators
the field $\A_{\hcnb}$ stands for either the photon or a gluon field,
which are treated on the same footing.  In SCET$_{\rm II}$, we treat
the photon as a collinear field in the $\nb$ direction, 
$\A_{{\bar c}\perp}^{(\rm em)} \sim \lambda$.  Since it appears only once in
each operator, we are free to make such a scaling assignment.  

In the above list of operators, we have separately indicated the mass
dimensions of fields in the $n$-hard-collinear, $\nb$-hard-collinear,
and soft brackets, respectively. In those cases where the only soft
field is the heavy-quark field, we have included it in one of the
hard-collinear sectors in such a way that both hard-collinear brackets
carry zero fermion number. SCET$_{\rm II}$ operators with zero fermion
number can be constructed from the building blocks in
Table~\ref{tab:build} which fulfill $[\lambda]\ge d$. Beyond $d=5$,
operators appear which cannot be arranged to have zero fermion number
in each hard-collinear sector.  For example, 
\be d=6: \qquad
\bar\X_{\hcn}\,\Gamma^\prime\,\X_{\hcnb}\,\,\bar\X_{\hcn}\,
\Gamma^\prime\,h \,, \qquad
 \bar\X_{\hcnb}\,\Gamma^\prime
\X_{\hcn}\,\,\bar\X_{\hcnb}\, \Gamma^\prime\,h\,, \dots \,.  
\ee 
In Table~\ref{tab:fermion} we have generalized the first column of
Table~\ref{tab:build} to include building blocks with non-zero fermion
number, by splitting the various fermion bilinears in two halves in
all possible boost-invariant ways.  With the exception of $\X_c$, all
building blocks again satisfy $[\lambda]\ge d$.  In fact, since
$\bar{\X}_c\Gamma^{\prime\prime}\X_c =0$, the operator constructed
from the $n$- and $\nb$-hard-collinear sectors must contain a factor
$(\nb\cdot\partial_c)^{-1}\sla{\nb}/2$, so that the bound
$[\lambda]\ge d$ is recovered in the final operator. SCET$_{\rm I}$
operators with $d\ge 6$ are therefore not relevant to a leading-power
analysis.

For the $n$-hard-collinear sector, we may use Table~\ref{tab:build} to list  
the leading-order matching relations onto operators with minimal collinear 
field content $\bar\X_c(\dots) \X_c$. This yields:  
\begin{align} \label{eq:ndimension}
& d_n=1:\;
\A_{hc\perp}
&&\hspace{-12pt}\to 
S_s \left( \frac{1}{\nb\cdot\partial_c n\cdot \partial_s}
\bigg[ \frac{1}{\nb\cdot \partial_c}\, \bar \X_c \frac{\nb\!\!\!/}{2}
 \Gamma^{\prime} \X_c\bigg]
 \A_{s\perp} \right) S_s^\dagger 
&&\!\!\!\!\!\sim \lambda^2 
\nonumber\\
& d_n=2:\; 
\left\{ 
\begin{matrix}
\A_{hc\perp} \A_{hc\perp}\,, \\[6pt]
{1\over \nb\cdot\partial_{hc}}\bar\X_{hc} {{\not\nb}\over 2}\Gamma^{\prime}
 \X_{hc} 
\end{matrix}
\right\} 
&&\hspace{-12pt}\to 
{1\over\nb\cdot\partial_c} \bar\X_c {\sla{\nb}\over 2}\Gamma^{\prime} \X_c
&&\!\!\!\!\!\sim \lambda^2 \nonumber\\
& d_n=3:\;
\bar\X_{hc} \Gamma^\prime h 
&&\hspace{-48pt}\to 
\left\{ 
\begin{matrix}
\frac{1}{\nb\cdot\partial_c n\cdot \partial_s}
\bigg[ \frac{1}{\nb\cdot \partial_c} \bar \X_c \frac{\nb\!\!\!/}{2}
 \Gamma^{\prime} \X_c\bigg]
\, \bar{\Q}_s\Gamma^{\prime\prime} \H_s \,,
 \\[12pt]
 \frac{1}{\nb\cdot \partial_c\,n\cdot \partial_s} 
\bigg[  \frac{1}{\nb\cdot \partial_c} \bar
  \X_c \frac{\nb\!\!\!/}{2} \Gamma^{\prime} \X_c\bigg] 
\bigg[ \frac{1}{n\cdot\partial_s} \bar \Q_s \frac{n\!\!\!/}{2} \Gamma^{\prime}
 \H_s\bigg]   \big\{
  \partial_{c\perp}^\mu,\, \A_{c\perp}^\mu,\, \partial_{s\perp}^\mu,\,
  \A_{s\perp}^\mu\big\} 
\end{matrix}
\right\} 
&& \!\!\!\!\!\sim \lambda^4 \nonumber\\[6pt]
& d_n=4:\;
\bar\X_{hc} \A_{hc\perp} \Gamma^{\prime} h 
&&\hspace{-12pt}\to
\bigg[ \frac{1}{\nb\cdot \partial_c}\, \bar \X_c \frac{\nb\!\!\!/}{2}
 \Gamma^{\prime} \X_c\bigg] \;
\bigg[ \frac{1}{n\cdot \partial_s}\, \bar{\Q}_s\,{\sla{n}\over 2}
 \Gamma^\prime \H_s\bigg]
&& \!\!\!\!\!\sim \lambda^4
\end{align}
Note the presence of the soft Wilson lines, $S_{s}(\dots)\,S_{s}^\dagger$, 
for the $d_n=1$ case in (\ref{eq:ndimension}).  These factors are 
required in order to preserve soft gauge invariance, and can be derived via 
the field redefinitions (\ref{eq:sgaugeinv}). 
For the case of flavor-singlet final states, we may again build
additional operators using the replacements (\ref{eq:singletop}),
(\ref{eq:singletop2}).  Also, in the cases $d_n=1$ and $d_n=3$, operators with 
collinear field content $\A_{c\perp}$ appear at one order lower in $\lambda$ 
than those listed in (\ref{eq:ndimension}),
and can combine with subleading Lagrangian interactions to yield leading-power 
contributions, cf.~(\ref{eq:Xcreplace}).  

Similarly, in the $\nb$-hard-collinear sector, using the analogue
of Table~\ref{tab:build}, we find the leading operators with
$\nb$-collinear field content $\A_{{\bar c}\perp}^{(\rm em)}$: 
\begin{align}\label{eq:nbdimension}
d_{\nb}=1:& \quad
\A_{\hcnb\perp}
&&\hspace{-12pt}\to 
\A_{{\bar c}\perp}^{(\rm em)}
&& \sim \lambda 
\nonumber\\
d_{\nb}=2:& \quad
\left\{ 
\begin{matrix} 
\A_{\hcnb\perp} \A_{\hcnb\perp}\,,\\[6pt]
{1\over  n\cdot\partial_{\hcnb}}  \bar\X_{\hcnb}
 {{\not n}\over 2}\Gamma^{\prime} \X_{\hcnb} 
\end{matrix}
\right\} 
&&\hspace{-12pt}\to 
\A_{{\bar c}\perp}^{(\rm em)} \A_{\sbr\perp} 
&& \sim \lambda^2 
\nonumber\\
d_{\nb}=3:& \quad
\bar\X_{\hcnb}\Gamma^{\prime} h 
&&\hspace{-12pt}\to
\A_{{\bar c}\perp}^{(\rm em)} \bigg[ {1\over\nb\cdot\partial_s} 
\bar\Q_{\sbr} {{\sla{\nb}}\over 2} \Gamma^{\prime} \H_\sbr \bigg]  
&& \sim \lambda^3 
\nonumber\\
d_{\nb} =4:& \quad
\bar\X_{\hcnb} \A_{\hcnb\perp} \Gamma^{\prime} h
&&\hspace{-12pt}\to
\A_{{\bar c}\perp}^{(\rm em)} \bar\Q_\sbr \Gamma^{\prime} \H_\sbr \,, \quad 
\A_{{\bar c}\perp}^{(\rm em)} \bigg[ {1\over \nb\cdot\partial_s}
 \bar\Q_\sbr {\sla{\nb}\over 2 } \Gamma^{\prime} \H_\sbr \bigg]
\big\{ \partial_{s\perp}\,, \A_{\sbr\perp} \big\}  
&& \sim \lambda^4 
\end{align}
Note that the soft Wilson lines appearing in the SCET$_{\rm II}$ 
building blocks in (\ref{eq:nbdimension}) are 
in the opposite direction compared to those in (\ref{eq:ndimension}). 

Returning now to (\ref{eq:scetigeneral}), we find that dimension-four
operators with $d=d_n+d_{\bar{n}}=3+1$ ($J^A$) or $d=1+3$ ($J^D$) can
contribute at leading power.  Similarly, at dimension five, those
operators with $d=4+1$ ($J^B$) or $d=2+3$ ($J^C$, $J^E$) can
contribute at leading power.  The operators with
$d=3+1$, $d=4+1$ and $d=1+1+3$ have been treated already in
Section~\ref{sec:currentops}. They correspond to the case where the
SCET$_{\rm I}$ operator contains the photon field.  The remaining
operators, with $d=1+3$ and $d=2+3$, represent new contributions
corresponding to emission of the photon from the spectator quark.

\section{Matching and factorization \label{sec:Matching}}

In the previous section, we have found all effective-theory operators
that can contribute to the decay amplitude at leading power. Our analysis was
concerned with the field content of the operators and the occurrence of
inverse derivatives in SCET$_{\rm II}$, but we have not yet specified their
color and Dirac structures. In this section, we present the relevant operators 
in all detail. We evaluate the Wilson coefficients necessary for the 
phenomenological discussion in Section~\ref{sec:phenomenology} and show that 
the resulting matrix elements can be brought into the form of the 
factorization theorem (\ref{eq:factorization}).

\subsection{SCET$_{\rm I}$ matching}

We collect here the relevant SCET$_{\rm I}$ operators as derived in
Section~\ref{sec:tableology}.  Again, we first consider the operators
representing photon emission from one of the current quarks and then
discuss those operators corresponding to emission from the
spectator quark.  We initially restrict our attention to flavor
non-singlet final-state mesons.  The additional operators that arise for
flavor-singlet final states are considered separately at the end.

\subsubsection{Photon emission from the current quarks}

Two SCET$_{\rm I}$ operators are relevant for the case of photon emission 
from the current quarks, given by the $d=3+1$ and $d=4+1$ entries in 
(\ref{eq:scetigeneral}).  For the first of these, we write
\begin{equation}\label{eq:JA}
J^A(x,s,a)=\bar\X_{\hcn}(x_++s\nb+x_\perp)\,(1+\gamma_5)\,{\calAslash}^{{\rm
    (em)}}_{\hcnb\perp}(x_-+a n+x_\perp)\,h(0)\, e^{-im_b v\cdot x}\,.
\end{equation}
In order not to overburden the notation, we refrain from indicating
the flavor of the light-quark field.  The dependence on the parameters
$s$ and $a$ arises because the $n$-hard-collinear fields are allowed
to live at arbitrary points on the $\nb$-light-cone, and the
$\nb$-hard-collinear fields at arbitrary points on the $n$-light-cone
(cf.\ the discussion in Section~\ref{sec:building}). Furthermore, the position 
arguments of the fields have been multipole expanded, as appropriate for a 
product of fields $\phi_{\hcn}\phi_{\hcnb}\phi_s$: 
\begin{equation}
\phi_{\hcn}(x)\,\phi_{\hcnb}(x)\,\phi_s(x) 
= \phi_{\hcn}(x_+ + x_\perp)\,\phi_{\hcnb}(x_-+ x_\perp)\,\phi_s(0)
+ \dots \,,
\end{equation}
yielding the peculiar $x$ dependence of the fields in (\ref{eq:JA}). 

We use translational invariance to set $x=0$ and suppress the position
argument in the following. The representation of the weak Hamiltonian for 
photon emission from the current quarks reads
\begin{equation}
{\cal H}_W^{\rm current} \to \int\! ds\int\!da\, \tilde{C}^A(s,a) J^A(s,a) 
+ \sum_{j=1,2} \int\!ds\int\!dr\int\!da\, \tilde{C}_j^B(s,r,a) J_j^B(s,r,a)
  + \dots  \,,
\end{equation}
with the ellipsis denoting terms not relevant to a leading-power analysis.   
Here 
\begin{equation}
\begin{aligned}
J^A(s,a)&=  \bar{\X}_{\hcn}(s \nb)(1+\gamma_5) \, 
{\calAslash}^{{\rm (em)}}_{\hcnb\perp}(a n) \, h(0) \,, \\
J^B_1(s,r,a)&= \bar{\X}_{\hcn}(s\nb)(1+\gamma_5) \, 
{\calAslash}^{{\rm (em)}}_{\hcnb\perp}(a n) \,
\calAslash_{\hcn\perp}(r \nb) \, h(0) \,, \\
J^B_2(s,r,a) &=  \bar{\X}_{\hcn}(s\nb)(1+\gamma_5) \,
\calAslash_{\hcn\perp}(r \nb)
 \, {\calAslash}^{{\rm (em)}}_{\hcnb\perp}(a n) \, h(0) \,. 
\end{aligned}
\end{equation}
We define 
Fourier-transformed Wilson coefficients as
\begin{equation}\label{eq:Fourier}
\begin{aligned}
C^A(E, E_\gamma) &= \int\! ds\int\!da\, e^{is\nb\cdot P}
 e^{ia n\cdot P_\gamma} \tilde{C}^A(s,a) \,, \\
C_i^B(E, E_\gamma, u)&=\int\!ds\int\!dr\int\!da\,
 e^{i (us+\bar{u}r)\nb\cdot P} e^{ia n\cdot P_\gamma} \,
 \tilde C_i^B(s,r,a) \,,
\end{aligned}
\end{equation}
where $E \equiv n\cdot v \nb\cdot P/2$ and $E_\gamma \equiv n\cdot
P_\gamma/(2n\cdot v)$. The quantity $\nb\cdot P$ is the large
component of the total outgoing $n$-hard-collinear momentum, and
similarly $n\cdot P_\gamma$ is the large component of the outgoing
photon momentum.  We will suppress these quantities in the arguments
of the Wilson coefficients in the following. The variable $u$ denotes
the fraction of the large component of the $n$-hard-collinear
momentum carried by the quark field, and $\bar{u}=1-u$ is the fraction
carried by the gluon field.  The Wilson coefficients receive
contributions from different weak-interaction operators, and we give
separate matching results, $\Delta_i C^A$ and $\Delta_i C^B_{1,2}$,
for the different $Q_i$ in (\ref{eq:Hweak}).  For $b\to s$ transitions
we have
\begin{equation}
\begin{aligned}
C^A(\mu) &= {G_F\over \sqrt{2}}\sum_{p=u,c} V^*_{ps} V_{pb} 
\bigg[ \sum_{i=1,2}{ C}_i(\muQCD) \Delta_i^pC^A(\muQCD,\mu) 
+ \sum_{i=3}^8 {C}_i(\muQCD) \Delta_iC^A(\muQCD,\mu) \bigg] \,.
\end{aligned}
\end{equation}
The same expression with $s\to d$ gives the coefficient for $b\to d$ 
transitions. Analogous expressions define $\Delta_i C^B_{1,2}$. We will 
concentrate on the phenomenologically most relevant operators, which are 
$Q_7$, $Q_1$, and $Q_8$. The scale $\muQCD$ is the scale at which QCD and
the effective weak Hamiltonian are matched onto SCET$_{\rm I}$, and $\mu$ is 
the renormalization scale in the effective theory.

The matching coefficients for $Q_7$ are obtained directly
from the form-factor analysis and are given as
\be
\begin{aligned}\label{eq:delta7}
  \Delta_7 C^A &= {e\, \overline{m}_b\, E_\gamma \over 4\pi^2}\left(
    -2C_{T1}^A
    +{1\over 2}C_{T2}^A + C_{T3}^A \right) , \\
  \Delta_7 C_1^B &= {e\, \overline{m}_b\, E_\gamma \over 8\pi^2 E}
  \left( {1\over
      2}C_{T6}^{B'} + C_{T7}^{B'} \right) , \\
  \Delta_7 C_2^B &= {e\, \overline{m}_b\, E_\gamma \over 8\pi^2 E}
  \left( {1\over 2}C_{T2}^{B'}
    + C_{T3}^{B'} \right) .
\end{aligned}
\ee
The tensor-current Wilson coefficients have been calculated through
one-loop order, for $C_{Ti}^A(E)$ in
\cite{Bauer:2000yr,Beneke:2004rc}, and for $C_{Ti}^B(E,u)$ in
\cite{Beneke:2004rc,Becher:2004kk}.  Explicit expressions for the
combinations appearing in (\ref{eq:delta7}) are listed in 
Appendix~\ref{sec:formulae}. 
In the above expressions, the $\overline{\rm MS}$ quark mass must
be evaluated at the QCD matching scale, i.e., $\overline{m}_b\equiv
\overline{m}_b(\muQCD)$. For the process $B\rightarrow V\gamma$, we have
$2E_\gamma= m_B(1- m_V^2/m_B^2)$ 
and $2E =m_B$, with $E_\gamma$ and $E$ defined after (\ref{eq:Fourier}).   

\begin{figure}
\begin{center}
\subfigure[]{\label{fig:figQ1}
\begin{tabular}{ccccc}
\psfrag{q}{$Q_1$}\includegraphics[width=0.18\textwidth]{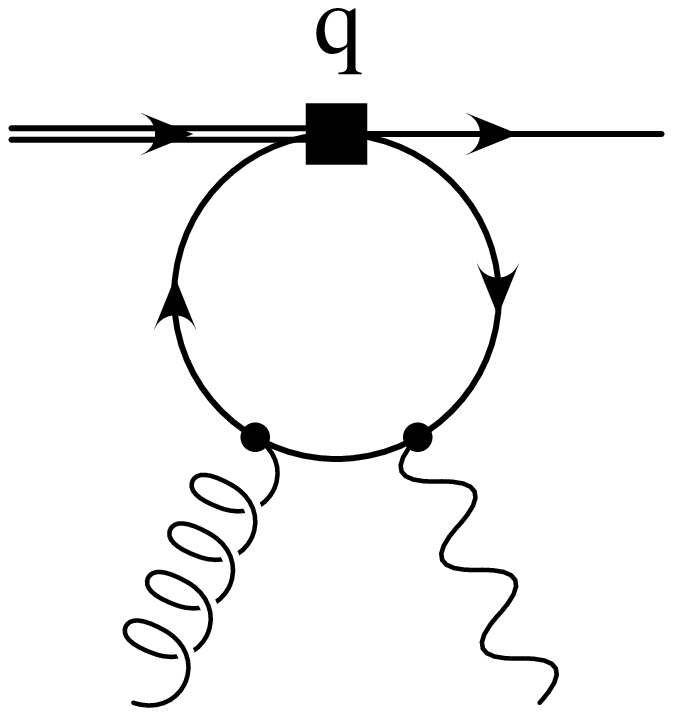} &&
\psfrag{q}{$Q_1$}\includegraphics[width=0.18\textwidth]{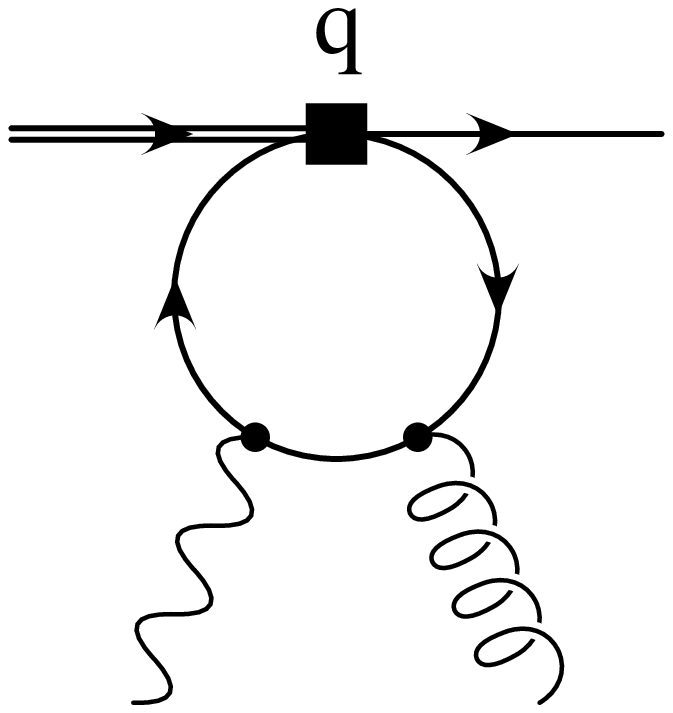} && 
\end{tabular}}
\subfigure[]{\label{fig:figQ8}
\begin{tabular}{c}
\psfrag{q}{$Q_8$}
\raisebox{0.083\textwidth}
{\includegraphics[width=0.24\textwidth]{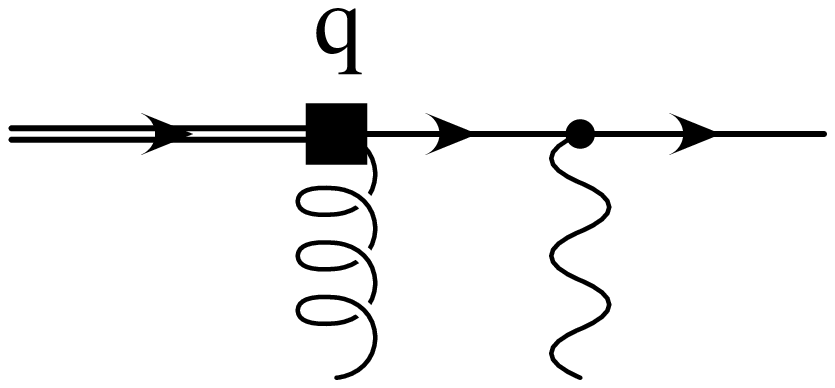}}
\end{tabular}
}
\end{center}
\vspace*{-0.5cm}
\caption{Leading-order QCD diagrams for the matching of $Q_1$ and $Q_8$ onto
 $J^B_i$. Other diagrams are power suppressed or vanish.\label{fig:Btype}}
\end{figure}

For the operators $Q_1^q$ ($q=u,c$) and $Q_8$, we may deduce the
one-loop matching onto $A$-type operators from results available in the 
literature~\cite{Greub:1996tg,Buras:2001mq}. We find
\be
\Delta_1^q C^A = {\alpha_s C_F\over 4\pi} G_1(x_q) \Delta_7 C^A \,, \qquad
\Delta_8   C^A = {\alpha_s C_F\over 4\pi} G_8 \Delta_7 C^A   \,,
\ee
where $x_q = \overline{m}_q^2/m_b^2$ (we set $m_u=0$).  
The expressions for $G_1(x)$ and $G_8$ are the same as those in
\cite{Bosch:2001gv}, and for convenience are reproduced in 
Appendix~\ref{sec:formulae}. 
The $B$-type matching is obtained from the diagrams in
Figure~\ref{fig:Btype}, from which we find
\be
\begin{aligned} \label{eq:deltac}
  \mbox{Figure~\ref{fig:figQ1}:}\qquad \Delta_1^q C_1^B(u) &=
  {E_\gamma\over 4\pi^2}\, \frac{2e}{3}\,
  f\!\left({\overline{m}_q^2\over 4\bar{u}E E_\gamma}\right) \,,\quad &
  \Delta_1^q C_2^B(u) &= - \Delta_1^q C_1^B(u) \,, \\
  \mbox{Figure~\ref{fig:figQ8}:}\qquad \Delta_8 C_1^B(u) &=
  {\overline{m}_b\over 4\pi^2}\, \frac{e}{3}\, {\bar{u}\over u} \,, &
  \Delta_8 C_2^B(u) &= 0 \,.
\end{aligned}
\ee
The expression for $f(x)$ is also given in Appendix~\ref{sec:formulae}. 

\subsubsection{Photon emission from the spectator quark} 

From Section~\ref{sec:tableology}, we also find leading-power
SCET$_{\rm I}$ operators corresponding to photon emission from the
spectator quark. These contributions arise from dimension-two
operators in the $n$-hard-collinear sector mapping onto purely
collinear fields, cf.~(\ref{eq:ndimension}).  The $n$-hard-collinear
fields must therefore transform as a color singlet in order for the
resulting operators to have non-zero matrix elements with the physical
meson states.  Also, for the chirality structure appearing in the
Standard Model, only a single Dirac structure is relevant.  Absorbing
a factor $1/(2E)$ into the Wilson coefficients, and using the
projection properties $\sla{v}h = h$, $\bar{\X}_{\hcnb}\sla{\nb}=0$,
the resulting four-quark operator takes the form
\be
\begin{aligned}\label{eq:JC1}
J^C_1(s,r,a) &= \bar{\X}_{hc}(s\nb)(1+\gamma_5) {\sla{\nb}\over 2}
 \X_{hc}(r\nb) \, 
\bar{\X}_{\hcnb}(an) (1+\gamma_5) {\sla{n}\over 2} h(0) \,. 
\end{aligned}
\ee

\begin{figure}
\begin{center}
\subfigure[]{\label{fig:Q1spectator}
\begin{tabular}{c}
\psfrag{q}{$Q_1$}\includegraphics[width=0.16\textwidth]{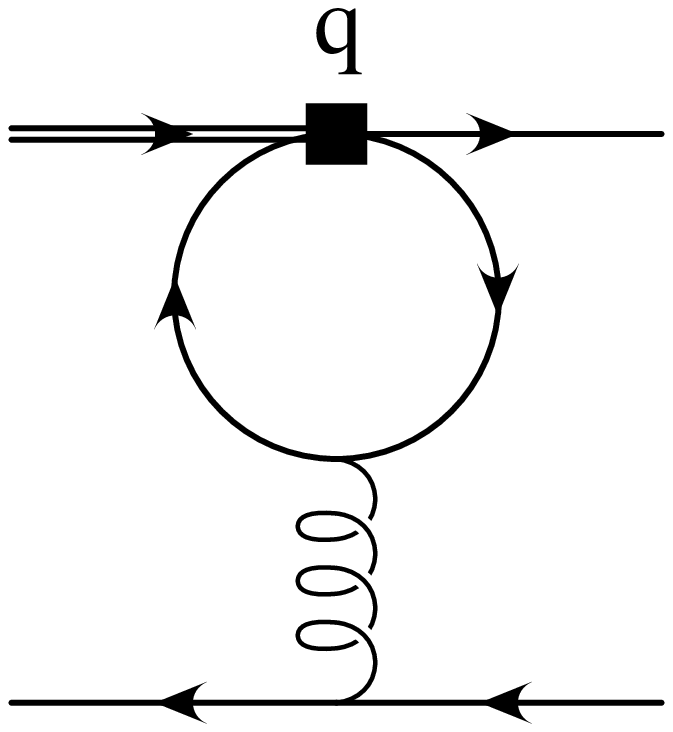}
\end{tabular}
}
\subfigure[]{\label{fig:Q8spectator}
\begin{tabular}{c}
\psfrag{q}{$Q_8$}\includegraphics[width=0.16\textwidth]{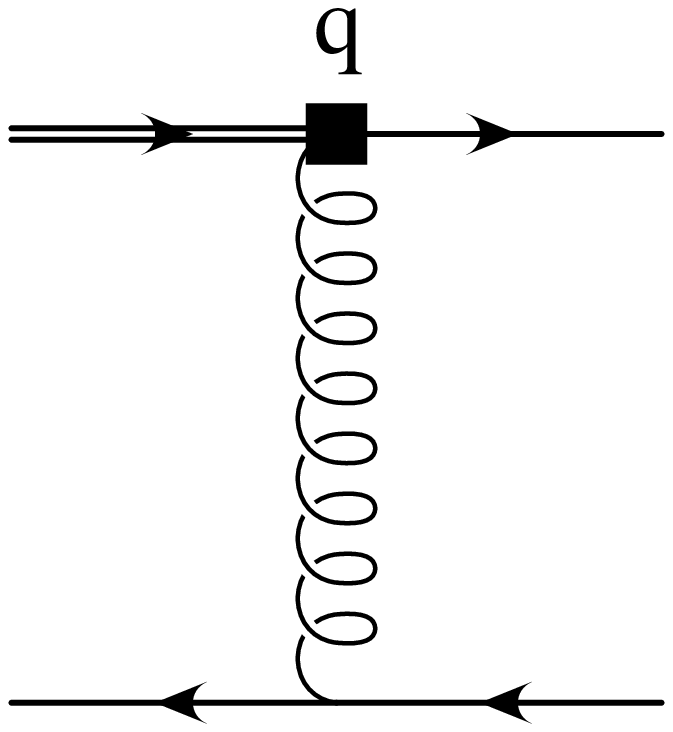}
\end{tabular}
}
\subfigure[]{\label{fig:Q1ann}
\begin{tabular}{c}
\psfrag{q}{$Q^u_{1,2}$}
\raisebox{0.0078\textwidth}{\includegraphics[width=0.16\textwidth]{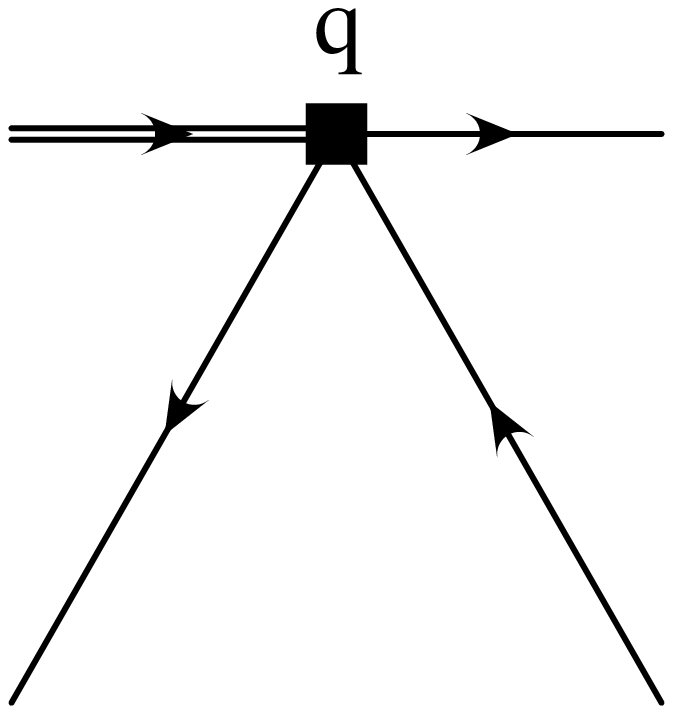}}
\end{tabular}
}
\end{center}
\vspace*{-0.5cm}
\caption{Leading-order QCD diagrams for the matching of $Q_{1,2}^p$ and
  $Q_8$ onto $J^C_i$.\label{fig:Ctype}}
\end{figure}

In the presence of New Physics, additional operators can appear in the
effective weak Hamiltonian. The class of non-standard operators
includes four-quark operators with scalar, pseudoscalar, or tensor
structures in place of the usual vector and axial-vector
structures~\cite{Borzumati:1999qt}.  In addition to $J^C_1$ in
(\ref{eq:JC1}), the following operators can then appear:
\begin{align}
J^C_2  &= \bar{\X}_{hc}(1+\gamma_5) {\sla{\nb}\over 2} \X_{hc} \, 
\bar{\X}_{\hcnb} (1-\gamma_5) {\sla{n}\over 2} h \,, &
J^C_3  &= \bar{\X}_{hc}(1-\gamma_5) {\sla{\nb}\over 2} \X_{hc} \, 
\bar{\X}_{\hcnb} (1+\gamma_5) {\sla{n}\over 2} h \,, \quad  \nonumber\\
J^C_4  &= \bar{\X}_{hc}(1-\gamma_5) {\sla{\nb}\over 2} \X_{hc} \, 
\bar{\X}_{\hcnb} (1-\gamma_5) {\sla{n}\over 2} h \,, &
J^C_5  &= \bar{\X}_{hc}(1+\gamma_5)\gamma_\perp^\mu {\sla{\nb}\over 2} \X_{hc}
 \, \bar{\X}_{\hcnb} (1+\gamma_5) {\sla{n}\over 2}\gamma_{\perp\mu} h \,,
 \quad \nonumber\\
J^C_6  &= \bar{\X}_{hc}(1-\gamma_5)\gamma_\perp^\mu {\sla{\nb}\over 2} \X_{hc}
 \, \bar{\X}_{\hcnb} (1-\gamma_5) {\sla{n}\over 2}\gamma_{\perp\mu} h \,. &&
\end{align}
The representation of the weak Hamiltonian for spectator-quark photon emission 
is then 
\be
\begin{aligned}
{\cal H}_W^{\rm spectator} &\to \sum_{k=1}^6 \int\!ds\int\!dr\int\!da\,
 \tilde{C}^C_k(s,r,a) J^C_k(s,r,a) + \dots \,.
\end{aligned}
\ee
In analogy with (\ref{eq:Fourier}) for the $B$-type 
operators, it is convenient to introduce the Fourier-transformed coefficients
\be
C_k^C(u)=\int\!ds\int\!dr\int\!da\, e^{i (us+\bar{u}r)\nb\cdot P}
 e^{ia n\cdot P_\gamma} \, \tilde C_k^C(s,r,a) \,.
\ee
The notation $n\cdot P_\gamma= 2E_\gamma n\cdot v$ anticipates that the
$\nb$-hard-collinear quark field matches onto the photon (and a soft
quark) in SCET$_{\rm II}$, see Figure \ref{fig:JCtoOC}. Clearly, $Q_7$
does not contribute to the matching onto $C$-type operators, and hence
\begin{equation}
\Delta_7 C^C_1 = 0 \,.
\end{equation}
Evaluating the first two diagrams shown in 
Figure~\ref{fig:Ctype} for the operators $Q_1^q$ and $Q_8$ yields
\be
\begin{aligned}
\mbox{Figure~\ref{fig:Q1spectator}:}\qquad  
\Delta^q_1 C^C_1 &= {2C_F\over N}{\alpha_s\over 4\pi}\left[
{2\over 3} + {2\over 3} \ln {4E E_\gamma\over \muQCD^2}-
G\!\left(\frac{\overline{m}_q^2}{4E E_\gamma},\bar{u}\right) \right] , \\
\mbox{Figure~\ref{fig:Q8spectator}:}\qquad  
\Delta_8 C^C_1 &= -{C_F\over N}{\overline{m}_b\over 2E}{\alpha_s\over \pi}
{1\over \bar{u}}\,.
\end{aligned}
\ee
The function $G(x,u)$ can be taken from \cite{Beneke:2003zv} and is
reproduced in Appendix~\ref{sec:formulae}.   
For the charged decay mode $B^- \to V^-\gamma$, the third diagram in 
Figure~\ref{fig:Ctype} also contributes:
\be
\mbox{Figure~\ref{fig:Q1ann}:}\qquad  
\Delta^u_1 C^C_1 = 2\,\delta_{qu} \,, 
\ee 
where $q$ refers to the flavor of the spectator quark inside the $B$ meson.  
 
\subsubsection{Flavor-singlet final states}

From (\ref{eq:scetigeneral}) we find two new types of SCET$_{\rm I}$ operators 
that can give rise to leading contributions. For the chirality structure 
appearing in the Standard Model, the following operators are relevant: 
\begin{equation}
\begin{aligned}
J^D(s,a) 
&=\bar{\X}_{\hcnb}(an)(1+\gamma_5)\,  \calAslash_{\hcn\perp}(s\nb) \, h(0)
 \,,\\
J^E(s,r,a) &=\bar{\X}_{\hcnb}(an)(1+\gamma_5)\, h(0)\,
(g_{\perp}^{\mu\nu}+ i\epsilon_{\perp}^{\mu\nu})\,
 \A^a_{\hcn\perp\mu}(s\nb) \A_{\hcn\perp\nu}^{a}(r\nb)  \,,
\end{aligned}
\end{equation}
with $g_\perp^{\mu\nu}$ and $\epsilon_\perp^{\mu\nu}$ as defined in 
(\ref{eq:gperp}). 
In writing the operator $J^E$ we have used the fact that at leading power the
$n$-hard-collinear fields match onto purely collinear fields (and no
soft fields) in SCET$_{\rm II}$, so that we may restrict attention to
color-singlet operators in both the $n$- and $\nb$-hard-collinear sectors.  
Note that the relative sign of the $\epsilon_{\perp}^{\mu\nu}$-term
in the operator $J^E$ is without significance. The operator with the flipped
sign is equivalent, if one also replaces $\tilde C^E(s,r,a)\rightarrow \tilde C^E(r,s,a)$.
Since the outgoing hadron is generated from gluonic degrees of
freedom, the operators $J^D$ and $J^E$ can only contribute for
flavor-singlet final-state hadrons.

As usual, we define
\be
\begin{aligned}
C^D &= \int\!ds\int\!da\, e^{i s\nb\cdot P}
 e^{ia n\cdot P_\gamma} \, \tilde C^D(s,a)  \,,\\
C^E(u) &= \int\!ds\int\!dr\int\!da\, e^{i (us+\bar{u}r)\nb\cdot P}
 e^{ia n\cdot P_\gamma} \, \tilde C^E(s,r,a) \,.
\end{aligned}
\ee
$Q_7$ does not contribute to the matching onto $D$- or $E$-type operators, 
and hence
\begin{equation}
\Delta_7 C^D = \Delta_7 C^E= 0 \,.
\end{equation}
The matching of the operator $Q_1$ onto $J^D$ and $J^E$ vanishes at zeroth
order in $\alpha_s$, and hence
\be
\Delta_1^q C^D = 0  \,, \qquad \Delta_1^q C^E =0  \,.  
\ee
For the matching of $Q_8$ onto $J^D$ and $J^E$, we find
\be
\Delta_8 C^D = -{ E\overline{m}_b  \over 2\pi^2} \,,
 \qquad \Delta_8 C^E =0 \,. 
\ee

\subsection{SCET$_{\rm II}$ matching \label{sec:scetiimatch}}

We now write down the operators in the final effective theory and
perform the matching of SCET$_{\rm I}$ onto SCET$_{\rm II}$.  The 
matching coefficients for this second step are called jet
functions.  We begin again with the flavor non-singlet case, considering
photon emission from the current quarks as well as from the
spectator quark. We then discuss the new ingredients needed for the treatment
of decays with flavor-singlet final states.

\subsubsection{Photon emission from the current quarks} 

The analysis in Section~\ref{sec:tableology} showed which operator
structures $J^A$ matches onto. The explicit form of these operators and
their leading-order jet functions are given in \cite{Lange:2003pk}.
However, since the non-factorizable part of the form factor can be
simply defined as the matrix element of the operator $J^A$, we do not
need to perform this second matching step explicitly.

The current operators $J^B_1$ and $J^B_2$ match onto
\begin{equation}
\begin{aligned}
  O^B_1(x=0,s,t) &= \bar{\X}_c(s
  \nb)\,(1+\gamma_5)\,{\calAslash}^{{\rm (em)}}_{\bar c\perp}(0)
  \,{\sla{\nb}\over 2} \X_c(0) \,\,
  \bar{\Q}_s(t n)\,(1-\gamma_5)\,{\nslash\over 2} \H_s(0) \,, \\
  O^B_2(x=0,s,t) &= \bar{\X}_c(s \nb)\,(1+\gamma_5) \,{\sla{\nb}\over
    2}\, \X_c(0) \,\, \bar{\Q}_s(t n)\,(1+\gamma_5)\,{\nslash\over
    2}\,{\calAslash}^{{\rm (em)}}_{\bar c\perp}(0)\, \H_s(0)\,,
\end{aligned}\label{eq:OB}
\end{equation}
and two operators with color structure $T^a\otimes T^a$, which have
vanishing meson matrix elements.  A consistent matching of SCET$_{\rm I}$ onto
SCET$_{\rm II}$ beyond tree-level involves evanescent operators that mix with 
the operators in (\ref{eq:OB})~\cite{Becher:2004kk}. Since we will be 
concerned primarily with an analysis at leading-order in RG-improved 
perturbation theory, and hence with matching coefficients only at tree-level, 
we do not list these operators here. The operators in (\ref{eq:OB}) correspond 
to the $d=4+1$ case in (\ref{eq:ndimension}), (\ref{eq:nbdimension}).  
At tree level, the inverse derivatives appearing in (\ref{eq:ndimension}) are 
accounted for via the relation
\be
{1\over i\nb\cdot\partial + i0} \phi(x) = -i \int_{-\infty}^0\! ds\,
 \phi(x+s\nb) \,,
\ee 
and similarly for $n\leftrightarrow \nb$.  Beyond tree level, the Wilson 
coefficients of the operators in (\ref{eq:OB}) also develop logarithmic 
dependence on the light-cone variables $s$ and $t$. As usual, we introduce the
Fourier-transformed coefficient
\be
D^B_i(\omega,u) \equiv \int\!ds\int\!dt\, e^{-i\omega n\cdot v t }
 e^{ius\nb\cdot P} \tilde{D}^B_i(s,t) \,.  
\ee
The Wilson coefficient of $O^B_1$ is a convolution of the SCET$_{\rm I}$ 
Wilson coefficient $C_1^B$ with a jet function ${\cal J}_\perp$,  
\begin{equation}\label{eq:convoluteJ}
D^B_1(\omega,u,\mu) = {1\over \omega} \int_0^1\!dy\,
{\cal J}_\perp\left(u,y,\ln{2E\omega\over\mu^2},\mu\right) C^B_1(y,\mu) \,.
\end{equation}
The
operator $O^B_2$ involves the jet function ${\cal J}_{\parallel}$.  At
tree level the two are identical,
\begin{equation}
{\cal J}_{\parallel}(u,v)_{\rm tree} = {\cal J}_{\perp}(u,v)_{\rm tree}
 = - {4\pi C_F \alpha_s\over N}{1\over 2E\bar{u}} \delta(u-v) \,.
\end{equation}
The one-loop results for the two jet functions
can be found in \cite{Becher:2004kk,Hill:2004if}.  
We may recall that in the form-factor analysis the hard-scale matching 
coefficients are constant at tree level, independent of momentum fractions.
In this case, up to hard-scale radiative corrections, 
expressions such as those appearing in (\ref{eq:convoluteJ}) collapse into 
a simple integral over the jet function.  
Convolution with the meson LCDAs then yields  
a universal function $H_M$, identical for all
form factors describing the same final-state meson $M$~\cite{Hill:2004if}.   
In contrast, for the $B\to V\gamma$ analysis we see from (\ref{eq:deltac}) 
that even at tree level the coefficients are momentum-fraction dependent, so 
that the approximate universality represented by $H_M$ cannot be utilized in 
this case. 

\subsubsection{Photon emission from the spectator quark} 

\begin{figure}
\begin{center}
\begin{tabular}{c}
\psfrag{q}{}
\includegraphics[width=0.32\textwidth]{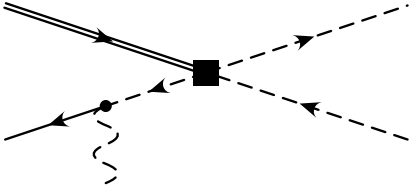} 
\end{tabular}
\end{center}
\vspace*{-0.5cm}
\caption{Matching of $J^C_i$ onto $O^C_i$. Dashed lines denote
  hard-collinear fields, full lines soft fields. Note that the quark
  that emits the photon is described by a hard-collinear field in the
  $\nb$-direction, while the other two quarks are collinear with the
  $n$-direction. \label{fig:JCtoOC}}
\end{figure}

For flavor non-singlet mesons, the only relevant SCET$_{\rm I}$
operators not already present in the form-factor analysis are $J^C_i$.
For the operator $J^C_1$, the corresponding SCET$_{\rm II}$ operator
is 
\begin{equation}
\label{eq:scetiiOC} O^C_1(x=0,s,t) = (n\cdot v)^2 \,
\bar{\X}_c(s \nb)\,(1+\gamma_5)\,{\sla{\nb}\over 2} \X_c(0) \,\,
\bar{\Q}_\sbr(t \nb)\,(1+\gamma_5)\, {\calAslash}^{{\rm (em)}}_{\bar
  c\perp}(0) {\nbslash\over 2} \H_\sbr(0) \,.  
\end{equation} 
The remaining $C$-type operators corresponding to non-standard
interactions are 
\begin{align}\label{eq:OC2}
O^C_2
&=   (n\cdot v)^2\, \bar{\X}_c\,(1+\gamma_5)\,{\sla{\nb}\over 2} \X_c \,\, 
     \bar{\Q}_\sbr\,(1-\gamma_5)\,{\calAslash}^{{\rm (em)}}_{\bar c\perp}
   {\nbslash\over 2}  \H_\sbr \,,  \nonumber\\
O^C_3
&=   (n\cdot v)^2\, \bar{\X}_c\,(1-\gamma_5)\,{\sla{\nb}\over 2} \X_c \,\, 
     \bar{\Q}_\sbr\,(1+\gamma_5)\, {\calAslash}^{{\rm (em)}}_{\bar c\perp}
   {\nbslash\over 2} \H_\sbr \,,  \\ 
O^C_4 
&=   (n\cdot v)^2\, \bar{\X}_c\,(1-\gamma_5)\,{\sla{\nb}\over 2} \X_c \,\, 
     \bar{\Q}_\sbr\,(1-\gamma_5)\, {\calAslash}^{{\rm (em)}}_{\bar
       c\perp} {\nbslash\over 2} \H_\sbr \,,  \nonumber\\
O^C_5
&=   (n\cdot v)^2\, \bar{\X}_c\,(1+\gamma_5)\gamma_\perp^\mu
   {\sla{\nb}\over 2} \X_c \,\, 
     \bar{\Q}_\sbr\,(1+\gamma_5)\, {\calAslash}^{{\rm (em)}}_{\bar
       c\perp}\gamma_{\perp\mu} {\nbslash\over 2} \H_\sbr \,, \nonumber\\ 
O^C_6 
&=   (n\cdot v)^2\, \bar{\X}_c\,(1-\gamma_5)\gamma_\perp^\mu
   {\sla{\nb}\over 2} \X_c \,\, 
     \bar{\Q}_\sbr\,(1-\gamma_5)\, {\calAslash}^{{\rm (em)}}_{\bar c\perp}
  \gamma_{\perp\mu} {\nbslash\over 2}\H_\sbr \,. \nonumber
\end{align}
Note that the soft building blocks in (\ref{eq:OB}) involve Wilson lines in 
the $n$-direction and a factor $\sla{n}$ next to $\bar{\Q}_s$, while the soft 
fields in (\ref{eq:scetiiOC}) and (\ref{eq:OC2}) involve Wilson lines in the 
$\nb$-direction and a factor $\sla{\nb}$ next to $\bar{\Q}_{\sbr}$. As a
result, the matrix elements of the soft parts of the $C$-type operators will 
involve the same $B$-meson distribution amplitude as the matrix element of 
$O^B_2$. We define Fourier-transformed Wilson coefficients (recall that 
$\nb\cdot v = 1/n\cdot v$)
\begin{equation}
D^C_i(\omega,u) \equiv \int\!ds\int\!dt\, e^{-i\omega \nb\cdot v t} 
e^{ius\nb\cdot P} \tilde{D}^C_i(s,t) \,, 
\end{equation}
and
\begin{equation}
\label{eq:scetiiDC}
D^C_i(\omega,u,\mu) = {e_q\over \omega} {\cal J}^C_{ij}
 \left(\ln{2E\omega\over \mu^2}, \mu \right) C^C_j(u,\mu) \,, 
\end{equation}
where $e_q \equiv 2e/3$ for an up-type quark, and $e_q\equiv -e/3$ for
a down-type quark. From the Feynman rules of SCET$_{\rm I}$ it follows that
${\cal J}^C_{ij}$ is proportional to the unit matrix,  
\begin{equation} 
{\cal J}^C_{ij} = \delta_{ij}\,{\cal J}^C \,. 
\end{equation}
To see this, we recall that 
the $n$-hard-collinear and $\nb$-hard-collinear parts of the operators
$J^C_i$ match independently onto SCET$_{\rm II}$. 
The different operators $J^C_i$, and also $O^C_i$, are distinguished only by 
the chirality of the fermion fields, and by 
the Dirac structure next to the heavy quark, both of which remain unchanged in 
the matching procedure. 
From the diagram in Figure~\ref{fig:JCtoOC}, we then find
\begin{equation}\label{eq:jetC}
 {\cal J}^C = 1 + \order(\alpha_s) \,.
\end{equation}

\subsubsection{Flavor-singlet final states} 

\begin{figure}
\begin{center}
\subfigure[]{\label{fig:singleta}
\begin{tabular}{cccc}
\psfrag{JD}[b]{$J_D$}\includegraphics[width=0.3\textwidth]{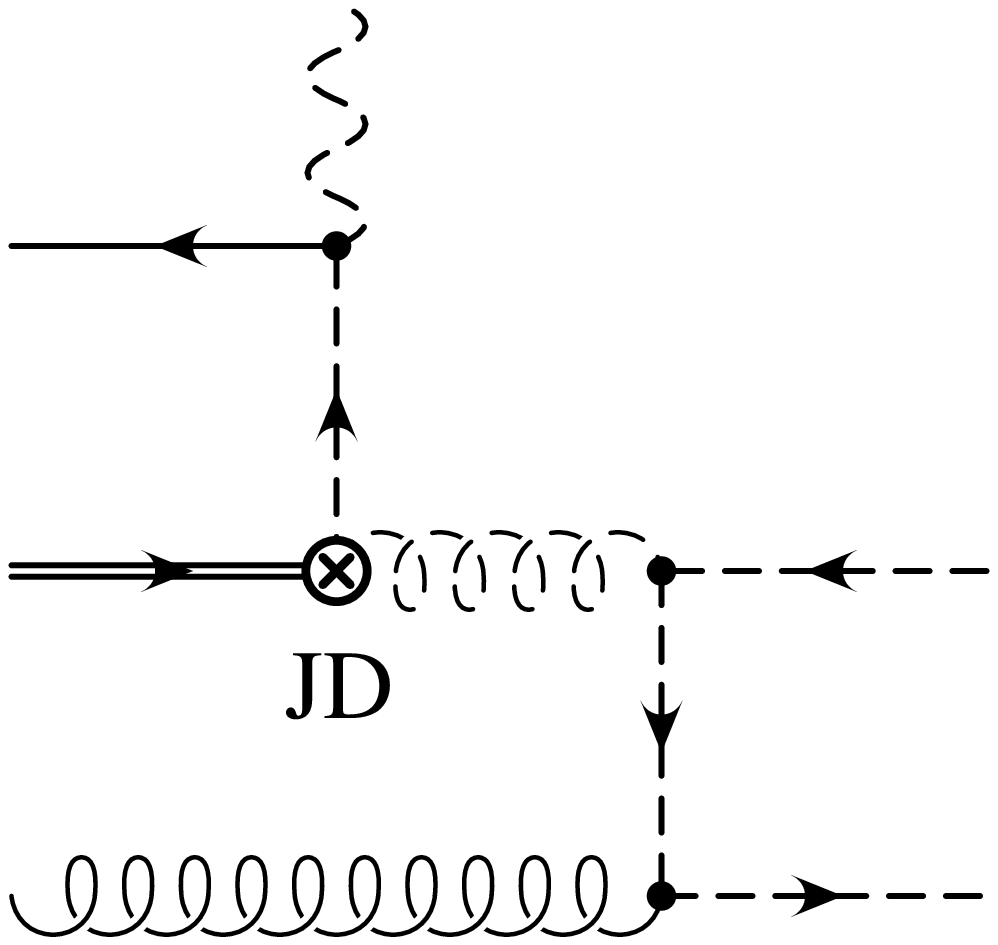} &&&
\end{tabular}
}
\subfigure[]{\label{fig:singletb}
\begin{tabular}{cccc}
\psfrag{JE}[b]{$J_E$}\includegraphics[width=0.3\textwidth]{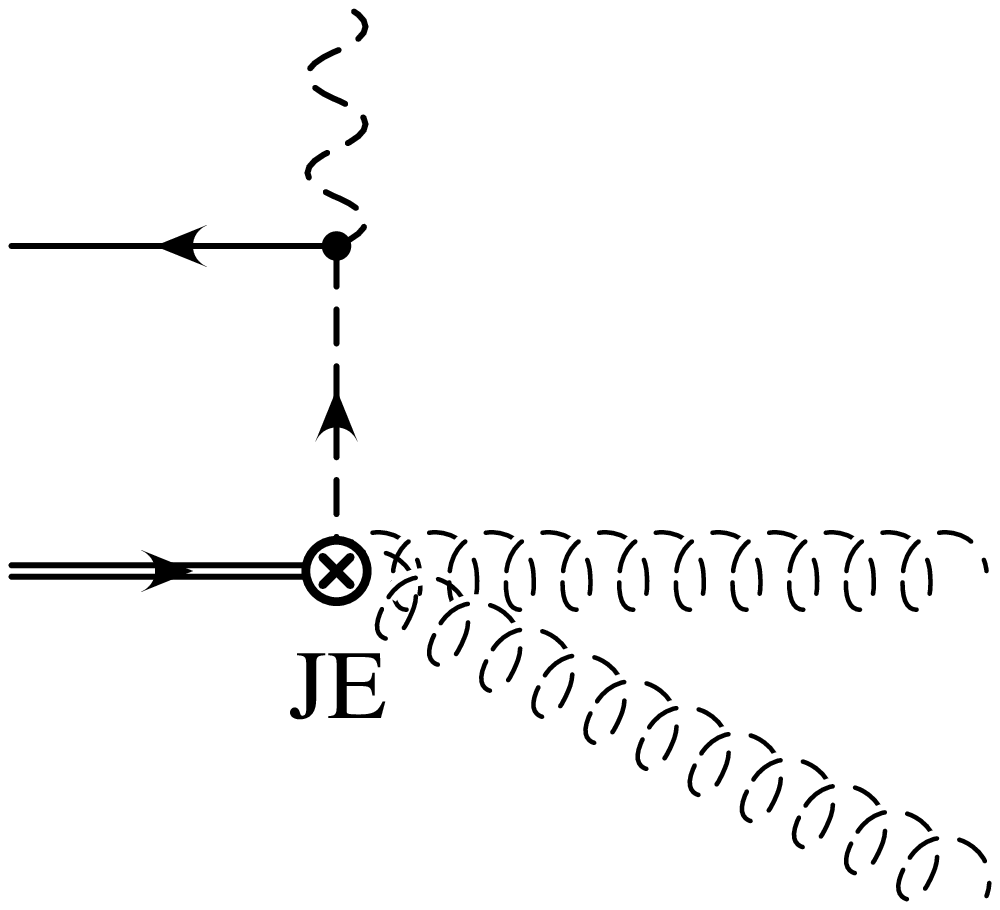} 
\end{tabular}
}
\end{center}
\vspace*{-0.5cm}
\caption{Leading-order SCET$_{\rm I}$ diagrams for the matching of the
  operators $J^D$ and $J^E$ onto SCET$_{\rm II}$ operators $O^D$ and
  $O^E$. Dashed lines denote hard-collinear fields, solid lines soft
  fields.}
\end{figure}

Finally, let us discuss the new ingredients involved when flavor-singlet 
final states are considered. The modifications in this case are of two types. 
Firstly, new operators appear in the matching of $A$- and $B$-type operators 
onto SCET$_{\rm II}$, corresponding to new contributions to form factors 
\cite{Beneke:2002jn}. The new $A$-type operators are related to those 
appearing already in the flavor non-singlet case by the replacements 
(\ref{eq:singletop}), (\ref{eq:singletop2}) and (\ref{eq:Xcreplace})
in (\ref{eq:fourA}), (\ref{eq:fourA2}) and (\ref{eq:twoA}), respectively. 
Similarly, the new $B$-type contributions are given by the replacement 
(\ref{eq:singletop2}) in (\ref{eq:fourB}). The symmetry relations obeyed by 
the $A$-type form-factor contributions remain unchanged in 
the flavor-singlet case; these contributions derive from  SCET$_{\rm I}$ 
currents $J^A_\Gamma = \bar{\X}_{\hcn} \Gamma h$,  and the symmetry relations 
follow directly from the projection properties of the spinor fields 
$\X_{\hcn}$ and $h$. The new $B$-type form-factor contributions are 
factorizable, involving the same $B$-meson LCDA, and the leading-twist 
two-gluon LCDA of the light meson. We concentrate here on the second new 
ingredient in the flavor-singlet case, namely the operators $J^D$ and $J^E$. 
These operators contribute {\it only} to flavor-singlet decays, and their 
contributions are unique to the radiative $B$-decay mode.  

The operator $J^D$ matches onto operators with collinear field content 
$\bar{\X}_c(\dots)\X_c$: 
\be\label{eq:ODops}
\begin{aligned}
O^{D}_1 &= (n\cdot v)^2\, \bar{\X}_c(s\nb) (1+\gamma_5) {\nbslash\over 2}
 \X_c(0) \,\, 
\bar{\Q}_{\sbr}(t^\prime\nb)(1+\gamma_5)  S_{\sbr}^\dagger(0) S_s(0)
{\calAslash}^{{\rm (em)}}_{\bar c\perp}(0) \calAslash_{s\perp}(tn)
  {\nbslash\over 2} \H_{s}(0)  \,, \\
O^{D}_2 &= (n\cdot v)^2\, \bar{\X}_c(s\nb) (1-\gamma_5) {\nbslash\over 2}
 \X_c(0) \,\, 
\bar{\Q}_{\sbr}(t^\prime\nb)(1+\gamma_5) S_{\sbr}^\dagger(0) S_s(0)
 {\calAslash}^{{\rm (em)}}_{\bar c\perp}(0)  \calAslash_{s\perp}(tn)
  {\nbslash\over 2} \H_{s}(0) \,,  
\end{aligned}
\ee
and also onto operators with purely gluonic collinear field content, given by 
the replacement (\ref{eq:singletop2}).  
The matching conditions in this case take the form 
\be\label{eq:DDdef}
D^D_i(\omega,\omega^\prime,u,\mu) = 
{e_q\over 2E\omega}{4\pi\alpha_s \over 2E\omega^\prime}\,
 {\cal J}^D_i\!\left(\ln{2E\omega\over \mu^2}, 
\ln{2E_\gamma\omega^\prime\over\mu^2}, u ,\mu \right) C^D(\mu) \,,
\ee
where we define 
\be
D_i^D(\omega,\omega^\prime,u) = 
\int ds \int dt \int dt^\prime e^{-i\omega n\cdot v t}
 e^{ i\omega^\prime \nb\cdot v t^\prime }
 e^{ius\nb\cdot P} \tilde{D}^C_i(s,t,t^\prime) \,.
\ee
The jet functions, obtained from the Feynman diagram in 
Figure~\ref{fig:singleta}, are given by 
\be
{\cal J}^D_1 = {1\over u N} + \order(\alpha_s) \,, \quad
{\cal J}^D_2 = {1\over \bar u N} +\order(\alpha_s) \,.
\ee
The operators $O^D_i$ are of leading power despite the fact that their soft 
part involves an additional gluon field. The matrix elements of the 
corresponding operators involve non-valence Fock states of the $B$ meson, but 
the presence of the extra gluon field 
is compensated by an additional inverse soft derivative. 
In a purely diagrammatic analysis, such a contribution can
easily be missed, while the operator analysis performed in the
previous section guarantees that all leading operators are included.
As discussed after (\ref{eq:ndimension}), $J^D$ also matches onto 
\be
\begin{aligned}
 O_3^D &= (n\cdot v) \, \bar{\Q}_{\sbr}(t\nb)(1+\gamma_5)
  S_{\sbr}^\dagger(0) S_s(0)
  \calAslash_{\bar{c} \perp}^{(\rm em)}(0) \calAslash_{c\perp}(0) \H_s(0) \,,
\end{aligned}
\end{equation}
which appears at one power in $\lambda$ lower than $O^D_1$ and $O^D_2$.
Similar to the A-type current in (\ref{eq:twoA}), this operator
describes an ``endpoint'' configuration of the two mesons where some of the
partons carry very small momenta. It can combine with subleading terms
in the SCET$_{\rm II}$ Lagrangian or subleading meson current operators
involving soft-collinear fields. The time-ordered product of $O^D_3$
with the subleading Lagrangians $ {\cal L}_{s+sc}$ and ${\cal L}_{c+sc}$
contains terms with the same field content as $O^D_1$ and $O^D_2$.
The interactions with soft-collinear fields ${\cal L}_{s+sc}$ and
${\cal L}_{c+sc}$ \cite{Becher:2003qh} are both suppressed by
$\lambda^{1/2}$ which makes the overall contribution leading power.
Its presence signals an infrared divergence at $\omega\to 0$ in
(\ref{eq:DDdef}), when the soft gluon in (\ref{eq:ODops}) becomes
soft-collinear. The analysis based on power counting of soft-collinear
modes provides a systematic procedure to determine the presence or
absence of such endpoint singularities. The $D$-type contribution
cannot be expressed in factorized form in terms of a finite
convolution integral over (generalized) meson LCDAs, and it is also
not related to the non-perturbative quantities appearing in the form
factor. Because these operators would contribute to $B^*\to P\gamma$,
we conclude that for flavor-singlet final states this decay mode does
not obey a factorization formula such as (\ref{eq:factorgeneral}).

The operator $J^E$ is related to $J^C$ by the replacement 
\be
\bar  \X_{hc} \frac{\nb\!\!\!/}{2\nb\cdot \partial_{hc}} \Gamma^{\prime}
 \X_{hc} \to 
\A_{hc\perp} \A_{hc\perp} \,.
\ee
It maps onto a SCET$_{\rm II}$ operator related to $O^C$ in 
(\ref{eq:scetiiOC}) by the corresponding replacement in (\ref{eq:singletop2}). 
The jet function in this case arises from the Feynman diagram in 
Figure~\ref{fig:singletb}, and is identical to ${\cal J}^C$ in 
(\ref{eq:scetiiDC}).  

\subsection{Matrix elements and factorization\label{sec:factor}}

The analysis of Section~\ref{sec:tableology} determined the SCET$_{\rm II}$ 
operator structures onto which the SCET$_{\rm I}$ current $J^A$ can be 
matched. Performing the matching explicitly and taking matrix elements yields 
expressions involving endpoint-divergent convolution 
integrals~\cite{Lange:2003pk}.  These infrared divergences indicate a 
sensitivity to endpoint momentum configurations, as verified by the presence 
of soft-collinear momentum regions at leading power.  
The contributions from such infrared momentum modes spoil factorization and 
cannot be calculated perturbatively.   Since we cannot reduce the $A$-type 
contribution to simpler hadronic quantities, we simply define 
the SCET$_{\rm I}$ matrix elements%
\footnote{
Since the quantity $\zeta_M(E)/\sqrt{m_{B^{(*)}}}$ is independent of
$m_b$, the heavy-quark flavor symmetry could be made manifest by 
extracting an additional factor 
$\sqrt{m_{B^{(*)}}}$ in the definition of $\zeta_M(E)$~\cite{Hill:2004rx}, 
similar to the definition of $F(\mu)$ in (\ref{eq:Fdef}) below.  
However, since we are concerned primarily with $B$ mesons, we will use the 
normalization (\ref{eq:zetadef}) that is commonly used in the literature.  
}
\be\label{eq:zetadef}
\langle M(p) | \bar{\X}_{hc} \Gamma h  | B^{(*)}(v) \rangle 
 = - 2E\,\zeta_M(E)\,  {\rm tr}\big[  \overline{{\cal M}}_M(n) \Gamma
 {\cal M}_{B^{(*)}}(v) \big] \,, 
\ee
where, as usual, $2E\equiv n\cdot v\,\nb\cdot p$. We have used the spinor 
wave-functions appropriate to the heavy-quark and large-energy limits:  
\begin{align}
{\cal M}_B(v) = {1+\sla{v}\over 2}(-\gamma_5) \,,
\quad&\quad  {\cal M}_{B^{*}}(v)= {1+\sla{v}\over 2} \sla{\eta} \,, \\
{\cal M}_P(n) = {\sla{n}\sla{\nb}\over 4}(-\gamma_5) \,,  
\quad\quad  {\cal M}_{V_\perp}(n) &= {\sla{n}\sla{\nb}\over 4}
 \sla{\eta}_\perp \,,
\quad\quad {\cal M}_{V_\parallel}(n) = -{\sla{n}\sla{\nb}\over 4}\,, \nonumber
\end{align}
where $\eta$ is the polarization vector in the case of vector mesons,
and $\overline{\cal M}\equiv \gamma^0{\cal M}^\dagger \gamma^0$. 

The same power-counting arguments in Section~\ref{sec:tableology} showed 
that subleading soft-collinear interactions are absent from the matrix 
elements of $B$-type operators.  
Let us note, however, that the SCET$_{\rm II}$ Lagrangian still contains leading-power
interactions of soft-collinear gluons with both soft and collinear
fields. In analogy to the decoupling of the soft-gluons in 
SCET$_{\rm I}$~\cite{Bauer:2001yt}, it is possible to perform field
redefinitions that remove the soft-collinear interactions from the
leading-order soft and collinear Lagrangians~\cite{Becher:2003kh}.
Under the same field redefinitions, the soft-collinear interactions also 
decouple from the operators $O^B_1$ and $O^B_2$, and therefore  
the matrix elements of these operators factorize at leading power;
the corresponding correlator diagrams consist of non-interacting soft and
collinear parts (see Figure~\ref{fig:matching}). 
The matrix elements of $O^B_1$ and $O^B_2$ can then 
be written as convergent convolution integrals over the meson LCDAs:
\be\label{eq:Fdef}
\begin{aligned}
\langle 0| \bar{\Q}_s(t n)\,{\sla{n}\over 2}\Gamma \H_s(0)
 |{B}^{(*)}(v) \rangle 
&= \frac{i F(\mu)}{2}\,\sqrt{m_{B^{(*)}}}\,{\rm tr}\bigg[{\sla{n}\over 2}
 \Gamma {\cal M}_{B^{(*)}}(v) \bigg]
\int_0^\infty
d\omega\,e^{-i \omega t n\cdot v}\,\phi_B(\omega,\mu) \,, \\
{ \langle M(p) | \bar\X_c(s\nb)\,\Gamma\,\frac{\nb\!\!\!/}{2}\X_c(0)
 |0\rangle }
&= \frac{i f_{M}(\mu) }{4}\nb\cdot p \,
{\rm tr}\bigg[ \overline{\cal M}_M(n)\, \Gamma  \bigg] 
 \int_0^1\!
du\, e^{i u s \nb\cdot p } \phi_ M(u,\mu) \,.
\end{aligned}
\ee
Let us note that the soft-collinear interactions are present, and do not 
decouple, for the operators
with color structure $T^a\otimes T^a$. The simple fact that an
operator can be written as a product of soft and collinear fields does
not guarantee factorization. 

We now collect the various elements and write down the leading-power 
decay amplitudes.  For the Standard Model prediction we have
\begin{align}
\langle V\gamma_L | {\cal H}_W | B(v) \rangle   &= 
 2m_B\,C^A(\mu) 
  \,\zeta_{V_\perp}\!\!\left(\frac{m_B}{2},\mu\right)  
+\frac{m_B^{3/2} F(\mu)}{2} \int_0^\infty\!
  \frac{d\omega}{\omega}\,\phi_B(\omega,\mu)
  \int_0^1\! du\,
   f_{V_\perp}\!(\mu)\phi_{V_\perp}(u,\mu) \nonumber\\
&\qquad \times \int_0^1\!dv {\cal
    J}_\perp\!\left(u,v,\ln{m_B\omega\over\mu^2},\mu\right)\,C^B_1(v,\mu)
   \nonumber\\
&\equiv   2m_B\left[C^A\,\zeta_{V_\perp}+\frac{\sqrt{m_B}F}{4} \phi_B\otimes
    f_{V_\perp}\phi_{V_\perp}\otimes {\cal J}_\perp \otimes C^B_1 \right] 
 , \nonumber\\
\langle V\gamma_R | {\cal H}_W | B(v) \rangle &=  0 \,,
 \label{eq:amplitude}\\
\langle P \gamma_L | {\cal H}_W | B^*(v) \rangle &=
 2m_{B^*}\left[ C^A \zeta_{P} +  \frac{\sqrt{m_{B^*}}F}{4} \phi_B
 \otimes f_P \phi_P \otimes e_q {\cal J}^C C^C_1 \right]
 , \nonumber\\
\langle P \gamma_R | {\cal H}_W | B^*(v) \rangle  &=  
-\frac{m_{B^*}^{3/2}F}{2} \phi_B \otimes f_P \phi_P \otimes
 {\cal J}_\parallel \otimes C^B_2  \,. \nonumber
\end{align}
The factors of 2 appearing in the
above matrix elements arise from evaluating the polarization
sums. For example, in the decay $B\to V\gamma$ the prefactor is
$\left(g_\perp^{\mu\nu} + i\epsilon_\perp^{\mu\nu}\right)
\varepsilon^{*}_{\mu}\eta^{*}_{\nu}=-2$ if both the photon and the
light meson have left-circular polarization, and zero otherwise. The metric
and epsilon tensor in the transverse plane were introduced in
(\ref{eq:gperp}). It is interesting to note that for right-circular photon 
polarization, the $B^*\to P\gamma$ amplitude is completely 
factorizable~\cite{Grinstein:2004uu}.   
The same is not true for $B^*$ decays with left-handed photon polarization, 
and here spectator emission gives rise to isospin violation at leading power. 

For flavor-singlet final states, new non-factorizable contributions, which are 
not already present in the form factors, arise from the $D$-type operators.  
These operators contribute to the process $B^*\to P\gamma$ for left-handed 
photon polarization, and we thus conclude that the amplitude for this process 
(in the flavor-singlet case) does not obey a factorization formula of the form 
(\ref{eq:factorgeneral}).  For right-handed photon polarization, on the other 
hand, the amplitude for $B^*\to P\gamma$  remains factorizable in the 
flavor-singlet case. 

In the presence of New Physics operators with a chirality structure
different from the Standard Model, one obtains additional contributions
\be\label{eq:nonstandard}
\begin{aligned}
\langle V\gamma_L | {\cal H}_W^{\rm NP} | B(v) \rangle  &= 
-m_B^{3/2}F  \phi_B \otimes f_{V_\perp} \phi_{V_\perp}
\otimes e_q {\cal J}^C C^C_5  \,, \\
\langle V\gamma_R | {\cal H}_W^{\rm NP} | B(v) \rangle  &= 
+m_B^{3/2}F  \phi_B \otimes f_{V_\perp} \phi_{V_\perp}
\otimes e_q {\cal J}^C C^C_6  \,. 
\end{aligned}
\ee
A contribution to the non-standard Wilson coefficients $C_5$ and $C_6$ would 
give a leading-power isospin-violating contribution to the $B\to V\gamma$ 
decay amplitude, with left- and right-handed photon polarization, 
respectively \cite{Kagan:2001zk}.  

The factorization formulas for the $B\rightarrow V$ (and $B^*\to P$) form 
factors at zero momentum transfer involve the same hadronic parameters as 
appear in (\ref{eq:amplitude}), but with
different Wilson coefficients $C^A$ and $C^B$.  For example,
\begin{align}\label{eq:vectorFF}
  \langle V(p',\eta) |\, \bar s\,\gamma^\mu\,b\, |  {B}(p)
  \rangle 
&= 2i\,\epsilon^{\mu\nu\rho\sigma}\eta^*_\nu
  p^{\prime}_\rho p_\sigma \, \frac{V(q^2)}{m_B+m_V}\\
  &= \frac{2i\epsilon^{\mu\nu\rho\sigma}\eta^*_\nu
  p^{\prime}_\rho p_\sigma}{m_B}  \left[ C^A_{V}\,
   \zeta_{V_\perp}(E)+\frac{\sqrt{m_B}F}{4} \phi_B\otimes
  f_{K^*_\perp}\phi_{K^*_\perp} \otimes {\cal J}_\perp\otimes C_V^B\right] \,,
\nonumber
\end{align}
where $q\equiv p-p^\prime$. Factorization theorems for all form factors are 
given in \cite{Hill:2004if}.  One can thus eliminate the non-factorizable 
piece $\zeta_{V_\perp}$ in (\ref{eq:amplitude}) in favor of a form
factor at $q^2=0$, and rewrite the resulting expression in the form of the
factorization theorem (\ref{eq:factorization}). Note that any choice
of the renormalization scale $\mu$ will lead to large perturbative
logarithms: the coefficients $C^A$ and $C^B$ contain logarithms of the
hard scale, $\ln(\mu^2/m_b^2)$, and the jet function contains logarithms of the
hard-collinear scale, $\ln(\mu^2/\omega m_b)$. One can resum these 
logarithms by solving the RG equations for the Wilson coefficients and the jet 
functions \cite{Hill:2004if}. The phenomenological impact of this resummation
will be discussed in Section~\ref{sec:phenomenology}.

\subsection{Light-quark masses} 

We have demonstrated factorization for $B\to V\gamma$ by expanding the 
weak Hamiltonian onto a complete basis of operators in the effective theory, 
and then isolating those contributions that cannot be absorbed into the 
$B\to V$ form factor.  For these contributions, we demonstrated the 
insensitivity to infrared momentum regions that would signal endpoint 
divergences in the hard-scattering convolution integrals.  
This infrared insensitivity in turn was demonstrated by the decoupling of the 
soft-collinear messenger modes that could potentially communicate between the 
soft and collinear sectors of the theory to spoil factorization.  
An interesting question to ask is how light-quark mass terms can affect the 
conclusions drawn from analyzing the massless case.  

Light quark masses of order $\Lambda$ can be 
expanded in the propagators of hard-collinear particles, 
\be
{1\over p_{hc}^2 - m_q^2} = {1 \over p_{hc}^2} 
\Big(1 + \order(\lambda) \Big)  \,, 
\ee
so that the quark masses can be ignored at leading power for such momentum 
regions. In contrast, the propagators of collinear or soft particles, 
\be
{1\over p_c^2 - m_q^2} \,, \quad 
{1\over p_s^2 - m_q^2} \,, 
\ee
cannot be expanded, so that the quark masses appear at leading power in the 
low-energy theory. For the region of soft-collinear momentum we have 
\be
{1\over p_{sc}^2 -m_q^2} = {1\over -m_q^2} \Big( 1 + \order(\lambda) \Big) \,, 
\ee
so that in the presence of such a light-quark mass, no pinch singularities 
arise in the fermion propagators from this region. The absence of the 
soft-collinear mode does not imply that all quantities in the low-energy 
theory factorize, but rather that the question of factorization has become
more subtle.  Individual Feynman diagrams for the soft and collinear modes 
contain divergences that are no longer regulated by dimensional 
regularization. Additional regulators may be introduced to make the diagrams 
individually well-defined, but such regulators link the soft and collinear 
sectors.  Demonstrating factorization then involves showing 
insensitivity to the additional regulator.  

We restrict ourselves here to the more tractable case where $m_q\ll
\Lambda$.  We recall that the fundamental object under study, the
correlator (\ref{eq:LSZ}), is free of infrared singularities, and so
has a smooth limit as $m_q\to 0$, keeping $\Lambda$ fixed.  The
dependence is not analytic, however, so that we cannot simply treat
the mass term as a perturbation, and work systematically to arbitrary
order.  The non-analyticity is associated with new regions that appear
in the presence of the mass term.  Such non-analyticities arise only
from the propagator denominators. In one-loop examples, we find that
the leading nonanalyticities are quadratic in the fermion masses as
long as $m_q\ll \Lambda$.  Terms linear in the mass appear only from
the numerator structure, with denominators described by the massless
case. Assuming that this property persists at higher order in
perturbation theory, it is
straightforward to show that the soft-collinear fields decouple in the
usual way from the new terms in the soft and collinear Lagrangians.
Factorization properties of the decay amplitudes are therefore
unchanged, except that the hadronic parameters ($\zeta_{V_\perp}$,
$\phi_{V_\perp}$, $\phi_B$) are modified by quark-mass effects. Beyond
linear order, the analysis may become more complicated.  If the
leading non-analyticities are quadratic in $m_q$, the
factorization formula (\ref{eq:factorization}) holds at least up to
terms of $\order(m_q^2/\Lambda^2)$.

\section{Phenomenology of \boldmath $B\to K^*\gamma$ \unboldmath  
\label{sec:phenomenology}}

From the decay amplitude, we obtain the following result for the
$B\to K^*\gamma$ branching fraction: 
\begin{equation}\label{eq:branching}
{\rm Br}( B\to K^*\gamma) = {\tau_B m_B \over 4\pi}\left(
  1-{m_{K^*}^2\over m_B^2}\right)\left| {\cal A} \right|^2 \,,
\end{equation}
where we introduce the notation 
\be
\begin{aligned}
{\cal A} &\equiv C^A \zeta_{K^*_\perp} +
  \frac{\sqrt{m_B}F}{4} \phi_B\otimes f_{K^*_\perp}\phi_{K^*_\perp}
 \otimes {\cal J}_\perp\otimes C_1^B \\
&\equiv {\cal A}_{\rm soft} + {\cal A}_{\rm hard} \,. 
\end{aligned}
\ee
Neglecting contributions proportional to $V_{ub}$, the Wilson coefficients are
\begin{align}
C^A&=\frac{G_F}{\sqrt{2}} V_{cs}^* V_{cb} \left[
  C_7+\frac{C_F \alpha_s}{4\pi} \Big( C_8 G_8
  +C_1 G_1(x_c) \Big) \right] \Delta_7 C^A +{\cal O}(\alpha_s^2)\,,\\ 
C_1^B&=\frac{G_F}{\sqrt{2}} V_{cs}^* V_{cb} \left[
  C_7+C_8\frac{\bar u}{3u}+C_1\frac{1}{3}f\!\left(\frac{\overline{m}_c^2}{4\bar u
  E E_\gamma}\right)
 \right]\Delta_7 C_1^B+{\cal O}(\alpha_s) \,.
\end{align}

In the above equations, we have suppressed the scale dependence of the
various quantities. The Wilson coefficients of the effective weak
Hamiltonian, $C_1$, $C_7$, and $C_8$, depend on the renormalization scale
$\muQCD$ in the ``full theory'' consisting of ordinary QCD and
the operators $Q_i$. The quantity $\Delta_7 C^A$ depends on $\muQCD$ 
as well as on the renormalization scale in SCET$_{\rm I}$,
$\mu$. This dependence on $\mu$ is canceled by the opposite
dependence of $\zeta_{K^*_\perp}(\mu)$. Since we will determine the
non-factorizable part ${\cal A}_{\rm soft}$ directly from a physical
form factor, it is simplest to choose the scale $\muQCD=\mu\sim m_b$ in 
this part. This choice guarantees the absence of
large logarithms in the perturbative expansion of $C^A$. Such a choice
is not appropriate for the factorizable part, ${\cal A}_{\rm hard}$,
but since the two parts are separately RG invariant, we are free to choose
different scales in the two parts \cite{Hill:2004if}.

Since we have performed two matching steps for the factorizable part,
a single choice for all renormalization scales inevitably leads to
large perturbative logarithms: 
the SCET$_{\rm I}$ matching coefficients depend on the hard
scale $m_b$, and the jet function on the hard-collinear scale
$\sqrt{\omega m_b}\sim \sqrt{\Lambda m_b}\approx 1.5\, {\rm GeV}$.
Also, the meson LCDAs are
typically given at a low renormalization scale $\mu \approx 1\, {\rm \,GeV}$. 
By solving the RG equation for the Wilson coefficients, one can sum up
perturbative logarithms of ratios of these scales, and match consistently 
onto the hadronic matrix elements at the low scale. 
Below we will give
the result obtained after resummation and compare
it to the fixed-order result.

\subsection{Non-perturbative input\label{sec:numerical}}

In order to evaluate the branching ratio, we need the value of
$\zeta_{K^*_\perp}$ at the kinematical point $q^2=0$, corresponding to 
maximum recoil energy of the $K^*$ meson, as well as the meson LCDAs 
$\phi_B$ and $\phi_{K^*_\perp}$ to evaluate the
factorizable part. Unfortunately, there is no direct experimental
information on these quantities available, so that we will rely on sum-rule
determinations.  
The value of $\zeta_{K^*_\perp}$ can be determined from any $B\rightarrow
K^*_\perp$ form factor at zero momentum transfer, because all such form factors
have the same non-factorizable part. We will use
the vector form factor $V$, which fulfills the factorization theorem
(\ref{eq:vectorFF}). This choice is convenient, since the factorizable
part of $V$ is $\order(\alpha_s)$ suppressed compared to other form factors, 
e.g.\ the tensor form factor $T_1$. In other words, 
$C^B_V$ vanishes at tree level, and we have
\begin{align}
\frac{m_B}{m_B+m_V}\,V^{B\to K^*}(q^2)= C^A_{V}(E,\mu)\,
\zeta_{K^*_\perp}(E,\mu)+{\cal O}(\alpha_s(m_b)\alpha_s(\sqrt{\Lambda m_b}))
 \,.
\label{eq:zeta}
\end{align}
The $B\rightarrow K^*$ form factors have been determined from
light-cone sum rules. The most recent evaluation gives $V(0)=0.411 \pm
0.045$ \cite{Ball:2004rg}. This value is compatible with, but somewhat lower
than the earlier result $V(0)=0.458 \pm 0.069$
\cite{Ball:1998kk}. Using the known one-loop expression for $C^A_{V}$
\cite{Bauer:2000yr,Beneke:2004rc}, the sum-rule determination yields 
\begin{equation}\label{eq:zetaV}
\zeta_{K^*_\perp}\equiv \zeta_{K^*_\perp}(E=\frac{m_B}{2},\mu=m_b)
 =0.40\pm0.04 \,.
\end{equation}
Let us check whether the sum-rule results for the axial form factor
$A_1$ and the tensor form factor $T_1$ give the same value of
$\zeta_{K^*_\perp}$, as required for consistency with the heavy-quark limit.  
The relations between the vector form factor $V$ and the axial form factor 
$A_1$ is especially simple, i.e.,\ 
\begin{align}
\frac{m_B^2}{(m_B+m_V)^2}\,\frac{V(0)}{A_1(0)}
&= 1 + \order(1/m_b) \,.
\end{align}
This form-factor relation does not receive perturbative corrections
\cite{Hill:2004if}. The sum-rule value for this ratio is very close to
unity, and the value $\zeta_{K^*_\perp}=0.39\pm0.05$ extracted from the axial
form factor is consistent with the value obtained from $V$. The
relation between the vector and tensor form factors is slightly more
complicated, since their Wilson coefficients $C^A$ and $C^B$ are
different.  In particular, the factorizable piece is not suppressed in
the case of the tensor form factor, for which the sum rule evaluation
gives $T_1(0)=0.33 \pm 0.04$~\cite{Ball:2004rg}.  Evaluating the
factorizable part using the hadronic input as given in
Table~\ref{tab:numerical}, and including the resummation effects as
discussed in Section~\ref{sec:resummation}, we obtain
$\zeta_{K^*_\perp}=0.37\pm 0.04$, again consistent with the value from
$V(0)$.

The $K^*$-meson LCDA is used as an input for the sum-rule evaluation of 
$V(0)$. Following \cite{Ball:2004rg}, we parameterize this function in terms 
of the lowest two Gegenbauer moments as 
\begin{equation}
\phi_{K^*_\perp}(u,\mu)=6u(1-u)\left[1+
 a_1(\mu)C_1^{3/2}(2u-1) + a_2(\mu)C_2^{3/2}(2u-1) \right] , 
\end{equation}
and we use $a_1(1\,{\rm GeV})=0.1\pm0.1$~\cite{Braun:2004vf}, $a_2(1\,{\rm GeV})=0.1\pm0.1$. For
the $K^*$ decay constant we use 
$f_{K_\perp^*}(1\,{\rm GeV})=170\pm 10\,{\rm MeV}$.  
For the $B$ meson, we take the model \cite{Braun:2003wx} (see 
\cite{Grozin:1996pq} for an alternative form)
\begin{equation}\label{eq:Bmodel}
\phi_B(\omega,\mu=1\,{\rm GeV})=\frac{4\lambda_B^{-1}}{\pi}
\frac{\omega\mu}{\omega^2+\mu^2}
\left[\frac{\mu^2}{\omega^2+\mu^2}-\frac{2(\sigma_B-1)}{\pi^2}\ln
  \frac{\omega}{\mu}\right] \,,
\end{equation}
with parameters $\lambda_B=460\pm 110\,{\rm MeV}$ and $\sigma_B=1.4\pm0.4$. 
To leading order in perturbation theory, and if no RG improvement is 
performed, the factorizable part depends only on the first inverse moment of 
the LCDA, defined as 
\begin{equation}
\lambda_B^{-1}(\mu)=\int_0^\infty \frac{d\omega}{\omega}
 \phi_B(\omega,\mu) \,. 
\end{equation}
Note that in the model (\ref{eq:Bmodel}), 
$\lambda_B(\mu=1\,{\rm GeV})=\lambda_B$. The quantity $F$ defined in 
(\ref{eq:Fdef})  is related to the $B$-meson decay constant,
$\sqrt{m_B} f_B = K_F(\mu) F(\mu)$, up to higher orders in $1/m_b$, with
\cite{Neubert:1993mb}
\begin{equation}
 K_F(\mu)=1+\frac{C_F \alpha_s(\mu)}{4\pi}\left(3\ln{m_b \over \mu} 
 - 2 \right) .
\end{equation} 
We use the value $f_B=200\pm 30\, {\rm MeV}$ for the $B$-meson decay
constant, which lies in the ball park of lattice and sum-rule
determinations of this quantity.  For the $b$-quark mass in the
$\overline{\rm MS}$ scheme, we use $\overline{m}_b(m_b)=4.25\pm
0.1\,{\rm GeV}$~\cite{Eidelman:2004wy}. For the charm-quark mass we
use $\overline{m}_c=1.1\pm 0.2\,{\rm GeV}$, a range of values that
corresponds to an $\overline{\rm MS}$-mass with a scale between
$m_c$ and $m_b$. 

Note that the branching ratio depends only weakly on the value of the
pole mass $m_b=4.8\,{\rm GeV}$, through the one-loop corrections to $C^A$.  The
pole mass can be eliminated in favor of a low-scale subtracted
$b$-quark mass. For the $B$-meson lifetime, we
use $\tau_B=1.60\,{\rm ps}$.  We use three-loop running for $\alpha_s$
with $\Lambda_{\rm QCD}^{n_f=5}=217\,{\rm MeV}$~\cite{Eidelman:2004wy}
and work with $n_f=4$ below the scale $\mu=m_b$ and $n_f=3$ below the
intermediate scale $\mu=\sqrt{\Lambda_h m_b}=1.55\,{\rm GeV}\approx
m_c$, where $\Lambda_h=0.5\,{\rm GeV}$ represents a typical hadronic
scale. Numerically, using $n_f=3$ or $n_f=4$ makes very little
difference in the results.

\begin{table}
\begin{center}
\begin{tabular}{|ll|ll|}\hline
$m_B$ & $5.28\,{\rm GeV}$ &
$\tau_B=(\tau_{B^+}+\tau_{B^0})/2$ & $1.60\, {\rm ps}$ \\ \hline
$\overline{m}_b(m_b)$ & $4.25\pm 0.1\,{\rm GeV}$ &
$m_b$ & $4.8\,{\rm GeV}$ \\ \hline
 $f_B$ & $200\pm 30\, {\rm MeV}$ &
 $\zeta_{K^*_\perp}$& $0.41\pm 0.04$   \\ \hline
$\lambda_B$ & $460 \pm 110\, {\rm MeV}$ & $\sigma_B$ & $1.4\pm0.4$  \\ \hline
$m_{K^*}$ & $894\, {\rm MeV}$ & 
$f_{K^*_\perp}(1\,{\rm GeV})$ & $170\pm 10 \, {\rm MeV}$ \\ \hline
$a_1(1\,{\rm GeV})$ & $0.1\pm 0.1$ & $a_2(1\,{\rm GeV})$ & $0.1\pm 0.1$ \\ \hline
$\Lambda_{\rm QCD}^{n_f=5}$ & $217\, {\rm MeV}$ & $|V_{cs}^*\,V_{cb}|$
 & $0.040\pm 0.002$ \\
\hline 
$\overline{m}_c$ & $1.1\pm 0.2\, {\rm GeV}$ &  $C_1(m_b)$ & $\phantom{-}1.108\;
 ({\rm LL})$ \\ \hline
$C_7^{({\rm eff})}(m_b)$ & $-0.320\; ({\rm LL})$, $-0.311\; ({\rm NLL})$
 & $C^{({\rm eff})}_8(m_b)$ &
$-0.151\; ({\rm LL})$ \\ \hline
\end{tabular}
\end{center}\vspace*{-0.3cm}
\caption{Numerical input values and uncertainties.
See \cite{Buras:1993xp} for the definition of 
$C^{({\rm eff})}_{7,8}$. Leading-log (LL) accuracy is sufficient for $C_1$ and 
$C^{({\rm eff})}_8$, while we need the next-to-leading-log (NLL) value for 
$C^{({\rm eff})}_7$ in $C^A$. \label{tab:numerical}}
\end{table}

\subsection{Resummation\label{sec:resummation}}

The anomalous dimensions of the SCET$_{\rm I}$ current
operators $J^B$ were calculated in \cite{Hill:2004if}. The anomalous
dimensions of the SCET$_{\rm II}$ four-quark operators are given by the
anomalous dimensions of the meson LCDAs. In both SCET$_{\rm I}$ and 
SCET$_{\rm II}$, the operators are non-local along one or more light-cone 
directions; their anomalous
dimensions are distributions that describe how the operators at
different light-cone coordinates mix among themselves. The necessary
steps to solve the evolution equations were spelled out in detail in
\cite{Hill:2004if} and we refrain from repeating the discussion here.
After resummation, the factorizable part of the amplitude takes the form
\begin{equation}\label{eq:resumm}
{\cal A}_{\rm hard} =\frac{\sqrt{m_B}F(\mu)}{4} \phi_B(\mu)\otimes
f_{K^*_\perp}(\mu)\phi_{K^*_\perp}(\mu)\otimes U_{\rm II}(\mu,\mu_i)
\otimes {\cal J}_\perp(\mu_i)\otimes U_{\rm I}(\mu_i,\mu_h)\otimes
C_1^B(\mu_h)\,, \quad
\end{equation}
where $U_{\rm I}$ evolves the coefficient $C^B_1$ from the hard scale
to the intermediate, hard-collinear scale, and $U_{\rm II}$ describes
the second evolution step down to the low scale of order $1\,{\rm
  GeV}$.  ${\cal A}_{\rm hard}$ is independent of the scales
$\muQCD$, $\mu_h$, $\mu_i$, and $\mu$, up to higher orders in
$\alpha_s$ at these scales when evaluated at fixed order in
perturbation theory. To estimate the uncertainty from higher-order
perturbative contributions, we will independently vary the hard and
the hard-collinear scales by a factor of $\sqrt{2}$ around their
central values $\mu_h=m_b$ and $\mu_i=\sqrt{\Lambda_h m_b}$ with
$\Lambda_h=0.5\,{\rm GeV}$. Throughout, we set $\muQCD=\mu_h$. Let us
stress again that we count $m_c$ as a hard scale and therefore do not
resum perturbative logarithms of ${m_c}/{m_b}$. In view of the fact
that $m_c$ is numerically rather close to the intermediate scale it
might be advantageous to count $m_c$ as hard-collinear instead of
hard, which would allow one to also resum such logarithms. However, an
effective-theory framework for such a treatment is not yet available.

To gauge the size of the corrections to the amplitude, we will express 
them in terms of the leading-order result
\begin{equation}
{\cal A}^{(0)}=C^{A(0)}\,\zeta_{K^*_\perp}
 =-\frac{G_F V_{cs}^* V_{cb}}{\sqrt{2}}\,\frac{e\,}{2\pi^2}\,E_\gamma\,
 \overline{m}_b(m_b)\,C_7^{LL}(m_b)\,\zeta_{K^*_\perp}\,.
\end{equation}
Including $\order(\alpha_s)$ corrections, and normalizing with respect to the 
leading-order result, the contribution of the non-factorizable part is
\begin{equation}\label{eq:AsoftNLL}
{\cal A}_{\rm soft}^{\rm NLL} = C^A\,\zeta_{K^*_\perp}={\cal A}^{(0)}
 \left[ (1.091\mp 0.052\mp 0.027) + i(0.062\mp 0.014\mp 0.016) \right] , 
\end{equation}
where the first uncertainty comes from varying $\muQCD=\mu_h$ and the second
from the variation of $m_c$ in the contribution from $Q_1$.  The larger values 
of ${\cal A}_{\rm soft}$ arise from the lower values of the scale $\mu_h$, 
which we indicate with the symbol ``$\mp$''. The
factorizable part of the amplitude is smaller, similar in size to the
$\order(\alpha_s)$ correction to the non-factorizable part. We find
\begin{equation}\label{eq:AhardLL}
{\cal A}_{\rm hard}^{\rm LL} ={\cal A}^{(0)} \left[
 (0.055\mp0.010\mp0.009\pm 0.019)
 + i(0.031\mp0.005\mp0.004 \pm 0.011) \right] .
\end{equation}
The first uncertainty comes from varying the intermediate scale $\mu_i$, the 
second from varying the hard scale $\mu_h$.   The third uncertainty is
estimated by varying the hadronic input parameters within the ranges
in Table~\ref{tab:numerical}. It is dominated by the uncertainty in
the $B$-meson LCDA and in ${m_c}/{m_b}$, see Table~\ref{tab:hadronic}.

Let us compare this result to what is obtained in a fixed-order calculation. 
Using a common scale $\mu_h=\mu_i=\sqrt{\Lambda_h m_b}$, we find
\begin{equation}
{\cal A}_{\rm hard}={\cal A}^{(0)}\left( 0.116 + i 0.062 \right).
\end{equation}
The above scale choice guarantees the absence of large logarithms in the jet 
function, but will lead to perturbative logarithms $\ln(\mu_i^2/m_b^2)$ 
in $C^B$. We could instead use a large scale $\mu_h=\mu_i=m_b$, which 
eliminates the logarithms from $C^B$ but induces logarithms 
$\ln(\mu_h^2/\Lambda_h m_b)$ in the jet function. 
This gives
\begin{equation}
{\cal A}_{\rm hard} = {\cal A}^{(0)} \left( 0.032 + i 0.019 \right) .
\end{equation}
With this scale choice, the result is more than a factor three smaller.
The fixed-order calculation suffers from large scale uncertainties, 
which are greatly reduced by performing RG improvement.

\begin{table}
\begin{center}
\begin{tabular}{|c||c|c|c|c|c|c|c|c|}\hline
 & $f_B$ & $\lambda_B$ & $\sigma_B$ & $f_{K^*_\perp}$ & $a_1$ & $a_2$ &
 $m_c/m_b$ 
\\ \hline
$10^3\Delta{\cal A}_{\rm hard}/{\cal A}^{(0)}$ & $\pm8\pm5i$ &
 $\mp14\mp8i$ & $\mp3\mp i$ & $\pm3\pm2i$ & $\pm5\pm i$ & $\pm5\mp i$ &
 $^{\phantom{+0}5-7i}_{-10+5i}$ 
\\ \hline
\end{tabular}
\end{center}\vspace*{-0.3cm}
\caption{Hadronic uncertainties in the evaluation of the factorizable
part ${\cal A}_{\rm hard}$ in units of $10^{-3}\,{\cal A}^{(0)}$. The 
parameters are varied within the bounds given in Table~\ref{tab:numerical}. 
We add the symmetrized errors in quadrature to estimate the total hadronic 
uncertainty for which we obtain $\Delta{\cal A}_{\rm hard}={\cal
  A}^{(0)}(0.019\pm0.011i)$.} 
\label{tab:hadronic}
\end{table}

The one-loop corrections to the jet function are beyond the accuracy
of our calculation, since at the same order also the effect of the two-loop 
running and the $\order(\alpha_s)$ corrections to $C^B$ would need to be
included. However, to get an idea of the impact of these corrections,
we evaluate ${\cal A}_{\rm hard}$ including the one-loop jet function on top 
of the leading-order evolution. The effect of perturbative corrections to the
jet function are moderate:
\begin{equation}\label{eq:1loopjet}
{\cal A}_{\rm hard}^{{\rm LL + 1-loop }\,\,
 {\cal J}_\perp}={\cal A}^{(0)} \left[
 (0.063\mp 0.007\mp 0.010\pm 0.023)
 + i(0.034\mp0.003\mp0.004 \pm 0.013) \right] ,
\end{equation}
with the errors arising from the same sources as in (\ref{eq:AhardLL}). The 
uncertainty from varying the renormalization scales $\mu_h$ and $\mu_i$ in the
factorizable part is very small, indicating that perturbative corrections to 
the above result are likely to be small.  

Combining (\ref{eq:AsoftNLL}) and (\ref{eq:1loopjet}), we obtain from 
(\ref{eq:branching}) the branching ratio
\begin{align}\label{eq:finalBr}
{\rm Br}(B\rightarrow K^*\gamma)&= 
\left[6.6 \pm 1.3_{\zeta_{K^*_\perp}} \pm 1.3_{\Lambda/m_b}\pm 0.7_{\rm CKM} \pm
  0.7_{\mu_i, \mu_h} \pm 0.4\right]\times 10^{-5}\,.
\end{align}
We have isolated the uncertainties associated with
$\zeta_{K^*_\perp}$, power corrections, the CKM prefactor, and the renormalization scale;
the latter uncertainty is dominated by the soft term
(\ref{eq:AsoftNLL}). The final
uncertainty in (\ref{eq:finalBr}) is associated with the remaining
input parameters. 
If $\zeta_{K^*_\perp}$ is determined from the tensor form factor, as
discussed after (\ref{eq:zetaV}), the branching ratio comes out lower,
with central value $5.7\times 10^{-5}$.

The uncertainties from power corrections are difficult to quantify. We
have followed standard practice and estimated these corrections by
assuming $\Lambda\approx 500\,{\rm MeV}$, which gives a $10\%$
uncertainty in the amplitude. In the case of charmless two-body decays
classes of enhanced power corrections have been identified which are
larger than this naive estimate \cite{Beneke:1999br,Beneke:2003zv}.
However, since these chirally enhanced and annihilation contributions
are absent in our case we believe the power corrections are of natural
size. In addition to the scale $m_b$, our results involve the
hard-collinear scale $\mu_{hc}\sim\sqrt{\Lambda m_b}$ and the charm
quark mass. The expansion of the amplitude in the inverse of the
hard-collinear scale is quadratic, so that these contributions are of
the same size as the $\Lambda/m_b$ corrections. The situation is
similar for the power corrections associated with the charm quark. We
have analyzed the diagram shown in Figure \ref{fig:Btype}a and find
that the corrections are $\Lambda^2/m_c^2$. 
Similar conclusions were reached in \cite{Bosch:2001gv} based 
on the endpoint behavior of light-cone wavefunctions.   
While we do not have a formal
proof that linear terms are always absent, this seems plausible
since we are integrating out a heavy charm quark. At
leading power in $\Lambda/m_b$, the charm quark corrections are
furthermore suppressed by $\alpha_s$ at the hard or hard-collinear
scale, which compensates for the fact that the Wilson coefficient $C_1$
multiplying these contributions is approximately 3 times larger than the dipole
coefficient $C_7$.

Although we find agreement with previous analytic results in the 
literature~\cite{Beneke:2001at,Bosch:2001gv}, 
there are some differences in the evaluation 
of the final branching fraction (\ref{eq:finalBr}) 
due to hadronic input parameters and to our RG analysis.  The largest
difference comes from the overall normalization given by the input
value of the form factor.  The remaining difference in the soft contributions
from $Q_1$ and $Q_8$ is due to our smaller value of the charm-quark
mass, and to our use of $\zeta_{K^*_\perp}$ for the soft
matrix element in place of $T_1$, which includes (higher-order)
hard-scattering terms unrelated to $Q_1$ and $Q_8$.  For the hard
contributions, the difference is accounted for by $\sim 1\sigma$
variations in the input values of $m_c$, $\lambda_B$ and $a_1$, and by
our inclusion of a complete leading-order RG analysis.~%
\footnote{
For example, in place of the quantity $a_7^c(K^*\gamma)$ in 
equation (55) of \cite{Bosch:2001gv}, we have 
\begin{equation*}
a_7^c(K^*\gamma) \to 
-0.320 + 0.009 + [\zeta_{K^*_\perp}/T_1(0)](-0.098-0.023 i) 
+ C_7^{LL} (0.055+0.031i - 0.042) \,.
\end{equation*}
The first two terms arise from the dipole coefficients $C_7^{LL}$ and 
$C_7^{NLL}$.  The third term represents the 
soft contribution of $Q_1$ and $Q_8$, from (\ref{eq:AsoftNLL}), 
and similarly the fourth term gives the hard contribution of $Q_1$ and $Q_8$, 
from (\ref{eq:AhardLL}). 
}

The $B\rightarrow K^* \gamma$ branching ratio has been accurately
measured by the CLEO \cite{Coan:1999kh}, Belle \cite{Nakao:2004th}, and 
BaBar \cite{Aubert:2004te} collaborations. An average of their
results gives ${\rm Br}(B^0\rightarrow K^{*0}\gamma)=(4.03\pm
0.26)\times 10^{-5}$ and 
${\rm Br}(B^+\rightarrow K^{*+}\gamma)=(4.01\pm 0.20) \times 10^{-5}$. At 
leading power, the factorization formula predicts that the branching ratios 
for the charged and neutral decay are identical.  At subleading power, photon 
emission from the spectator quark breaks the isospin symmetry. This effect 
was estimated in \cite{Kagan:2001zk}. 

Our result for the decay rate is 65\% larger than the
experimental result, or $1.2\,\sigma$ with the errors in (\ref{eq:finalBr}). 
If we were to take such a discrepancy seriously, it would be difficult
to attribute the difference in the exclusive decay to New Physics, 
given that the prediction for the inclusive $b\to s\gamma$ decay agrees well
with the experimental result \cite{Neubert:2004dd}.
In principle, it is possible that New Physics affects the two decay modes
differently: spectator emission is suppressed by $1/m_b^3$ in the
inclusive decay, while it can be leading order in the exclusive decay
if New Physics is present. However, the presence of such operators
would typically lead to large isospin asymmetries, as we discuss
below in Section~\ref{sec:asymm}.  
It is also not plausible that higher-order perturbative
effects could account for the difference. Either the sum-rule result
for $\zeta_{K^*_\perp}$ is $\sim 30\%$ too large, or there are power
corrections of this size which violate the factorization theorem (or
some combination of these possibilities).  If we instead 
use the experimental result for the
branching fraction to determine the non-factorizable part of the form
factor, we obtain $\zeta_{K^*_\perp}=0.31 \pm 0.02$.

\subsection{Isospin and CP asymmetries from New Physics \label{sec:asymm}}

We now consider the effect of the $C$-type operators, which arise from
spectator emission. As discussed earlier, the two operators
$O^C_5$ and $O^C_6$ that contribute to $B\rightarrow K^*\gamma$ have
vanishing Wilson coefficients in the Standard Model. Consequently,
spectator emission is power suppressed in the Standard Model, and the New
Physics effects associated with these operators can lead to isospin
and CP asymmetries that are enhanced over the Standard Model
predictions.

Because the jet-functions of the $C$-type operators are proportional
to the charge of the spectator quark, the presence of these operators
induces an asymmetry between the charged and neutral decay modes:
\begin{align}
\label{eq:ios}
&\Gamma(B^+\rightarrow K^{*-}\gamma) - \Gamma(B^0\rightarrow K^{*0}\gamma)
 \nonumber\\
&\quad \approx {m_B^3 f_{K^*_\perp}^2 f_B^2 \over 16\pi \lambda_B^2 } \left[ 
e_u^2 (|\hat C^u_5|^2+|\hat C^u_6|^2)-e_d^2 (|\hat C^d_5|^2+|\hat C^d_6|^2)
 \right]\\  
&\quad = \frac{1.5 \times 10^{-3}}{\tau_B}
 \left( |\hat C^u_5|^2+ |\hat C^u_6|^2-\frac{1}{4} |\hat C^d_5|^2
 -\frac{1}{4} |\hat C^d_6|^2 \right) \times {\rm TeV}^4 \,.   \nonumber
\end{align}
In general, the Wilson coefficients $C^{u,d}_{5,6}$  are
functions of the light-cone momentum fraction of the light current
quark and are convoluted with the LCDA of the light meson: ${\hat C}\equiv
\int_0^1 du\, \phi_{K^*_\perp}(u) C(u)$.  If the New Physics takes the
form of four-quark operators at the hard scale, the coefficients will
be constant and ${\hat C}_{5,6}=C_{5,6}$. If the New Physics is
isospin symmetric, then $C^{u}_{i}=C^{d}_{i}$ and the difference
(\ref{eq:ios}) is positive. However, in general the operators entering
the charged and neutral decays can have different Wilson coefficients.
Experimentally, the difference between the decay rates, normalized by the 
average lifetime $\tau_B$ as in (\ref{eq:ios}),  is
$(-3\pm 3)\times 10^{-6}$, where we take a weighted average of the 
values from \cite{Coan:1999kh}, \cite{Nakao:2004th} and \cite{Aubert:2004te}. 
The most recent estimate of the difference in the Standard Model is 
$(-5 \pm 3 )\times 10^{-6}$~\cite{Beneke:2004dp}.
From (\ref{eq:ios}), we find that the Wilson coefficients evaluated at 
renormalization scale $\mu\sim m_b$ must obey 
$|\hat{C}_{5,6}| \lesssim 5 \times 10^{-2}\,{\rm TeV}^{-2}$, meaning that
even with present precision one is able to probe New Physics effects at
scales of several TeV.
 
Let us now turn to the time-dependent CP asymmetry. The $B^0$ and
$\bar B^0$ mesons can decay into $|K_L^*\,\gamma_L\rangle$ or
$|K_R^*\,\gamma_R\rangle$. These two states can be combined into a
CP-even and a CP-odd final state, and since the spins are not measured, the 
experiments give the sum of the two rates. Expressed in terms of the 
amplitudes ${\cal A}_{L,R}={\cal A}(B^0\rightarrow K^* \gamma_{L,R})$ and 
$\bar{\cal A}_{L,R}={\cal A}(\bar{B}^0\rightarrow K^* \gamma_{L,R})$, the 
time-dependent CP asymmetry is
\begin{align}
  A_{CP}&=\frac{\Gamma(B^0(t)\rightarrow K^*\gamma)-\Gamma({\bar
      B}^0(t)\rightarrow K^*\gamma)}{\Gamma(B^0(t)\rightarrow
    K^*\gamma)+\Gamma({\bar B}^0(t)\rightarrow K^*\gamma)}\\
  &=\frac{\left[|{\cal A}_L|^2+|{\cal A}_R|^2-|\bar {\cal A}_L|^2
   -|\bar {\cal A}_R|^2\right]
    \cos(\Delta m_B t)-2\,{\rm Im}\left[ \displaystyle
    \frac{q}{p}(\bar {\cal A}_L
      {\cal A}_L^*+\bar {\cal A}_R {\cal A}_R^*)\right]\sin(\Delta m_B
    t)}{|{\cal A}_L|^2+|{\cal A}_R|^2+|\bar {\cal A}_L|^2
 +|\bar {\cal A}_R|^2}\,.\nonumber
\end{align}
The coefficients $p$ and $q$ relate the mass to the flavor eigenstates: 
$|B_{H,L}\rangle=p |B^0\rangle\pm q|\bar B^0\rangle$. In deriving the
above expression, we have assumed $|q/p|= 1$. This holds to
good approximation, since the width difference in the $B_d$ system is
very small.

In the Standard Model ${\cal A}_L$ and $\bar {\cal A}_R$ vanish to leading 
power in $1/m_b$.  The coefficient of $\sin(\Delta m_B t)$ is thus power
suppressed.  The prefactor of $\cos(\Delta m_B t)$ also happens to be
small in the Standard Model. The reason is that, up to terms which are
doubly Cabibbo suppressed, the $b\rightarrow s$ amplitude has only a
single weak phase, so that there is no CP violation in the decay. The
direct CP asymmetry from the Cabibbo suppressed terms is
\begin{align}
{|{\cal A}_R|^2 - |\bar{\cal A}_L|^2 \over 
|{\cal A}_R|^2 + |\bar{\cal A}_L|^2}
&= 2\, {\rm Im}\left( {V_{ub} V_{us}^* \over V_{cb}V_{cs}^*} \right) {\rm Im} 
\left( {\langle K^*\gamma |\,{ C}_1 Q_1^u + { C}_2 Q_2^u + \sum_{i=3}^8
  { C}_i Q_i |\bar B^0\rangle  \over 
 \langle K^*\gamma |\, { C}_1 Q_1^c + { C}_2 Q_2^c + \sum_{i=3}^8 { C}_i Q_i |\bar B^0\rangle} \right)
 +\order(\lambda_C^4)\\
&=\eta \lambda_C^2\left[ -  0.14 \pm 0.03_{\Lambda/m_b}\pm 
    0.03_{\mu_i,\mu_h} \pm 0.04_{m_c} \pm 0.02
    \right]+\order(\lambda_C^4)  \,.\nonumber
\end{align}
Power corrections are estimated as $20\%$ of the leading result.  
Uncertainties associated
with scale variation and the charm-quark mass are indicated explicitly, 
and the final uncertainty is due to the remaining input parameters.  
At leading power, the asymmetry is identical for neutral and charged $B$ 
mesons. Here $\lambda_C=V_{us}\approx 0.22$ is the sine of the Cabibbo angle, 
and the Wolfenstein parameter $\eta$ is related to the imaginary part of
$V_{ub}$. The CP violation in the Standard Model is thus negligible.
New Physics can change these predictions rather dramatically. If the
New Physics has operators that induce leading-order contributions to the 
decay amplitudes ${\cal A}_L$ and
$\bar {\cal A}_R$, large CP-violation effects from the interference of mixing
and decay can be observed, even if the New Physics operators do not
have a new CP-violating phase~\cite{Atwood:1997zr}. In the presence of New Physics
operators with additional phases, asymmetries in the decay can also
occur. Note that the operators $O^C_5$ and $O^C_6$ can contribute in
both cases. These operators are suppressed by $1/m_b^3$ in the
inclusive decay. A difference in the exclusive and inclusive direct CP
asymmetries could be explained by the presence of such operators.

Recently, both BaBar \cite{Aubert:2004pe} and Belle
\cite{Ushiroda:2005sb} have performed measurements of the
time-dependent CP asymmetry in the $B\rightarrow K^*\gamma$ decay.
Within large errors, their results are consistent with a vanishing CP
asymmetry.

\section{Discussion and conclusions\label{section:summary}}

We have established the factorization formula (\ref{eq:factorization})
for $B\to V\gamma$ decays, which provides the basis for the
phenomenological analysis of the decays $B\to K^*\gamma$ and
$B\to\rho\gamma$. To perform the diagrammatic analysis of the
factorization properties of the amplitude, we studied a current
correlation function from which the $B\to K^*\gamma$ amplitude can be
extracted.  The diagrams contributing to the correlator are expanded
around the heavy-quark limit using the strategy of regions. The
different momentum regions are represented by corresponding fields in
the effective theory. Similar to the case of two-body decays, such as
$B\to \pi\pi$, the decay $B\to K^*\gamma$ involves energetic partons
propagating in two directions. It is then necessary to introduce collinear 
fields along the light-meson direction as well as the photon direction,
making the effective theory analysis more involved than in the case of
the heavy-to-light form factors.  A large number of possible operator
structures appear and it becomes crucial to have an efficient way of
identifying the relevant operators both in SCET$_{\rm I}$ and
SCET$_{\rm II}$.  There are three complications that make the
construction of the operators non-trivial: (i) the power counting in
the two effective theories is different and the power of an operator
in SCET$_{\rm II}$ does not impose a strong constraint on the SCET$_{\rm I}$ 
operators that match onto it; (ii) the SCET$_{\rm II}$
operators can contain inverse soft derivatives which count as inverse
powers of the expansion parameter, and (iii) two collinear
sectors are present. We have set up an efficient formalism to construct the
operators which addresses all three issues.  We classify the
SCET$_{\rm I}$ operators by their dimension instead of their power in
$\lambda$ and derive a constraint on the maximum dimension of the
SCET$_{\rm I}$ operators beyond which they cannot match onto leading
SCET$_{\rm II}$ operators.  We construct the SCET$_{\rm II}$ operators
from gauge-covariant and boost-invariant building blocks.
The use of these building blocks makes it simple to identify how many
inverse soft derivatives can occur.  Finally, we separately match the 
collinear fields from the two sectors to account for the structure of the
SCET$_{\rm I}$ Lagrangian. 
Once the operators are identified, it becomes possible to make
all-orders statements concerning factorization by identifying which
classes of operators are insensitive to infrared momentum regions.
Our analysis shows that the $B\to K^*\gamma$ amplitude indeed takes
the form of the factorization formula (\ref{eq:factorization}).

The basis of effective-theory operators which we constructed for the
factorization proof can be used to analyze the effects of New
Physics in the decay $B\to V\gamma$.  These operators also
describe decays with a flavor-singlet final-state meson, and the related 
decay processes $B^*\rightarrow P\gamma$ of $B^*$ vector mesons.  The 
New Physics operators yield
calculable contributions at leading power to the isospin asymmetry and
the time-dependent CP asymmetry in $B\to K^*\gamma$, so that measurements
of these asymmetries provide useful constraints on the corresponding
New Physics operators.  For flavor non-singlet final-state mesons, we
found that the $B^*\rightarrow P\gamma$ amplitudes obey the
generalized factorization formula (\ref{eq:factorgeneral}). For the
flavor-singlet case, however, the $B^*\to P\gamma$ decay amplitude 
does {\it not}
obey a factorization formula of the type (\ref{eq:factorgeneral}),
highlighting the non-trivial nature of the $B\to V\gamma$ factorization
formula.  More precisely, only the $B^*\to P\gamma_L$ mode violates
the factorization formula.  The power counting of soft-collinear
fields makes apparent the sensitivity to infrared momentum regions at
leading power, and consequently the breakdown of factorization in this
case.  In contrast, the $B^*\to P\gamma_R$ mode is completely
factorizable, involving only the second term of
(\ref{eq:factorgeneral}), and the same power-counting arguments show
that the entire amplitude in this case is perturbatively calculable as
a convergent convolution integral over meson LCDAs.

The effective theory approach allows us to disentangle the different
scales in these decay processes and to resum perturbative logarithms
of these scales by solving the RG equations in the effective theory.
We reanalyzed the $B\to K^*\gamma$ branching ratio with recent values
for hadronic input parameters taken from light-cone sum rules, and
presented the first analysis of the hard-scattering terms to
leading-order in RG-improved perturbation theory. We also evaluated
the impact of perturbative corrections to the jet function, which are
potentially significant because they arise at a low scale $\mu\approx
1.5~{\rm GeV}$. Parts of these corrections were identified in the
diagrammatic analysis of \cite{Descotes-Genon:2004hd}, which isolated
large logarithms occurring in the hard-scattering kernels.  The result
for these logarithms is precisely reproduced by integrating the
one-loop jet function ${\cal J}_\perp$ in
\cite{Becher:2004kk,Hill:2004if} over the tree-level hard-scale
coefficient.  In our analysis the leading logarithms are resummed and
the remaining one-loop jet-function correction increases the
factorizable part of the amplitude by approximately
15\%. Depending on which QCD form factor is used to determine the
SCET quantity $\zeta_{K^*_\perp}$, our result for the $B\to K^*\gamma$ 
branching fraction differs
from the experimental value by $1\sigma-2\sigma$. The discrepancy is
smaller than in earlier evaluations, mostly because the new sum-rule
values for the $B\rightarrow K^*_\perp$ form factors are somewhat
smaller than earlier results.

The jet function appearing in $B\to V\gamma$ is identical to that appearing 
in $B\to V_\perp$ form factors. The universality of the jet function gives 
rise to new symmetry relations between different form factors when 
perturbative corrections at the hard scale are neglected. The reason is that 
the SCET$_{\rm I}$ Wilson coefficients for the form factors are constant at
tree level. In this approximation, the integral of the jet function over meson 
LCDAs yields a universal quantity, identical for all form factors
describing the same final-state meson~\cite{Hill:2004if}. The same integral 
over the jet-function also appears in processes such as $B\to\pi\pi$ at the
same level of approximation~\cite{Bauer:2004tj}. In contrast, for 
$B\to V\gamma$ the tree-level hard-scale coefficient is not a constant.  
Here, even in this leading-order approximation 
the universal hadronic parameters must be taken as the meson LCDAs,
and the perturbative expansion of the jet function is essential to
retain predictive power in the factorization formula.

In summary, we have presented the first factorization analysis for a
charmless $B$ meson decay that addresses both the construction of the
operator basis and the factorization properties of the matrix elements
in a systematic way.  The same techniques can be used to establish
factorization for hadronic decays such as $B\to\pi\pi$.

\subsection*{Acknowledgments}
We are grateful to Ben Pecjak for collaboration in the early stages of this 
work, and to Martin Beneke for useful discussions.
We would like to thank the Institute for Advanced Study (Princeton, NJ) for
hospitality and support during the fall term 2004. The research of 
M.N.\ is supported by the National Science Foundation under Grant PHY-0355005, 
and by the Department of Energy under Grant DE-FG02-90ER40542. Fermilab is 
operated by Universities Research Association Inc.\ under Contract 
No.~DE-AC02-76CH03000 with the U.S.\ Department of Energy. The research of 
R.J.H. is supported by the Department of Energy under Grant DE-AC02-76SF00515.

\appendix
\section{Matching coefficients\label{sec:formulae}}

In the matching onto the SCET$_{\rm I}$ current operators, the following
combinations of Wilson coefficients appear:
\begin{align}
2C_{T1}^A -{1\over 2}C_{T2}^A - C_{T3}^A &= \nonumber\\
 2 + {\alpha_s C_F \over
  4\pi}\,&\bigg[  - 4 \ln^2\!\!\frac{\mu }{x\, m_b}-10 \ln\!\frac{\mu}{m_b}
 - 4 \ln\!\frac{\muQCD}{x\, m_b} - 4 {\rm Li}_2(1 - x)-12 -
 \frac{\pi^2}{6} \bigg]+{\mathcal O}(\alpha_s^2) \,, \nonumber\\
{1\over 2}C_{T6}^{B'} +C_{T7}^{B'} &= 2x+{\mathcal O}(\alpha_s) \,, \\
{1\over 2}C_{T2}^{B'} + C_{T3}^{B'} &= {\mathcal O}(\alpha_s)\,, \nonumber
\end{align}
where $x=2E/m_b$. The A-type
coefficients can be found in \cite{Bauer:2000yr, Beneke:2004rc}. The
${\mathcal O}(\alpha_s)$ term for the $B$-type coefficients are also
known, see \cite{Beneke:2004rc,Becher:2004kk}. However, they appear only at
${\mathcal O}(\alpha_s^2)$ in the decay rate, since they are
multiplied by the jet function, which is proportional to $\alpha_s$.

We collect the known results from the literature for the matching 
coefficients appearing in Section~\ref{sec:Matching}. For the matching of 
$Q_1$ and $Q_8$ onto A-type current operators, we need the functions
\begin{align}
G_8 &=\frac{8}{3}\,\ln\frac{\muQCD}{m_b}+\frac{11}{3} +
\frac{2\pi i }{3} - \frac{2 \pi^2}{9} \nonumber\\
G_1(x) &= -\frac{104}{27}\ln\frac{\muQCD}{m_b} -\frac{833}{162}
 -\frac{20i\pi}{27}  +\frac{8\pi^2}{9} x^{3/2} \nonumber\\
        & +\frac{2}{9} \bigg[ 48+30i\pi-5\pi^2-2i\pi^3 -36\zeta (3) 
          +\left( 36+6i\pi-9\pi^2\right)\ln x \nonumber\\
        & \qquad\;\;\, +\left( 3+6i\pi\right) \ln^2\! x+\ln^3\! x \bigg] x 
           \nonumber\\
        & +\frac{2}{9} \bigg[ 18+2\pi^2 -2i\pi^3 
         +\left( 12-6\pi^2 \right)\ln x +6i\pi\ln^2\! x+\ln^3\! x\bigg] x^2 
           \nonumber\\
        & +\frac{1}{27} \bigg[ -9+112 i\pi-14\pi^2
          +\left(182-48i\pi\right)\ln x-126\ln^2\! x\bigg] x^3 +
          \order(x^4) \,, 
\end{align}
which were defined in \cite{Bosch:2001gv} and deduced from the results
of \cite{Greub:1996tg,Buras:2001mq}. Here $x$ is a ratio of squared masses, 
$x_q=m_q^2/m_b^2$, for a quark of flavor $q$. 
The matching of $Q_1$ onto $B$- and $C$-type operators involves the functions
\begin{equation}
f(x) =\left\{\begin{aligned} &1  + 4x\left[{\rm
        arctanh}(\sqrt{1-4x})-i \frac{\pi}{2}\right]^2 \,;
 &&\text{ for } x < 1/4 \,, \\
&1-4x\,\arctan^2\frac{1}{\sqrt{4x-1}} \,; &&\text{ for } x\geq 1/4 \,,
\end{aligned}\right.
\end{equation}
and \cite{Beneke:2003zv}
\begin{align}
 G(x,u) &= -4\int_0^1\!dv\,v(1-v) \ln[x-v(1-v)u] \nonumber\\
   &= \frac{2(12x+5u-3u\ln x)}{9u}
    - \frac{4\sqrt{4x-u}\,(2x+u)}{3u^{3/2}}
    \arctan\sqrt{\frac{u}{4x-u}} \,. 
\end{align}

\end{document}